\documentclass[superscriptaddress, nofootinbib, preprintnumbers, twocolumn]{revtex4}

\usepackage{amsmath}	
\usepackage{amsthm}		
\usepackage{amssymb}	
\usepackage{eufrak}
\usepackage{datetime}
\usepackage{graphicx}
\usepackage{xcolor}
\usepackage{verbatim}
\usepackage{bm}
\usepackage{float}

\usepackage[colorlinks=true, citecolor=midblue, linkcolor=midblue, urlcolor=midblue]{hyperref}

\usepackage{setspace}

\definecolor{grey}{rgb}{0.4,0.4,0.4}
\definecolor{dullmagenta}{rgb}{0.4,0,0.4}
\definecolor{darkblue}{rgb}{0,0,0.4}
\definecolor{midblue}{rgb}{0,0,0.5}
\definecolor{midred}{rgb}{0.5,0,0}
\definecolor{orange}{rgb}{1,0.5,0}
\definecolor{lightbrown}{rgb}{0.75,0.5,0.25}
\definecolor{tan}{cmyk}{0.14,0.42,0.56,0}
\definecolor{djunglegreen}{cmyk}{0.99,0,0.52,0}
\definecolor{lightgreen}{rgb}{0,1,0}
\definecolor{olivegreen}{cmyk}{0.64,0,0.95,0.40}
\definecolor{midgreen}{rgb}{0.0,0.675,0.0}
\definecolor{darkgreen}{rgb}{0,0.5,0}

\newcommand{\ie}{i.e.}

\smallskipamount

\settimeformat{ampmtime}

\usepackage{float}

\usepackage{amsmath}	
\usepackage{amsthm}		
\usepackage{amssymb}	
\usepackage{tikz}
\usetikzlibrary{decorations.pathmorphing, decorations.markings, positioning}
\usepackage[compat=1.1.0]{tikz-feynman}
\usepackage{orcidlink}
\usepackage{multirow}

\begin{document}

\title{GUT Color-Triplet Around the Corner: \\
The Story of an Overlooked Lightness} 
\author{Gia Dvali}
\email{georgi.dvali@lmu.de}
\affiliation{
	Arnold Sommerfeld Center,
	Ludwig-Maximilians-Universit{\"a}t,
	Theresienstra{\ss}e 37,
	80333 M{\"u}nchen,
	Germany,}
\affiliation{
Max-Planck-Institut f\"ur Physik, Boltzmannstr. 8, 85748 Garching, Germany}

\author{Manuel Ettengruber\,\orcidlink{0000-0001-7331-6370}}
\email{manuel-meinrad.ettengruber@ipht.fr}
\affiliation{Universit{\'e} Paris--Saclay, CNRS, CEA, Institut de Physique Theorique, 91191, Gif-sur-Yvette, France
}
\author{Anna Jankowsky}
\email{annasky@mpp.mpg.de}
\affiliation{
	Arnold Sommerfeld Center,
	Ludwig-Maximilians-Universit{\"a}t,
	Theresienstra{\ss}e 37,
	80333 M{\"u}nchen,
	Germany,}
\affiliation{
Max-Planck-Institut f\"ur Physik, Boltzmannstr. 8, 85748 Garching, Germany}

\author{Tongxuan Zhang}
\email{Tongxuan.Zhang@lmu.de}
\affiliation{
	Arnold Sommerfeld Center,
	Ludwig-Maximilians-Universit{\"a}t,
	Theresienstra{\ss}e 37,
	80333 M{\"u}nchen,
	Germany,}
\affiliation{
Max-Planck-Institut f\"ur Physik, Boltzmannstr. 8, 85748 Garching, Germany}


\begin{abstract}
  In traditional approach to the doublet-triplet splitting problem in grand unified theories (GUTs) it is assumed that color-triplet partner of the Higgs doublet must be superheavy in order not to mediate proton decay. However, it has been proposed~\cite{dvali1992} that the splitting, instead of the mass sector, can occur among the Yukawa couplings. In this scenario the color-triplet emerges as a long-lived $T$-particle with the mass around the TeV scale. This framework provides a direct low-energy window into GUT physics.
  In the present paper, we study an explicit  
  supersymmetric $SO(10)$  GUT which realizes this mechanism, including effects of 
  gravity-mediated supersymmetry breaking. We show how the shift 
  of the heavy fields, triggered by
  supersymmetry-breaking, generates a tiny coupling between the $T$-particle and the Standard Model fields and explore the resulting phenomenology. 
 Depending on the order of gravitino mass in which the coupling emerges, the lifetime of the $T$-particle is either around a second or exceeds the age of the Universe. 
   In our analysis, we validate some earlier constraints on (super)gravity vacua imposed   by the $S$-matrix: based on the 
  Goldstino Argument, we show that none of the GUT vacua have positive energy densities, whereas the Minkowski vacua are safe against tunneling. We also highlight the role of the species scale cutoff in sensitivity of the low energy observables towards the GUT physics.


\end{abstract}

\maketitle




\section{Introduction}
\label{sec:Introduction}

 The universal prediction of grand unified 
 theories is the existence 
 of a color-triplet partner, $T$, of the electroweak Higgs doublet, H.   The very concept of unification of the Standard Model into a compact covering group demands that 
 $H$ and $T$ belong to the same irreducible representation 
 of the GUT symmetry. For example, in the minimal 
 $SU(5)$ GUT, $H$ and $T$ reside  in the Higgs $5$-plet transforming as a fundamental representation of $SU(5)$.  
  In supersymmetric extensions of GUTs, the Higgs scalar
  content gets doubled - $H, T, \bar{H},\bar{T}$ - and they also acquire the fermionic partners.
 These are introduced in form of chiral superfields
 which shall be denoted by the same symbols 
 as their scalar components. The fermionic partners of the colored triplets shall be denoted by 
 $\tilde{T}, \tilde{\bar{T}}$ and be referred to as 
 ``triplinos".  
 The colored states, including their supersymmetric partners, shall be called the $T$-particles.  
      
 This model-independent feature of the GUT framework  comes with a puzzle, usually referred to as the
 \textit{doublet-triplet splitting 
 problem}.  The point is that in the limit of an unbroken 
 GUT symmetry,  the $T$-particles share the Yukawa couplings (to the Standard Model fermions) with their Higgs doublet partners.
 However, unlike the Higgs doublets, the 
 $T$-particles are \textit{leptoquarks}, as they 
 trigger both quark-quark and quark-lepton conversions.  Due to this, the exchange of $T$-particles can mediate the baryon and lepton number violating processes. In particular, they induce the proton decay. 
 
  The traditional approach to this problem  assumes 
  that  the breaking of GUT symmetry splits 
  the masses of doublet and triplet in the way that 
  suppresses the proton decay. 
  Namely, in this picture, while the mass of the doublet is 
   $\sim 100$ GeV, the triplet mass is of order the GUT scale, 
   $M_{\textrm{GUT}} \sim 10^{16}$ GeV. The explicit mechanism of the mass-splitting is model-dependent.    
 For example, it can be achieved at the expense
    of a fine-tuning, due to  group theoretic reasons, 
    or dynamically.  
    Regardless of its origin, 
    the GUT-scale mass of the color-triplet would 
    eliminate any prospect of its direct detection as well as its cosmological relevance.  
    
    However, an alternative scenario was put forward in \cite{dvali1992}, with further  detailed  construction of explicit models given in \cite{Dvali:1995hp}. The new idea of these proposals was that the doublet-triplet splitting   
  problem can be shifted from the masses to the couplings. 
   That is, the breaking of GUT symmetry, instead of masses,  splits the  Yukawa couplings of  $T$ and $H$ in such a way that the color-triplet is effectively decoupled from the Standard Model fermions. 
    At the same time, 
    there is no need for an extraordinary mass 
    splitting between the triplet and the doublet fields. 
     The tree-level masses can remain degenerate, 
     with moderate mass splitting induced by higher order effects such as  
   the RG running.

   As a result, in this scenario, the color-triplet partners 
   $T, \bar{T}$ of the Higgs doublets $H,\bar{H}$ emerge as light particles 
    with their masses in the TeV range.
    At the same time, the couplings of $T$-particles with quarks and leptons are extraordinarily 
    suppressed.   
        Due to these features, the phenomenological 
   signatures of $T$-particles 
   are rather spectacular.   In particular, thanks to their color charge, they can be produced in colliders and form 
   long-lived hadronic bound states with ordinary quarks.    
    Thus, the decoupled color-triplet theory~\cite{dvali1992, Dvali:1995hp} 
    opens up a possible low-energy window 
   to grand unification, something that is impossible 
    in the standard approaches to the doublet-triplet splitting problem. 
 
 This scenario
 has been further refined in a number of papers. 
  In particular, the implementation of the same idea within $SU(5)$ theory was discussed in \cite{Dvali:1996hs}.  
 As shown in \cite{Dvali:2024jhs}, the existence of light color-triplets with extremely suppressed Yukawa couplings provides a novel 
 avenue for correlated low-energy searches of GUT scale physics. In this picture, the processes mediated by the $T$-particles, such as 
 the proton decay and neutron ``disappearance", 
 are correlated with its direct production and subsequent decay in particle collisions at $\sim $TeV energies. This opens the prospect of  observable correlated signatures between three types of experimental searches, at particle colliders, proton decay searches, and studies of the cold neutrons.   
 Discussions of various aspects of the decoupled-triplet scenario can be found in~\cite{dvali1992, Dvali:1995hp, Dvali:1996hs,  Gogoladze:1995sd, Dvali:1997qv, Berezhiani:1998hg, Bajc:2002bv, Rakshit:2003wj, Dorsner:2006ye, Belotsky:2008se, Berezhiani:2011uu, Dvali:2020uvd, Haba:2002if, Dorsner:2024seb, Dvali:2024jhs}.

However, the phenomenological and cosmological implications of the $T$-particles crucially hang 
on the values of their Yukawa couplings to the Standard Model superfields. 
In the original realizations of the decoupled color-triplet scenario \cite{dvali1992, Dvali:1995hp}, in the supersymmetric vacuum, the 
$T,\bar{T}$-superfields  have vanishing Yukawa couplings with the Standard Model chiral superfields. This is due to the group-theoretic structure of the supersymmetric vacuum in which the $SO(10)$ symmetry is broken to the Standard Model symmetry group.
The tiny non-zero Yukawa couplings can only be generated after supersymmetry breaking. 
In the previous articles it was assumed that 
the coupling is generated at the order 
$\frac{m_{3/2}}{M_{\textrm{GUT}}}$, where 
$m_{3/2}$ is the gravitino mass, which also sets 
the mass splitting among the Standard Model  
fields and their superpartners in scenarious with  gravity-mediated supersymmetry breaking. 
This is because SUSY-breaking can induce the shift of the vacuum expectation values (VEVs) of the GUT fields that connect 
 $T$-particles with the Standard Model fermions. 

 However, this assumption has never been verified explicitly.  This determines one of the goals of the present paper. Namely, we study the generation of the $T$-couplings via gravity-mediated SUSY breaking.  As usual, in this scenario  SUSY is broken in some hidden sector without any direct coupling to the Standard Model in the superpotential. The breaking is mediated to the visible sector via (super)gravity. 
 Within this framework, we quantify the
 generation of the $T$-couplings by employing the knowledge previously obtained in 
\cite{Dvali:2020uvd}.

  Namely, in~\cite{Dvali:2020uvd} a universal argument based on Goldstino properties was 
  introduced which imposes certain general constraints on the VEVs in supergravity. We shall refer to it as the \textit{Goldstino Argument}. In particular, this argument shows that in gravity-mediated supersymmetry breaking the VEVs of the heavy GUT Higgs fields become shifted relative to their supersymmetric values by the amount that is bounded from above by the gravitino mass, i.e., 
  \begin{equation} \label{shiftVEV}
  \Delta ({\rm VEV}_{\textrm{GUT}})  \lesssim m_{3/2} \,.    
  \end{equation} 
 In the absence of a protective symmetry, the shift saturates the above bound and is of order $m_{3/2}$. In contrast, in the case of a protective symmetry, the shift may occur in higher order in $\frac{m_{3/2}}{M_{\textrm{GUT}}}$, or not at all.  
   
 Taking into account this general effect,  we show that it induces a tiny coupling of the color-triplet to the Standard Model fields, which is bounded from above by $\frac{m_{3/2}}{M_{\textrm{GUT}}}$.  This implies that the color-triplet is an extremely long-lived particle with distinct phenomenological signatures and cosmological implications.  We give explicit scenarios in which the Yukawa couplings of $T$-particles occur in the order $\frac{m_{3/2}}{M_{\textrm{GUT}}}$ or higher. We refer to these scenarios as the 
 Model~I and the Model~II, respectively. 
 In these scenarios the lifetimes of $T$-particles are vastly different.  
 
 In cases represented by the Model~I, in which the triplet Yukawa couplings materialize in the order $\frac{m_{3/2}}{M_{\textrm{GUT}}}$, the lifetime of the $T$-particle is around a second. 
 
 In contrast, the generation of the coupling 
 in the order $\left(\frac{m_{3/2}}{M_{\textrm{GUT}}} \right )^2$, or higher, endows the $T$-particle with a lifetime largely exceeding the age of the Universe. 
 An explicit scenario of this type is provided by the Model~II.  
  
 Next, we study various aspects of the vacuum structure. 
This includes identifying the supersymmetric vacuum with the correct symmetry-breaking pattern as well as
studying its modification induced by the effects of supergravity and by gravity-mediated supersymmetry-breaking.  
We discuss the energy splitting between various vacua and absence of tunneling among them. 

   Our study of the vacuum structure of the theory reinforces, within explicit GUT framework, certain previously proposed consistency criteria of (super)gravity theories. 
    The first one is the \textit{no-de Sitter} proposal put forward in ~\cite{Dvali:2013eja, Dvali:2014gua, Dvali:2017eba}. This criterion states that a consistent theory of gravity  excludes  any stable or  meta-stable vacuum with positive energy density.  A deep physical foundation of this constraint is traced to the consistency of $S$-matrix formulation of quantum gravity, currently the only existing formulation both in field theory as well as in string theory~\cite{Dvali:2020etd}.  
    In the present paper, using the \textit{Goldstino Argument} of~\cite{Dvali:2020uvd}, we show how the supergravity GUT enforces the \textit{no-de Sitter} condition on the theory vacua.  
    
     Another proposal confirmed within the validity of our analysis is the 
  criterion of  \textit{Minkowski safety} proposed in~\cite{Dvali:2011wk}. According to it, 
   in a consistent theory of gravity no tunneling  from the Minkowski vacuum can take place.  That is, the tunneling rate must be strictly zero.  
   This requirement, too, is linked  with the consistency of the $S$-matrix formulation. 
   
   Again, taking into account the \textit{Goldstino Argument}, we show  that  up to accuracy $\sim \frac{m_{3/2}}{M_{\rm GUT}}$  the theory is in no-tunneling regime, in agreement with the  \textit{Minkowski safety} criterion.

In the present paper, we also show that the degrees of freedom 
  that guarantee the decoupling of the color-triplet field from the Standard Model fermions, at the same time, can eliminate the problematic relations between the fermion masses characteristic of grand unification. 

The paper is organized as follows. 
   In section \ref{sec:Model} we display an explicit $SO(10)$-model, the Model~I, realizing the light color-triplet scenario. An alternative version of the model, the Model~II, is given in appendix \ref{appD}. 
Section \ref{sec:theory} is a review of the general features of the decoupled light color-triplet, stating its mass range, the decay-rate and the minimal lifetime.  
  In particular, it is shown that the predicted lifetimes of the $T$-particles in 
  Model~I and Model~II can differ by more than $26$ orders of magnitude.    
The collider phenomenology of $T$-particles is outlined in section \ref{sec:Meson-decays}, differentiating among the $T$-number-preserving and $T$-number-violating decays. 
In section \ref{sec:cosmobounds} we discuss the cosmological bounds and summarize our results, presenting the relevant parameter space of the $T$-particles. An example of a parameter choice leading  to a cosmologically-stable  $T$-meson is given in section \ref{sec:parameterexample}.
In section \ref{sec:vacua} we perform the analysis of the vacuum 
structure and extract effects that are applicable to a generic supergravity GUT framework. In particular, we validate the
\textit{no de Sitter} as well as the  \textit{Minkowski safety} criteria. 
We conclude in section \ref{sec:conclusion}.

\section{An Explicit Model: Model~\textrm{I}}
\label{sec:Model}

\noindent For a better illustration of the idea, we give some explicit examples of the model that realizes the light color-triplet scenario \cite{dvali1992}.  In this section, we present the Model~\textrm{I}. 
We split the discussion of the model in two stages. First, performing the explicit minimization, we 
identify the supersymmetric vacuum in which the GUT symmetry is broken down to the Standard Model and the couplings (but not the masses) of the Higgs doublet and its color-triplet partner are split.
The doublet has ordinary Yukawa couplings, necessary for giving the masses to quarks and leptons, whereas 
triplet is decoupled. 

We then take into account the supersymmetry breaking effects mediated by supergravity. This breaking can trigger the shift of the GUT VEVs. 
The universal Goldstino argument of~\cite{Dvali:2020uvd}
shows that the shift 
is bounded from above by (\ref{shiftVEV}). The detailed discussion of this effect is given in the Appendix \ref{appB}. 

Applying the above general mechanism, we show that in the model presented in the current section (the Model \textrm{I})
the heavy GUT VEVs are shifted by order of the gravitino mass $m_{3/2}$.  This shift endows the color-triplets $T,\bar{T}$ with the tiny Yukawa couplings to quark and lepton superfields in the  order $\frac{m_{3/2}}{M_{\textrm{GUT}}}$. 
An alternative model, the Model~\textrm{II}, is presented in the appendix \ref{appD}. As shown there, in the Model~\textrm{II}, for the minimal K\"ahler function, no Yukawa coupling for $T,\bar{T}$
is generated in the order $\frac{m_{3/2}}{M_{\textrm{GUT}}}$. However, the non-minimal 
terms in K\"ahler can change the story. 

 Before moving on to the description of explicit models, we briefly formulate our  criterion of naturalness for the superpotential. As it is well known, the  superpotential is protected from perturbative loop corrections. This introduces the notion of \textit{technical naturalness}, which implies that one does not need to justify the form of the superpotential by any additional symmetry. Over decades, this criterion has been widely applied to supersymmetric model building.

In our effective field theoretic (EFT) construction we shall restrict ourselves to an intermediate version of naturalness with the following essence. 
 On the one hand, we shall not require a justification of the form of the superpotential by an exact symmetry. 
 On the other hand, we shall demand that the
 low energy predictions of the model must be stable against small \textit{relative} variations of the parameters.  In particular, the zero values of superpotential mass-terms and  couplings are justified, but not the fine tunings among different masses and/or couplings.  This criterion is weaker than the 
 requirement of a full symmetry justification  but is stronger than the  ordinary technical naturalness, and it has been practiced in supersymmetric model building.

We now move to the description of the Model~\textrm{I}. 
As usual, per generation, the Standard Model fermions 
 are embedded in the chiral superfields transforming 
 in $16$-dimensional representation of the 
 $SO(10)$ group.  We denote these chiral superfields by the same symbols as their fermionic components. 
  It is useful to give the reduction in terms of 
  representations of the $SU(5)$ subgroup, 
  \begin{equation} 
  16_F = 10_F + \bar{5}_F + 1_F\,,
  \end{equation}
 where the embeddings are 
\begin{equation}
    \bar{5}_F = \begin{pmatrix} d^c_r & d^c_g & d^c_b & e^- & -\nu \end{pmatrix}
\end{equation}
and
\begin{equation}
    10_F = \frac{1}{\sqrt{2}} \begin{pmatrix} 0 & u^c_b & -u^c_g & -u_r & -d_r \\ -u^c_b & 0 & u^c_r & -u_g & -d_g \\ u^c_g& -u^c_r & 0 & -u_b & -d_b \\ u_r & u_g & u_b & 0 & e^+ \\ d_r & d_g & d_b & -e^+ & 0 \end{pmatrix}.
 \end{equation}
Here, the superscript $c$ denotes anti-particles and $r,g,b$ are the three colors of QCD.
All fermions are written in the left-handed Weyl basis.
In addition, the $SU(5)$ singlet $1_F$ is identified with the left-handed anti-neutrino superfield  $\nu^c$.  

   We shall require that the model is invariant under the $R$-parity as well as under a matter-parity type discrete symmetry which acts as:
$16_F \rightarrow -16_F$ and all the Higgs chiral superfields invariant.

  The GUT symmetry breaking is achieved by 
 the Higgs fields transforming in $54$, $45$ and $16,\bar{16}$-dimensional representations of $SO(10)$.   We employ three Higgs superfields transforming in $45$-representation, which we denote by $45_A\equiv A_{ij}, 45_B \equiv B_{ij}$ and $45_C \equiv C_{ij}$ respectively. 
  Various VEVs split the symmetry breaking job in the following patterns:
\begin{equation} \label{eq:SO10breaking}
\begin{split}
    SO(10) &\stackrel{\langle 54 \rangle}{\xrightarrow{\hspace*{1cm}}} SU(4)_c \times SU(2)_L \times SU(2)_R \\
  SO(10) &\stackrel{\langle 45_A \rangle}{\xrightarrow{\hspace*{1cm}}} SU(4)_c \times SU(2)_L \times U(1)_R \\
   SO(10) &\stackrel{\langle 45_B \rangle}{\xrightarrow{\hspace*{1cm}}} SU(3)_c \times U(1) \times SU(2)_L \times U(1)_R \\
  SO(10)  &\stackrel{\langle 16 + \bar{16} \rangle}{\xrightarrow{\hspace*{1cm}}} SU(5) 
       \;,
    \end{split}
\end{equation}
indicating that the 
$SO(10)$-symmetry is broken to the Pati–Salam subgroup
$G_{PS} \equiv SU(4)_c\times SU(2)_L\times SU(2)_R$ by a 54-dimensional Higgs field, while an adjoint $45_A$ acquires the VEV in the 
(1,1,3) direction of $G_{PS}$, breaking 
$SU(2)_R$ to $U(1)_R$.  The $45_B$-plet additionally breaks $SU(4)_c$ down to $SU(3)_c\times U(1)$.
At the same time, $16+\bar{16}$ pair acquires VEVs on the  components that leave invariant 
the $SU(5)$-subgroup of $SO(10)$. 
 The superposition of the above VEVs breaks the symmetry down to the Standard Model gauge group, 
 $SU(3)_c\times SU(2)_L \times U(1)_Y$. 

We must note that, while group-theoretically a set of 
$54, 16,\bar{16}$  or $45,16,\bar{16}$ representations can break the $SO(10)$-symmetry down to the Standard Model, they are insufficient for achieving such globally-supersymmetric vacuum with fixed GUT scale VEVs within a renormalizable superpotential. 

In such a case, as shown in earlier constructions~\cite{Dvali:1995hp}, 
in order to obtain the   desired pattern of symmetry breaking and realize the decoupled color-triplet scenario~\cite{dvali1992}, together with $\bar{16},16$ 
we need at least one $54$ and a set of $45$-plet superfields.

However, since, as we shall argue, in our approach the renormalizability  
is not a constraint, we shall consider both possibilities. Namely, in the Model I, we shall restrict our analysis by a renormalizable Higgs superpotential, whereas in the Model II we shall drop this requirement. This allows to reach the same goal without using a $54$-plet Higgs.

Hence, in the Model I we have
the following Higgs sector:
\begin{equation}
    \Sigma_{ij} \equiv 54\;, \; A_{ij}\,,  B_{ij}\,,  C_{ij}\, \equiv \,45\,,\; \chi_\alpha\equiv 16\,,\,\bar{\chi}_\alpha \equiv \bar{16}\,, 
\end{equation}
where the $SO(10)$ tensor and spinor indexes are 
denoted by Latin and Greek letters, respectively.   
In addition, of course, there is the ten-dimensional chiral multiplet $10_{Hi}=
\begin{pmatrix}
    T+\bar{T}\\
    H+\bar{H}
\end{pmatrix}
$ which houses the electroweak Higgs doublets as well as their color-triplet partners. 
The characteristic feature of the light color-triplet scenario  \cite{dvali1992} is that this multiplet couples separately neither to the GUT Higgs fields nor to $16_F$-plets that house the Standard Model fermions. Instead, it couples simultaneously to both sectors via an effective  high-dimensional operator (see below).

We choose the superpotential of the heavy GUT Higgs superfields in the following form:
\begin{equation} \label{WGUT}
\begin{split}
    W&=\frac{M}{2}\Sigma^2+\frac{\lambda}{3}\Sigma^3+\alpha\Sigma AB+\tilde{M}AB\\
    &+\frac{1}{2}M_AA^2+\beta A\bar{\chi}\chi+ M_{\chi} \bar{\chi}\chi \\
    &+ \frac{1}{2}\,M_C C^2 \, + \, \gamma \, C [A,B]
    \;,
\end{split}
\end{equation}
where \begin{itemize}
    \item $M$, $\tilde{M}$, $M_A$, $M_C$ and $M_{\chi}$ are mass parameters 
    \item $\lambda$, $\alpha$, $\beta$ and 
    $\gamma$ are couplings
    \item $\Sigma^2 \equiv \Sigma_{ij}\Sigma_{ji}$
    \item $\Sigma AB \equiv \Sigma_{ik}A_{km}B_{mi}$
    \item $A\bar{\chi}\chi\equiv A_{ij}\bar{\chi}_\alpha \sigma_{\alpha \beta}^{ij}\chi_\beta$
    \item $\bar{\chi}\chi\equiv \bar{\chi}_\alpha\chi_\alpha$
    \item $C[A,B] \equiv C_{ij}[A,B]_{ji}$
\end{itemize}
and $\sigma_{\alpha \beta}^{ij}$ are the $SO(10)$-generators in the spinor representation, with spinor indices $\alpha,\beta =1,...,16$.  The purpose of the last line in (\ref{WGUT}) is to remove the  accidental degeneracies of the supersymmetric vacuum.

Now, the central point of the decoupled triplet scenario 
is that the Yukawa couplings of the 
Higgs doublets $H,\bar{H}$ emerge from the following high-dimensional operator ~\cite{dvali1992},
 \begin{equation} \label{Y5D}
       \frac{1}{\Lambda} \, 10_{Hi}\;A_{ij}\;16_F \;\gamma_{j} \; 16_F\; ,
    \end{equation}
 where $\gamma_j$-s are the $SO(10)$ gamma matrixes  and  $SO(10)$-spinor indices are not shown explicitly.  The generation indices are also suppressed, since the mechanism works for arbitrary number of fermion generations. 
 The mass parameter $\Lambda$ plays the role of the
 regulator scale. Its origin will be discussed separately.

 The operator (\ref{Y5D}) can be viewed as originating from a 
fundamental theory above the cutoff scale $\Lambda$, 
or being a result of integrating out heavy 
chiral superfields with masses 
$\sim M_{\textrm{GUT}}$, as described in the original articles~\cite{dvali1992, Dvali:1995hp}.
This shall be discussed in more details later.

 The operator (\ref{Y5D}) ensures that the Yukawa couplings are generated only through a non-zero VEV of the $45$-plet $A_{ij}$. However, since this VEV breaks the GUT symmetry, the 
 couplings of the weak doublets $H,\bar{H}$  and their color-triplet partners $T,\bar{T}$ are split.  Moreover, the 
 VEV of $A_{i,j}$ in the color direction ($i,j \in [1,6]$) 
 is zero (see below), rendering the $T$-particles decoupled from the superfields in $16_F$.

\subsection{SUSY Vacuum}

           The straightforward analysis presented in Appendix A \ref{appA}, shows that the theory possesses a
           supersymmetric vacuum
          delivering the breaking \eqref{eq:SO10breaking}.
           In this vacuum, the VEVs of the superfields are given by the following Ansatz: 
 \begin{equation} \label{PATTERN} 
    \begin{split}
    C_{ij}& = 0 \,, \\
 \Sigma_{ij}&=\text{diag}(\stackrel{6 \times}{2,...};\stackrel{4\times}{-3,...})\Sigma\,, \\
 A_{ij} & =\text{diag}(0, 0, 0, A\epsilon,A\epsilon)\,, \\
B_{ij}& =\text{diag}(\tilde{B}\epsilon, \tilde{B}\epsilon, \tilde{B}\epsilon, B\epsilon,B\epsilon)\,, ~ {\rm with} ~ \epsilon \equiv \begin{pmatrix}
              0 & 1\\
              -1 & 0
          \end{pmatrix}  \,,\\
 \chi_{\alpha}&=\bar{\chi}_{\alpha}=\chi \vert +++++ \rangle \,,~ {\rm is }\\& SU(5)-{\rm invariant~components}\,. \\          \end{split}  
  \end{equation}
 The VEV of the $SO(10)$-spinor representation satisfies  $\sigma_{2i-1,2i}\vert +++++ \rangle=\vert +++++ \rangle$, where $\sigma_{2i-1,2i}\,,~ 
 i\in [1,\;5]$ are $SO(10)$ Cartan sub-algebra generators in the spinor representation.

 In terms of the superpotential parameters the VEVs are,
  \begin{equation} \label{SUSYVEVs} 
    \begin{split}
    \Sigma &=\frac{\tilde{M}}{3 \alpha} \,, \\
    \chi^2 & =-\frac{M_A M_{\chi}}{4\beta^2} \,,\\
     B & =\frac{5}{\alpha}\frac{4\beta}{M_{\chi}}\left(\lambda\frac{\tilde{M}^2}{9\alpha^2}-\frac{M\tilde{M}}{3 \alpha}\right)\,, \\
     \tilde{B}& =\frac{3}{20 \beta}\frac{M_A M_{\chi}}{\tilde{M}} \,,\\
     A & =\frac{M_{\chi}}{4 \beta} \,. \\
    \end{split}  
  \end{equation}
   All  VEVs are fixed and the non-zero ones are of order $M_{\textrm{GUT}}$. \\

Plugging the VEV of the $45$-plet $A_{ij}$ in (\ref{Y5D}), we see that the Higgs doublets acquire the ordinary Yukawa couplings with the 
MSSM quark and lepton multiplets, with 
the effective Yukawa coupling constants set by the ratio $Y_F \sim A/\Lambda$. 
 At the same time, the color-triplet components, 
 $T, \bar{T}$, that reside in $10_{Hi},~i\in [1,6]$, decouple. Despite having no mass-splitting, the color-triplets are unable to mediate the proton decay and are harmless.  
 This addresses the doublet-triplet splitting problem without the need of mass-splitting~\cite{dvali1992}.

\subsection{Couplings of $T$-particles Induced by Gravity-Mediated SUSY Breaking}
    
    As shown in Appendix \ref{appB}, using the Goldstino Argument of \cite{Dvali:2020uvd} and taking into account the gravity-mediated SUSY-breaking effects, the VEV of the 45-plet is shifted to:
\begin{equation} \label{AtildeVEV}
 A_{ij}=\text{diag}(\tilde{A}\epsilon, \tilde{A}\epsilon, \tilde{A}\epsilon, A\epsilon,A\epsilon)  
\end{equation}
 where in the Model I, we have $\tilde{A} \sim m_{3/2}$. \\

 It is important to notice, however, that in other versions of the decoupled triplet theory  the shift can be much smaller. 
 An explicit model of this type, the  Model II,  is given in the Appendix \ref{appD}.  As we will show, in this model, for a minimal K\"ahler
metric, no shift of $\tilde{A}$ is generated in the order $\sim m_{3/2}$.

    As the result of the shift of $\tilde{A}$, 
   the triplets $T,\bar{T}$,  (residing in $10_{Hj}$ with $j\in [1,\;6]$)
   acquire non-zero Yukawa coupling with the Standard Model fermions, 
        \begin{equation} \label{AtildeVEVT}
            (T + \bar{T})_{10}\frac{\langle \tilde{A}_{45} \rangle}{\Lambda} (16_F \times16_F)\;,
        \end{equation}
where, in the Model I,
\begin{equation} \label{AtildeVEV1}
    \frac{\langle \tilde{A}_{45} \rangle}{\Lambda}\sim \frac{m_{3/2}}{M_{\textrm{GUT}}} \;\color{black}.
\end{equation}
Hence, the Yukawa coupling is tiny. 

In terms of the 
Standard Model superfields, the couplings read:
\begin{equation} \label{TYkawa}
    \begin{split}
          \: &g_{Tud} \: \bar{T} \, u^c d^c \; + g_{TQL} \: \bar{T}\,  Q L \;   +   \\ 
        +\;  &g_{Tue} \: T \, u^c e^c  \;+ g_{TQQ} \:  T \, Q Q \; +  g_{Td\nu} \: T \, d^c \nu^c  \,,
    \end{split}
\end{equation}
where $L \equiv (\nu, e)$ and $Q \equiv (u,d)$
denote the $SU(2)_L$-doublet superfields containing 
quarks and leptons.

 In the absence of other operators, per generation, the Yukawa couplings of the color-triplets would be universal and correlated with the ones of the  Higgs doublets,  
\begin{equation} \label{HYkawa}
    \begin{split}   
        & Y_d' \:  \bar{H} \, Q d^c \; + Y_e' \:  \bar{H} \, L e^c \; + Y_u' \: H \,  Q u^c\; +  \\
        +&  Y_{\nu}' \: H \, L \nu^c\,,       
  \end{split}
\end{equation}
as
\begin{equation} \label{HTYkawa}
    \begin{split}  
g_{Tud} &= g_{TQL} = g_{Tue} = g_{TQQ} = g_{Td\nu}= \\ &=
\frac{\tilde{A}}{A} \left( Y_d' =
Y_e'  
=Y_u' = Y_{\nu}' \right) \;. 
    \end{split}
\end{equation}
 The equality among the Yukawa couplings 
 of the Higgs doublets, 
 $ Y_d' =  Y_e'  =  Y_u'  =  Y_{\nu}'$, is a standard problem of the minimal 
 $SO(10)$ as it leads to the wrong fermion mass spectrum. The perturbative RG running is not sufficient for getting the right spectrum. 
 Ordinarily, this problem can be cured by the addition of higher-dimensional operators. In this respect, 
 the presence of effective high-dimensional operators
 in the Yukawa sector can be viewed as a generic feature of GUTs.  
 
 This strategy fits into the decoupled triplet scenario~\cite{dvali1992} rather naturally.    
 For example, for eliminating the unwanted relations among the fermion masses, it suffices to add operators with 
 additional insertions of the $45$-plet Higgs fields, e.g.,  
  \begin{equation} \label{Y6D}
  \begin{split}
      &  10_{Hj}\;A_{jk}\;A_{mn} \:16_F \;\gamma_{k} \gamma_m\gamma_n\; 16_F\;   \frac{1}{\Lambda^2} \, +  \\
    \: &  10_{Hj}\;A_{jk}\;A_{mn} A_{ls}\:16_F \;\gamma_{k} \gamma_m\gamma_n \gamma_l\gamma_s\  16_F\;   \frac{1}{\Lambda^3} \, +  \\
 &  10_{Hj}\;A_{jk}\;B_{mn} \:16_F \;\gamma_{k} \gamma_m\gamma_n\; 16_F\;   \frac{1}{\Lambda^2} \,  +  \\
  &  10_{Hj}\;A_{jk}\;B_{mn} B_{ls}\:16_F \;\gamma_{k} \gamma_m\gamma_n \gamma_l\gamma_s\  16_F\;   \frac{1}{\Lambda^3} \, + ... \,. 
        \end{split} 
    \end{equation}
 Since, the VEVs of $45$-plets break  
 $SU(4)_c$ and $SU(2)_R$ symmetries of the Pati-Salam group, the fermion mass relations of the minimal $SO(10)$ are shattered.  
    
Notice that the insertions of the type (\ref{Y6D}), will also reshuffle  the equality (\ref{HTYkawa}) among the Yukawa couplings of $T,\bar{T}$  superfields. Therefore, in what follows, we shall treat them as independent parameters. 

\subsection{The Origin of the Cutoff Scale $\Lambda$}

 We wish to briefly discuss the origin of the scale 
$\Lambda$. 
First, as shown in the original articles ~\cite{dvali1992, Dvali:1995hp}, 
     the operators of the type (\ref{Y5D}) and (\ref{Y6D}) can be obtained by integrating out the additional heavy chiral superfields with the masses $\Lambda \sim M_{\textrm{GUT}}$. 
     Some examples are discussed in the Appendix \ref{appC}.

However, the scale $\Lambda$
can in fact be the fundamental cutoff of the theory. Remarkably, this cutoff is not very far from the GUT scale.  
The reason is that in any theory propagating $N$ particle species, gravity imposes a universal non-perturbative upper bound on the cutoff scale - the \textit{species} scale - given by the expression
~\cite{Dvali:2007hz, Dvali:2007wp, Dvali:2008fd, Dvali:2008ec, Dvali:2010vm, Dvali:2009ne, Dvali:2008jb, Dvali:2009ks, Dvali:2012uq, Dvali:2020wqi, Dvali:2021bsy},  
\begin{equation}\label{eq:speciesscale}
    \Lambda = \frac{M_P}{\sqrt{N}},
\end{equation}
where $M_P$ is the Planck mass. 

Although an interested reader can consult the proofs given in the 
 references cited above, we would like to make the following important point about the 
 physical nature of $\Lambda$. 
 The species scale is a fully non-perturbative  fundamental scale that cannot be removed by any field redefinition or 
resummation of perturbative series. It is a direct consequence of features that are exact in semi-classical gravity and black hole physics.  Any significant  departure from these features  signals the entrance into the  strong quantum gravity regime.  In particular, the following features support the fundamental nature of the species cutoff 
unambiguously:  1) the break-down of black hole no-hair features 
and of thermality of Hawking radiation~\cite{Dvali:2007hz, Dvali:2007wp, Dvali:2008fd, Dvali:2010vm}; 2) saturation of black hole information storage capacity by the species entropy~\cite{Dvali:2008fd, Dvali:2020wqi};    
3) the equality between Bekenstein-Hawking and entanglement entropies
 ~\cite{Dvali:2008jb}\footnote{In fact, this equality automatically solves 
the puzzle previously known as the \textit{species problem}.}; 
4) non-perturbative saturation of unitarity by multi-particle scattering amplitudes at the point of \textit{optimal truncation}~\cite{Dvali:2020wqi}.  
The species scale is also indicated by perturbative arguments
(see, e.g.,  \cite{Dvali:2001gx, Veneziano:2001ah}).


Correspondingly, in general, the high-dimensional operators suppressed by the species scale $\Lambda$ are expected to be present in the low energy theory.  However, we must make the following clarifying remarks.  

First, this does not imply that the operators suppressed by higher scales cannot exist. The examples are provided by the perturbative low energy couplings of graviton and gravitino which are suppressed by $M_P$, regardless of  $\Lambda$. However, at the same time, in all processes that can reliably be treated within the effective field theory the momentum-transfer among particles is universally restricted by the species scale $\Lambda$ (\ref{eq:speciesscale}).     

 Secondly, the operators regulated by 
 $\Lambda$ can obey the selection rules governed by the structure of the UV theory operating above the cutoff. In the presents paper, we shall not enter into guessing these rules. Instead, we simply focus on the operators of our interest that can dramatically change the low-energy predictions of the GUT theory. 
The operator (\ref{Y5D}) belongs to this category.

 Now, already in the minimal $SU(5)$ GUT the number of species is 
$N > 100$, whereas in any realistic supersymmetric $SO(10)$ GUT, it exceeds several hundred.
 This places $\Lambda$ rather close to the 
 would-be perturbative GUT scale. 
 This has number of important implications.   
 
  In particular, in the light of the low cutoff scale  $\Lambda$, the unification scale $M_{\textrm{GUT}}$ becomes less sensitive to the low energy spectrum of the theory. This is because the gauge couplings are set by the 
  supersymmetric gauge kinetic function, 
  $f$, through the following interaction, 
  \begin{equation} \label{GaugeK}
      \int d^2\theta \, {\rm tr} (f\, W_{a} W_{a})\,,
  \end{equation}
  where the integral is taken over the chiral superspace, and $W_{a, ij}$ is the standard chiral superfield in 
  adjoint representation of $SO(10)$
  containing among its components the  $SO(10)$ gauge field strength as well as gauginos.  Here $a=1,...,4$ is a Lorentz spinor index and $i,j=1,2,...,10$, as previouly, are the $SO(10)$ tensor indices.

  Now, in general, $f$ is expected to be a holomorphic function of  all chiral superfields, including 
  the GUT Higgses, $\Sigma, A, B, \chi, \bar{\chi}$ as well as the hidden sector ones, scaled with the proper powers of  $\Lambda$. 
  The 
  dependence on GUT VEVs will lead to additional splitting of the gauge couplings, not accounted by the perturbative RG running with the participation of the light superfields.  This splitting largely liberates the low energy spectrum from the unification constraints. It also provides more freedom to the unification scale. 

  In addition, the non-minimal operators of the type (\ref{GaugeK}) will also split the masses of gauginos. A group theoretic analysis of this splitting for various representations can be found in  \cite{Chamoun:2001in}.  
    The novelty in the present discussion is the realization that the strength of the splitting can be enhanced by the low species scale (\ref{eq:speciesscale}) 
    ~\cite{Dvali:2007hz, Dvali:2007wp, Dvali:2008fd, Dvali:2008ec, Dvali:2010vm}.   
    
  For example, consider a term in the gauge kinetic function, 
  \begin{equation} \label{Xsplit}
    \Delta f_{ij} =  \frac{X\Sigma_{ij}}{\Lambda^2}\,,
  \end{equation}
  involving the $54$-plet $\Sigma_{ij}$
  and the hidden sector superfield $X$ that breaks supersymmetry through its $F_X$-term, 
  $F_X \sim m_{3/2}M_P$.  This induces the following  
  $SO(10)$-violating splitting in the gaugino masses, 
  \begin{equation} \label{splitGaugino} 
  \Delta m_{1/2} \sim \frac{F_X M_G}{\Lambda^2} \sim 
  m_{3/2} \frac{N M_{\textrm{GUT}}}{M_P}\,. 
  \end{equation}
  Given that in $SO(10)$ GUT the number of species is $N \sim 10^2-10^3$, the expression (\ref{splitGaugino}) implies an order-one splitting, which is rather significant.  Correspondingly, in the light of the high-dimensional operators suppressed by the species cutoff (\ref{eq:speciesscale}), the gaugino masses can be considered free of minimal $SO(10)$ constraints.

\section{General features of the Light Colored Higgs Triplet: mass range, decay rate and minimal lifetime} \label{sec:theory}

  Since we assume no direct coupling between 
  $10_{H}$ and the GUT Higgs fields in the superpotential, 
 in the SUSY limit, the supermultiplets $T, \bar{T}$ are mass-degenerate with $H,\bar{H}$. Their common SUSY mass can be non-zero if the superpotential already contains the non-zero $\mu$-term in unbroken SUSY limit,
 \begin{equation} \label{muterm}
     W_{\mu} = \frac{1}{2} \mu  10_{Hi}10_{Hi} =   \mu (\bar{T}T + \bar{H}H)\,.
 \end{equation}
 A non-zero $\mu$-term can also be generated by supersymmetry-breaking effects, for example, 
 via the Giudice-Masiero mechanism \cite{Giudice1988} (see below),  by a low-energy VEV of an $SO(10)$ singlet superfield \cite{Dvali:1995hp}, by a perturbative loop-correction mechanism of~\cite{Dvali:1996cu}
  or via a shift (\ref{shiftVEV}) of the heavy VEVs \cite{Dvali:2020uvd}.  For generality of our analysis, we shall allow a $\mu$ term to exist  in the superpotential without specifying  its precise origin. 
 The $\mu$ term is necessary for giving masses to the fermion 
 components, which are represented by the left-handed Weyl fermions $\tilde{T}$ and $\tilde{\bar{T}}$. Due to non-zero $\mu$ term, the triplinos form a massive Dirac fermion, 
 $\psi_{T} \equiv \tilde{T} + \tilde{\bar{T}}^*$,  
with the mass  $\tilde{m}_{T} = \mu $.   In unbroken SUSY theory, at the GUT scale,  this fermion is degenerate in mass with the Higgsino. 
 
After supersymmetry breaking, the mass matrix for the $T,\bar{T}$  scalar components has the form, 
 \begin{equation} \label{massTTT}
\hat{{\mathcal M}}_T =\bordermatrix{
	& T^*	& \bar{T}	\cr
T	& m_{T}^2 	& B_{\mu T}	\cr
\bar{T}^*	& B_{\mu T}^*	& m_{\bar{T}}^2 \cr
}\,.
\end{equation}
 The analogous mass matrix for the electroweak Higgs doublets is
  \begin{equation} \label{massHHH}
\hat{{\mathcal M}}_H =\bordermatrix{
	& H^*	& \bar{H}	\cr
H	& m_{H}^2	& B_{\mu H}	\cr
\bar{H}^*	& B_{\mu H}^*	& m_{\bar{H}}^2 	\cr
}\,.
\end{equation}

 For the minimal K\"ahler function, at the GUT scale, the masses are correlated in the following way, 
\begin{eqnarray}  \label{CorrMin} 
m^2_{T} &=& m^2_{\bar{T}} =  m^2_{H} = m^2_{\bar{H}} = 
 m_{3/2}^2+\mu^2   \nonumber \\
 B_{\mu H} &=& B_{\mu T} = B_{\mu}  \,. 
 \end{eqnarray}
The diagonal entries are real and positive, and the 
off-diagonal $B_{\mu T}$-terms can also be taken real and positive 
without any loss of generality.  

  In the minimal case, the mass matrix (\ref{massTTT}) has two eigenstates 
  \begin{equation} \label{PMTT}
  T_{\pm} = \frac{1}{\sqrt{2}}( 
 T \pm \bar{T}^*)\,,
 \end{equation} 
  with corresponding eigenvalues $m^2_{\pm} = 
m_{3/2}^2+\mu^2 \pm  B_{\mu}$.

Of course, the relations (\ref{CorrMin}) are changed by the RG running, but  within the minimal model the evolution is calculable and the structure and the feature 
$m^2_{T} = m^2_{\bar{T}}$ are maintained due to an effective symmetry $T \rightarrow \bar{T}^*$. In the minimal case, this symmetry is violated only by the Yukawa couplings  
(\ref{TYkawa}), which are suppressed by 
$ \frac{m_{3/2}}{M_{\textrm{GUT}}}$ and are
negligible.

However,  
once we go beyond the minimal K\"ahler function, the picture changes. The generic  K\"ahler
function can contain terms of the following form,
 \begin{eqnarray} \label{K2}
\Delta K &=& X^*X \left (\frac{\Sigma_{ij}}{\Lambda_1^3} +  \frac{A_{ij}}{\Lambda_2^3} + \frac{B_{ij}}{\Lambda_3^3} \right )  10_{Hi}^*10_{Hj}  
  +  \nonumber \\
 &&X^*X \frac{\Sigma_{ij}}{\Lambda_4^3}  10_{Hi}10_{Hj}  \,
  +  \, {\rm h.c.} \,  + \nonumber \\ 
&& X^*\frac{\Sigma_{ij}}{\Lambda_5^2} 10_{Hi} 10_{Hj} \, +  \,  {\rm h.c.} \, + \, ... \,,
\end{eqnarray} 
where $X$ is a hidden sector superfield that breaks SUSY by 
the VEV of its $F$-term, $F_X \sim m_{3/2} M_P$ and  
 $\Lambda_{a}\, ~ a = 1,2,...$ are the cutoff scales. As  already explained,  these are bounded from above by the species scale  (\ref{eq:speciesscale}), 
 $\Lambda = M_P/\sqrt{N}$,  and shall be taken to be of the 
 same order.  
 
 After inserting the VEVs of $F_X$ and of the GUT superfields, 
 the operators of non-minimal 
 K\"ahler generate the following $SO(10)$-non-invariant contributions to the masses.  The first line in (\ref{K2}) shifts the diagonal masses by the amounts
 \begin{equation} \label{shiftMT}
 \Delta m^2_{T} \sim  \Delta m^2_{\bar{T}} \sim   m_{3/2}^2 \frac{M_P^2M_{\rm GUT}}{\Lambda^3} \sim  m_{3/2}^2 N^{3/2} \frac{M_{\rm GUT}}{M_P}\,.
  \end{equation}
The second line in (\ref{K2}) contributes into the $B_{\mu}$-terms 
\begin{equation}  \label{shiftBmu}
 \Delta B_{\mu T} =   m_{3/2}^2 \frac{M_P^2M_{\rm GUT}}{\Lambda^3} \, \sim \, m_{3/2}^2 N^{3/2} \frac{M_{\rm GUT}}{M_P}\,.
 \end{equation}
Finally, the third line in (\ref{K2}) shifts the $\mu$-term by a contribution similar to the Giudice-Masiero mechanism, except, in the present case the effect is enhanced by the species scale as,  
 \begin{equation}  \label{shiftmu}
\Delta \mu_{T} =  m_{3/2}
\frac{M_PM_{\rm GUT}}{\Lambda^2} \sim m_{3/2} N \frac{M_{\rm GUT}}{M_P} \,.
 \end{equation}
Since the shift is not $SO(10)$-invariant we have used the subscript $_T$ for denoting the $\mu$-term of the color-triplets. Analogous shifts are of course experienced by the parameters in the 
doublet mass-matrix (\ref{massHHH}).

 As already discussed, the number of species supplied by 
 the $SO(10)$-model  is of order a few hundred. Correspondingly, 
 the shifts (\ref{shiftMT}), (\ref{shiftBmu}), (\ref{shiftmu}),  
  substantially modify the minimal relations (\ref{CorrMin}).  Hence,  at the level of the 
low energy EFT,  the mass matrixes 
 (\ref{massTTT}) and (\ref{massHHH}) shall be treated as the 
 generic $2\times 2$ matrixes, with the entries bounded by an order of magnitude $\sim m_{3/2}^2$.  
 
 Of course, the boundary conditions at the GUT scale must be such that the weak doublets $H, \bar{H}$ condense around 
 the weak interaction scale, $\sim 100$GeV. 
  At the same time,  both eigenstates of the color-triplet mass matrix
 (\ref{massTTT}), which  are given by 
  \begin{equation} \label{TTprime}
  \begin{split}
  &T' \equiv   
  \cos\theta_T \, T + \sin\theta_T \, \bar{T}^*\,, \\
 &\bar{T}' \equiv  -\sin\theta_T \,  T^* + \cos\theta_T \, \bar{T}\,, 
 \end{split}
 \end{equation} 
 with the mixing angle 
  \begin{equation} \label{angleTT}
 \theta_T = \frac{1}{2} \arctan \left ( \frac{2B_{\mu T}}{m_T^2 - 
 m_{\bar{T}}^2} \right ) \,,
 \end{equation} 
 must  have positive eigenvalues. This gives the condition, 
 \begin{equation} \label{angleTT}
  \frac{1}{2} \left({\rm tr} \hat{{\mathcal M}}_T \pm 
  \sqrt{{\rm tr} \hat{{\mathcal M}}_T^2 - 4 {\rm det } \hat{{\mathcal M}}_T} \right ) > 0  \,. 
 \end{equation}   
 The two signs correspond to two mass eigenvalues which we shall denote by $ m_{T'}^2$
 and $m_{\bar{T}'}^2$, respectively.

  
Thus, below the GUT scale,  in the decoupled color-triplet 
model, 
the spectrum of MSSM is enhanced by three new color-triplet
particles:  one Dirac fermion, $\psi_T$,
and two complex scalars $T',\bar{T}'$ given by (\ref{TTprime}).  
All three states have extremely suppressed Yukawa couplings with the MSSM chiral superfields. 
   The Yukawa couplings of $T',\bar{T}'$ scalars with ordinary quarks and leptons are 
   \begin{equation} \label{TPYukawa}
    \begin{split}
          \: &\,  T^{'} \sin\theta_T  \left ( g_{Tud} \: \, u^c d^c \; + g_{TQL} \, \: Q L  \right )^{\dagger} \;   +   \\ 
       &  T' \cos\theta_T  \left (g_{Tue} \: \, u^c e^c  \;+ g_{TQQ} \: \, Q Q \; +  g_{Td\nu} \: \, d^c \nu^c  \right )  \, + \\
          \: &\bar{T}' \cos\theta_T \left (g_{Tud} \: \, u^c d^c \; + g_{TQL} \:  \,  Q L  \right ) \;   -   \\ 
       & \bar{T}^{'} \sin\theta_T \left ( g_{Tue} \: \, u^c e^c  \;+ g_{TQQ} \:  \, Q Q \; +  g_{Td\nu} \:  \, d^c \nu^c \right )^{\dagger}  \,,
    \end{split}
\end{equation}
 where quarks and leptons, 
 which are represented as left-handed Weyl fermions
 ($Q,u^c,d^c,L,e^c,\nu^c$), 
 are denoted by the same symbols as their chiral superfields. 
 The generation and symmetry labels are suppressed. 
  As already discussed, the Yukawa coupling constants are set by the 
 ratio of the VEVs $\tilde{A}/A$  in (\ref{AtildeVEV}),
 whereas their flavor structure is determined by the high dimensional operators (\ref{Y6D}). 
 
 The couplings of triplino  to the Standard Model fermions 
 takes place through its Weyl components,  $\tilde{T}, \tilde{\bar{T}}$,
 \begin{equation} \label{tildeTYukawa}
    \begin{split}
          \: \tilde{\bar{T}}  \:  [ &g_{Tud} \: (\tilde{u^c} d^c + 
 u^c \tilde{d^c})         
          \; + g_{TQL} \: \, (\tilde{Q} L + Q \tilde{L}) ]  \;   +   \\          
           \tilde{T}  \: [& g_{Tue} \: \, (\tilde{u^c} e^c  + u^c \tilde{e^c}) \;+ g_{TQQ} \:  \, (\tilde{Q} Q + Q\tilde{Q}) \; +  \\          
        &  g_{Td\nu} \:  \, (\tilde{d^c} \nu^c + d^c \tilde{\nu^c}) ]\,,
    \end{split}
\end{equation}
where $\tilde{Q},\tilde{u^c},\tilde{d^c},\tilde{L},\tilde{e^c},\tilde{\nu^c}$ denote the supersymmetric partners of the ordinary fermions. 

  The low energy theory possesses a global $U(1)$-symmetry which acts on $T$-particles  in the following manner,  
   \begin{equation} \label{TnumberS}
    \begin{split}  
   &  T \rightarrow  {\rm e}^{i\alpha}  T\,, ~\: \tilde{T} \rightarrow  {\rm e}^{i\alpha}  \tilde{T}\,, \\
     &  \bar{T}  \rightarrow  {\rm e}^{-i\alpha}  \bar{T} \,, ~ \: \tilde{\bar{T}} \rightarrow  {\rm e}^{-i\alpha}  \tilde{\bar{T}}\,,     
    \end{split}
\end{equation}
where $\alpha$ is an arbitrary constant phase.
 We shall refer to it as the $T$-number symmetry denoted by 
 $U(1)_T$.  Since the soft mass terms (\ref{massTTT}) fully respect this symmetry, it is inherited by the mass eigenstates: 
     \begin{equation} \label{TnumberBar}
   T' \rightarrow  {\rm e}^{i\alpha}  T'\,, ~   \bar{T}'  \rightarrow  {\rm e}^{-i\alpha}  \bar{T}' \,.     
\end{equation}
 The $U(1)_T$-symmetry is just an approximate symmetry of the low energy theory. In particular, it is explicitly broken by the  Yukawa couplings, $g_{Tud}\,,  g_{TQL}\,, g_{Tue} \,, g_{TQQ} \,, g_{Td\nu}$, 
 which are bounded from above by $\sim \frac{m_{3/2}}{M_{\rm GUT}}$. 
 Therefore,  $U(1)_T$-symmetry  is exact in the limit $M_{\rm GUT} \rightarrow \infty$. 
  Due to this, the $U(1)_T$-symmetry provides some useful selection rules for the $T$-particle processes.   
  In particular, we can classify the $T$-particle decay 
 processes according to this symmetry.
 
 An example of a $T$-number preserving decay process  
is a transition of a heavier $T$-particle into its superpartner
and a gaugino, 
       \begin{equation} \label{TDecayT}
    T_{\rm particle}   \rightarrow   T_{\rm partner} + 
    {\rm gaugino} \,.     
\end{equation}
  If kinematically allowed, such transitions can lead to the decays of $T$-particles as well as of their bound states with rates unsuppressed by powers of $1/M_{\rm GUT}$.

  On the other hand, the $T$-number violating decays 
  are always suppressed by the powers of the Yukawa couplings and thus can occur only in order 
  $ \frac{m_{3/2}^2}{M_{\rm GUT}^2}$ or higher.

 In order of magnitude, the masses of all $T$-particles are bounded from above by the scale $m_{3/2}$. However, some can be significantly lighter. With this mass range in mind, we shall now outline the general collider constraints on $T$-particles.

\begin{figure}[t]
    \includegraphics[width=1\linewidth]{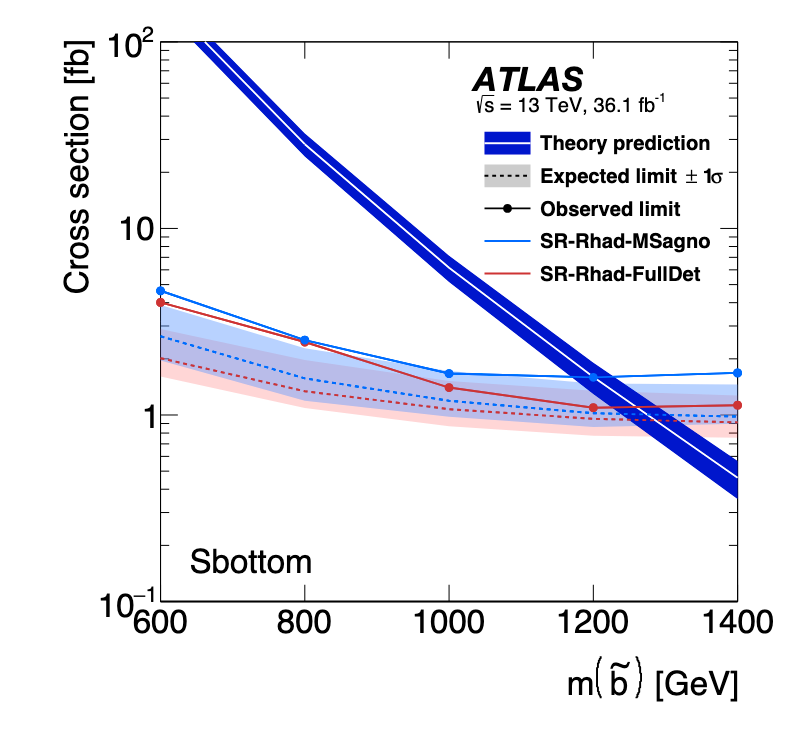}
    \caption{Expected (dashed lines) and observed (marked solid lines) upper cross-section limits in the 
sbottom R-hadron searches. The shaded
light-blue/light-red bands represent the $\pm1 \sigma$ uncertainties in the expected limits. The theory prediction along with its $\pm1 \sigma$ uncertainty is shown as a
white line and a dark-blue band, respectively. Reprinted from \cite{ATLAS:2019gqq}.
 \normalsize} 
 \label{fig:LHCbound}
\end{figure}

  As far as the production in particle collisions is concerned, the 
 $T$-particles share similarities with other colored states in MSSM, such as squarks and gluinos.  Therefore, as a guideline, we use the paper by the ATLAS collaboration \cite{ATLAS:2019gqq}, where the cross section limits in the sbottom $R$-Hadron searches have been derived (see Fig. \ref{fig:LHCbound}). The $R$-Hadron searches target the bound states between colored Standard Model states and colored SUSY particles.  Since such bound states share some phenomenological 
 similarities with the bound states formed by the $T$-particles ($T$-Hadrons), searches for $R$-Hadrons can serve as a benchmark for doing preliminary estimates of the experimentally allowed parameter space for $T$-particles. However, we must stress that these similarities only concern a part of the signatures, due to specifics of  $T$-particles such as their extremely long lifetime and 
 very distinct decay channels (see below).  
 
\begin{figure}[t]
    \includegraphics[width=1\linewidth]{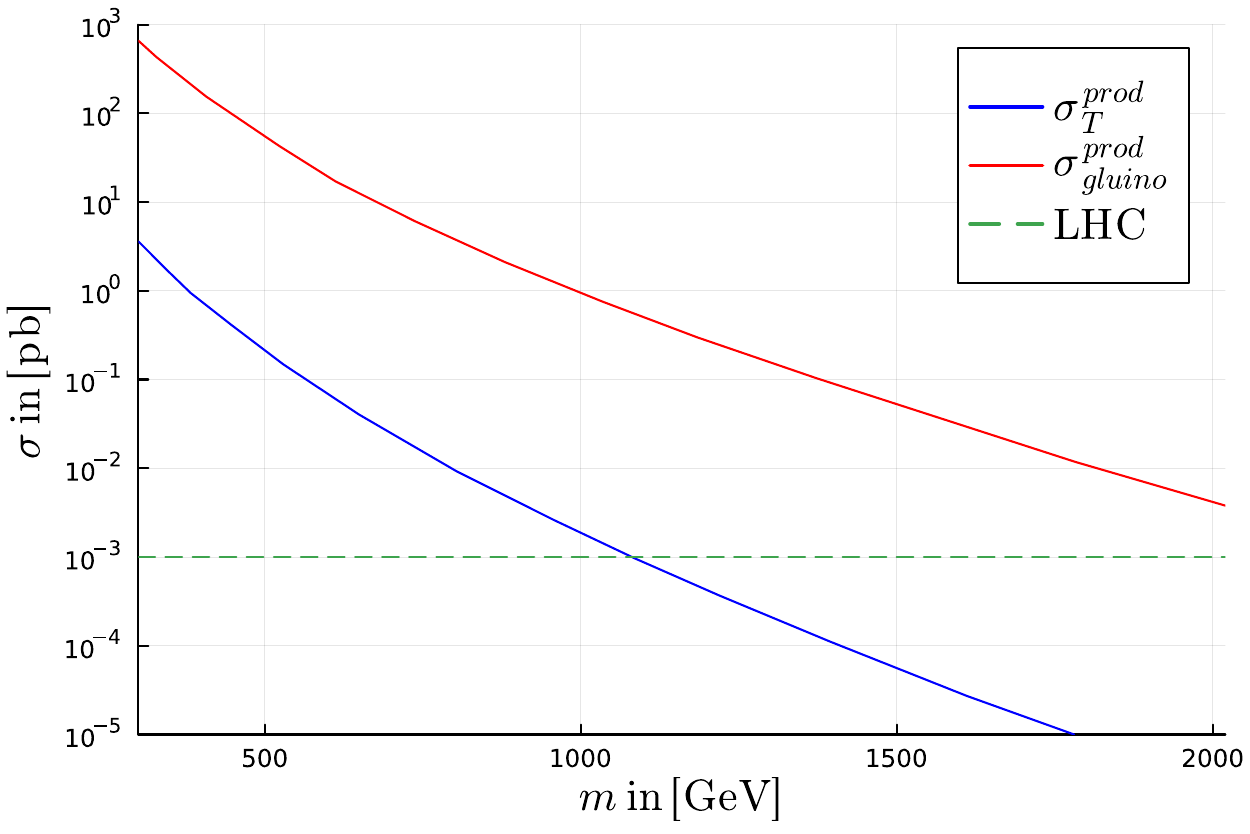}
    \caption{Production cross sections for a colored scalar in the fundamental representation of QCD depicted in blue \cite{Preuss:2021ebd} and, in comparison, the production cross section of a gluino in red \cite{Borschensky:2014cia} for $\sqrt{s} = 13 \textrm{ TeV}$. 
The dashed line marks a first-estimate LHC bound for colored scalars \cite{ATLAS:2019gqq}; cf. FIG. \ref{fig:LHCbound}.}  
 \label{fig:Val}
\end{figure}

The authors of~\cite{Preuss:2021ebd} analyzed LHC bounds on scalar pair production cross sections for multiple representations, as illustrated in FIG~\ref{fig:Val}. Within the relevant mass range, the exclusion cross section is approximately $1\,\text{fb}$. We reinterpreted this analysis to the present model of decoupled 
color-triplet by modifying their framework (Figures~2 and~7 of Ref.~\cite{Preuss:2021ebd}). The resulting lower bound on the 
mass of a $T$-particle is $1.1\,\text{TeV}$, shown in FIG~\ref{fig:Val}.

For the fermionic triplino a dedicated data analysis is presently lacking. Nevertheless, a reinterpretation of current LHC data for a variety of different possible new states is given in \cite{Koren:2024xof} and reports for a color triplet fermion with a fractional charge of $|Q| = 1/3$ a lower bound of $\tilde{m}_T \geq 1.8 \textrm{ TeV}$.  We notice that the same authors also give a slightly higher bound, 
$ \geq 1.4 \textrm{ TeV}$ on the scalar triplet.

 Since most of our analysis shall be limited by order of magnitude estimates, in which we place the masses of $T$ particles in few TeV range, for illustration purposes we shall use the lower bound of ~\cite{Preuss:2021ebd}.

   As discussed, the important effect of SUSY-breaking is the possibility of generating the Yukawa couplings 
  of $T$-particles with ordinary MSSM fields. 
   This is due to the shift of the heavy VEVs induced by supersymmetry breaking.  As shown in \cite{Dvali:2020uvd}, 
in the leading order in $\frac{m_{3/2}}{M_{\textrm{GUT}}}$-expansion, these 
VEVs can be shifted maximum by $\sim m_{3/2}$,  as expressed in the equation \eqref{shiftVEV}. 

 As discussed in details in the previous section as well as in the 
 appendix (\ref{appB}), the shift of the VEV of the 
 $\tilde{A}$-component of the $45$-plet $A_{ij}$
induces the Yukawa couplings of $T$-particles (\ref{AtildeVEVT}). 
However, the magnitude of the shift and correspondingly of the 
obtained coupling is model dependent and is bounded from above by $\sim \frac{m_{3/2}}{M_{\textrm{GUT}}}$. In the Model I, described by the superpotential (\ref{WGUT}), the shift of the 
coupling saturates this bound (\ref{AtildeVEV1}).  

On the other hand, as discussed in the appendix (\ref{appD}),
 in the Model II, in which the  superpotential of the GUT Higgs sector is given by  
(\ref{W2GUT}), for minimal K\"ahler
the shift can occur  only at higher order
in $\frac{m_{3/2}}{M_{\textrm{GUT}}}$. 
Therefore, in the two versions of the theory the lifetimes of the lightest $T$-particles, and correspondingly their observational signatures, are dramatically different.

The decay rate of the lightest scalar $T$-particle into quarks and leptons is 
\begin{equation}
    \Gamma (T) \approx \frac{\lambda^2}{16 \pi} 
    m_T \;,
    \label{lifetimeeq}
\end{equation}
 where we have introduced an effective parameter $\lambda$, for accounting for various decay channels.  For example, for $T'$ we have 
 \begin{equation} \label{lambda}
    \begin{split}
          \: & \lambda^2 \equiv  \cos^2(\theta_T) ({\rm tr} |g_{Tue}|^2 \:  + {\rm tr} |g_{Td\nu}|^2 + 2 {\rm tr} |g_{TQQ}|^2) + 
          \\ 
        \;  & \sin^2(\theta_T) ({\rm tr} |g_{Tud}|^2 \: + 2 {\rm tr} |g_{TQL}|^2)\,,
    \end{split}
\end{equation}
where the traces are taken over the final state fermions  participating in the given vertex. 
 Of course, one must take into account that depending on the precise mechanism of neutrino mass generation,  $\nu^c$, may not be a kinematically accessible final state.

 As discussed, the color-triplet Yukawa couplings are  suppressed relative to the ones of the Higgs doublets by a factor $\frac{\tilde{A}}{A}$ coming from the ratio of the VEVs in (\ref{AtildeVEV}). The flavor structure
  and correlations with the Yukawa couplings of the 
Higgs doublets are model dependent. This is because the simple minimal-$SO(10)$ relation
 (\ref{HTYkawa}) must be split by some high-dimensional operators of the type (\ref{Y6D})  for achieving the phenomenological consistency of the fermion masses. 
 Nevertheless, for an order of magnitude estimate, we can take 
 the relations (\ref{HTYkawa}) as a guideline.  Then,  the strongest Yukawa couplings of the $T$-particles will be of the order of the  Higgs Yukawa couplings with 
  the third generation fermions times the factor 
 $\frac{\tilde{A}}{A}$.  

 Since from the current constraints, the masses of scalar triplets are above TeV,  they can decay into all Standard Model fermions. 
 In Model I, taking for a rough estimate 
 $\lambda^2 \sim \left(\frac{m_{3/2}}{M_{\textrm{GUT}}}\right)^2$,  
 $m_{3/2} \sim 10^3$ GeV,  and the $T$-scalar mass at its 
 lower experimental bound, $m_T = 1.1\,\text{TeV}$, we get the following lifetime,
 \begin{equation} \label{minilt}
  \tau \sim \mathcal{O}(1\text{--}10)\,\text{s}\,. 
 \end{equation}

The small mass and the long lifetime 
of the color-triplet state gives rise to very distinct phenomenological signatures, which shall be investigated in later sections.   

Now, if the shift of the $\tilde{A}$-VEV in (\ref{AtildeVEV}) 
and, correspondingly, the 
Yukawa couplings of the $T$-particles are generated in next order in $\frac{m_{3/2}}{M_{\textrm{GUT}}}$, the decay rate becomes, 
\begin{equation}
    \Gamma (T) \approx \frac{\lambda^2}{16 \pi} m_T \sim 
    \frac{1}{16\pi}
    \left(\frac{m_{3/2}}{M_{\textrm{GUT}}} \right)^4 m_T \;.
    \label{lifetimeeq1}
\end{equation}
 For  $m_{3/2} \sim 10^3$GeV, 
 this is suppressed relative to the equation (\ref{lifetimeeq})
by a factor $\sim 10^{-26}$.  Correspondingly,  the lifetime is increased to 
\begin{equation} \label{maxilt}
    \tau \sim 10^{26}\, {\rm sec} \,,
\end{equation}
which largely exceeds the age of the Universe. Such a color-triplet 
can be considered to be stable from the cosmological point of view. 


For illustrative purposes, a comparison of the two models is presented in Table~\ref{tab:model_comparison}.

  \begin{table*}[htbp]
  \centering
  \caption{Overview of the two models - symmetry protection, benchmark estimates of particle lifetimes, and dominant phenomenology for $T$-particles.}
  \label{tab:model_comparison}
  \vspace{0.5em}
  \renewcommand{\arraystretch}{1.6}
  \setlength{\tabcolsep}{10pt}
  \resizebox{\textwidth}{!}{%
    \begin{tabular}{|l|l|l|p{4.5cm}|l|}
      \hline
      Scenario & Symmetry-protection & Yukawa couplings to SM fermions & $\tau_T$ & Dominant phenomenology \\
      \hline\hline
      \textbf{Model I} & Approximate $U(1)_T$ & $\text{Yukawa} \sim m_{3/2}/M_{\rm GUT}$ & $(1 - 10)\text{ s}$ & Stopped $T$-hadrons, late decays, BBN \\[0.5em]
      \hline
      \textbf{Model II} &$U(1)_T$  & $\text{Yukawa} \sim (m_{3/2}/M_{\rm GUT})^2\text{ or smaller}$ & $(\gtrsim 10^{26})\text{ s}$ & Effectively stable relic \\
      \hline
    \end{tabular}%
  }
\end{table*}
        
\section{Signatures of $T$-particles}  
\label{sec:Meson-decays}

As already discussed, the key prediction of the presented  framework is the existence of the light color-triplets in form of the scalars $T',\bar{T}'$ and the fermions (triplinos)  $\tilde{T}, \tilde{\bar{T}}$ which form a Dirac fermion  $\psi_T\equiv \tilde{T} + \tilde{\bar{T}}^* $.

  Once produced in a particle collision, the $T$-particles will form colorless bound states. These can be of the ``meson" or ``baryon" type. Notice that the names do not refer to a spin, but to a color composition. The $T$-mesons are formed by constituents of the opposite color. That is, they represent the color-anti-color bound states. For example,   
\begin{equation} \label{eq:Tmeson}
    (\bar{T}'u)^+, \;(T' u^c)^-, \; (\bar{T}'d)^0,\; (T'd^c)^0 ,\; (T'\tilde{\bar{T}})^0, \;  (\bar{T}'\tilde{T})^0 ,\; ... \,, 
\end{equation}
where the superscript indicates their electric charge. 
For clarity, we wrote the fermionic constituents in terms of 
their Weyl components.

Instead, baryon types are formed as invariant composites of three colored constituents. For example, 
\begin{equation} \label{eq:Tbaryon}
    (\tilde{T}dd)^-,\; (\tilde{\bar{T}}d^cd^c)^+, \;(\tilde{\bar{T}}u^cu^c)^-,\; (\tilde{T}uu)^+\;.
\end{equation}

 Of special interest are the bound states formed 
entirely by $T$-particles without the participation of ordinary quarks, such as the pure-$T$-mesons, 
 \begin{equation} \label{PTmeson}
  (T'\tilde{\bar{T}})^0, \;  (\bar{T}'\tilde{T})^0 ,\; ... \,, 
\end{equation}
 and the pure-$T$-baryons,  
 \begin{equation} \label{PTbaryon}
      (\tilde{T} T' \bar{T}^{'*})^{-} \,, ~ \; (\tilde{\bar{T}}^* T' \bar{T}^{'*})^{-} \,, ... \,.
\end{equation} 

 For definiteness, we shall focus on $T$-mesons and shall refer to the mixed and pure-$T$ composites as $Tq$ and $TT$ mesons, respectively. 
 
 In order not to confuse the reader with our brief notations, we note that $Tq$ or 
 $TT$-meson can denote 
 several objects. However, each time the meaning 
 should be clear from the context.
 For example, in terms of the original chiral superfields  $T\equiv (T,\tilde{T}) \;\text{and} \;\bar{T}\equiv (\bar{T},\tilde{\bar{T}})$,
 the neutral mesons of $TT$-type can come from the following ``composites" $TT^*,\, \bar{T}^*\bar{T}$ or 
 $T\bar{T}$. These, in the component form, contain
 various spin-$0$,  e.g.,
 \begin{equation} \label{Ttilde2} 
  {\rm spin}=0: ~~TT^{*},~\, \bar{T}^*\bar{T},\,~ T\bar{T},\,~ 
 \tilde{\bar{T}} \tilde{T} \, 
 \end{equation}
 and spin-$1/2$, e.g., 
 \begin{equation} \label{Ttilde1}
 {\rm spin}=\frac{1}{2}:~~
 \tilde{T}T^{*},\, ~ 
 \tilde{\bar{T}}\bar{T}^{*}, ~\,  \tilde{\bar{T}}T,~\, 
 \tilde{T}\bar{T} \,,
 \end{equation}
  mesons. Here, for clarity, we listed the components in terms of the original
 chiral superfields. Of course, as already discussed, after taking into account the 
 effects of soft supersymmetry breaking, 
 the mass eigenstates are $T',\,\bar{T}'$ and 
 $\psi_T$. Correspondingly, the above composites 
 get translated as proper superpositions of the mass eigenstates.  
 Nevertheless, in some cases it is useful to describe processes in terms of the original supersymmetric components.

 The phenomenologically important patterns of
 $T$-hadron decay then can be split into three general categories:

  {\it 1)}  The decay of the $Tq$-type hadron due to the decay of the constituent quark via the ordinary weak interaction \cite{Barbieri:1992yy}. We refer to this as a $T$-number-conserving $\beta$-decay.  Correspondingly, such decays are expected to have timescales characteristic to weak interactions.

 {\it 2)} The decay of a $Tq$ or a
 $TT$ hadron via a direct decay of a constituent $T$-particle.
   Depending on the mass spectrum, such decays can be realized in different patterns, which will be discussed separately. 
  In particular, a constituent $T$-particle (a fermion or a boson) can decay into its supersymmetric partner (a boson or a fermion) and either a gaugino or a gravitino.
  It can also undergo a three-body decay into its partner, a Standard Model fermion (a quark or a lepton) and a corresponding sfermion (a squark or a slepton).  All of the above decays are $U(1)_T$-conserving.  

 If such decays are kinematically forbidden,
the $T$-particles can undergo the $U(1)_T$-violating decay via the Yukawa couplings to the
Standard Model fermions. 
 As already discussed,  depending in which order in $\frac{m_{3/2}}{M_{\rm GUT}}$ the Yukawa couplings of $T$-particles  (\ref{TYkawa})  are generated,  such decays correspond to the lifetimes given by eq. \eqref{minilt}  or eq.(\ref{maxilt}).   \\
  
   {\it 3)} The $TT$-type mesons can undergo a $U(1)_T$-conserving decay into a pair of gauge or gravitational superfields. 

\begin{figure}[htbp]
  \centering
  \includegraphics[width=0.8\linewidth]{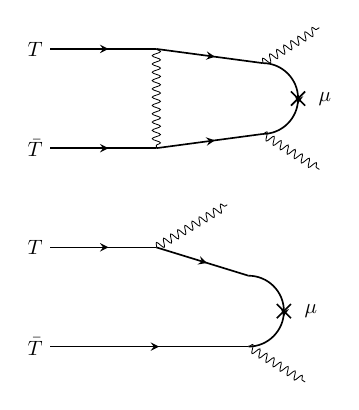}
  \caption{Superdiagrams describing the processes of 
  $T \bar{T}$-meson decay into a pair of gauge bosons and $(T\tilde{\bar{T}})^0$-meson  decay into a gauge boson and a gaugino.}
  \label{fig:TTdecay}
\end{figure}

For example, a neutral $T\bar{T}$ super-meson can decay into a pair of gauge superfields. 
  This process is described by the second super-diagram in Fig. \ref{fig:TTdecay}.
 Notice that such a transition takes place already at the level of unbroken SUSY. 
   At the component level it describes various processes. 

   One of them is the decay of the  bound state of two fermions $\tilde{T} \tilde{\bar{T}}$ into 
  a pair of the gauge fields, as depicted in
  Fig.~\ref{fig:TTdecay}. To estimate the decay width of the constituent annihilation, we use the wavefunction at the origin of a Coulombic (S)-wave bound state,
\begin{equation}
    |\psi(0)|^2\sim R_T^{-3}\sim (\alpha_s M_T)^{3}\;, 
\end{equation}
where $R_T$ denotes the Bohr radius of the Coulombic $TT$ bound state, $M_T$ the mass of the $TT$-type meson, and $\alpha_s\sim 0.1$. Therefore, the corresponding decay time is
\begin{equation}\label{TT2GaugeBosons}
    \tau\sim \frac{1}{\alpha^2 \alpha_s^3 M_T}\;,   
\end{equation}
where $\alpha$ denotes the 
fine structure constant for the corresponding gauge interaction. 
Taking $M_T\sim {\rm TeV}$ and $\alpha\sim 10^{-2}$, we obtain $\tau\sim 10^{-20} $ sec.

\begin{figure}[htbp]
  \centering
  \includegraphics[width=0.8\linewidth]{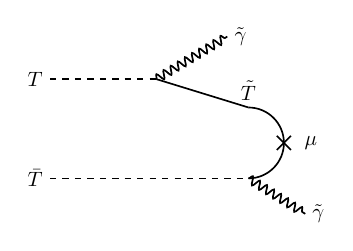}
  \caption{Superpartner diagram describing the decay into a pair of gauginos.}
  \label{fig:decay_gauginos}
\end{figure}

    Another component  of the same super-diagram describes the 
decay of  the $T'\bar{T}'$-meson into a pair of gauginos, as described by Fig. \ref{fig:decay_gauginos}. The decay time can still be estimated by eq.\eqref{TT2GaugeBosons} provided this decay channel is kinematically allowed.

\begin{figure}[htbp]
  \centering
  \includegraphics[width=0.8\linewidth]{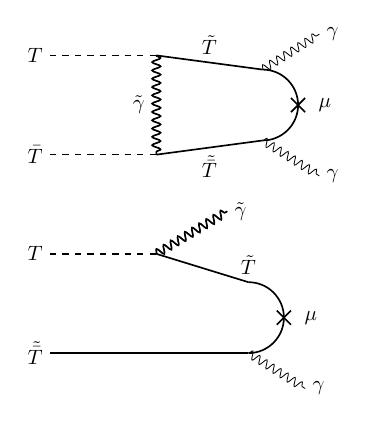}
  \caption{The component representation of the processes described 
   in Fig.  \ref{fig:TTdecay} via superdiagrams. 
   $\gamma$ and $\tilde{\gamma}$ denote the gauge boson 
   (photon) and a gaugino (photino) respectively.}
  \label{fig:TTdecayComp}
\end{figure}

 Notice that the $T' \bar{T}'$-meson can also 
 decay into a pair of the gauge bosons at the expense of intermediate gauge superfield exchange. 
 The supergraph describing such a process  
 is given by the first diagram  in Fig. \ref{fig:TTdecay} and its component form is given in Fig.  \ref{fig:TTdecayComp}. 
  The decay time satisfies
 \begin{equation}\label{TT2Gauginos}
     \tau < \frac{1}{\alpha^4\alpha_s^3 M_T}\;, 
 \end{equation}
where the loop suppression is not shown explicitly. For $M_T\sim{\rm TeV}$, $\tau<10^{-16}$ sec. 

 The bottom line of the discussion is the following. A $TT$-meson containing an even number of triplinos can undergo a $U(1)_T$-conserving decay process with the rate set by the mass of the bound state, $m_T$, and the powers of the gauge couplings, without 
 any extra suppression by the GUT scale or the Planck scale. 

Analogous processes can take place by replacing the gauge superfields with the gravitational ones. For example, 
 a $T\bar{T}$ or a $\tilde{T}\tilde{\bar{T}}$ meson can decay into a pair of gravitons or gravitinos. The corresponding decay time is 
\begin{equation} \label{TT2gg}
    \tau \sim \frac{M_P^4}{\alpha_s^3 M_T^5}\,,
\end{equation}
 which for $M_T \sim $ TeV is $\tau \sim 10^{40}$ sec.
 
The story with a bound state containing an  odd number of triplinos can be very different.
 In particular, the $U(1)_T$-conserving decay of a $T\tilde{T}$-meson containing one scalar and one fermionic component, e.g., $(T'\tilde{\bar{T}})^0$ or $(\bar{T}'\tilde{T})^0$,
depends on the mass of the $T$-meson versus other supersymmetric particles as well as on the nature of LSP.

If the $T\tilde{T}$-meson is heavier than 
the corresponding gaugino, $M_T > M_{\tilde{\gamma}}$, 
it can decay into a gauge field and the gaugino. 
In component form, this is shown by 
the second diagram in Fig. \ref{fig:TTdecayComp}. 
  The lifetime of the $T$-mesons is given by 
 \begin{equation} \label{TT2ggaugino}
     \tau \sim \frac{1}{\alpha^2 \alpha_s^3 M_T}\;.  
 \end{equation}

  If the $T\tilde{T}$-meson is lighter than a coupled-to-it gaugino, $M_T < M_{\tilde{\gamma}}$,  and at the same time is sufficiently heavier than a Standard Model sfermion
   (a squark or a slepton), it can decay into it and its partner fermion (a quark or a lepton), via the virtual gaugino exchange. This is shown in Fig. \ref{fig:Tmesondecay}. The corresponding lifetime  scales as, 
   \begin{equation} \label{TT2fftilde}
    \tau \sim  \frac{M_{\tilde{\gamma}}^2}{\alpha^2 \alpha_s^3 M_T^3} \,. 
   \end{equation}
 For $M_{\tilde{\gamma}}\sim M_T \sim$ TeV,
 both (\ref{TT2ggaugino}) and (\ref{TT2fftilde})
 give $\tau \sim 10^{-20}$ sec.

 Of course, in case of a proper hierarchy among 
the scalar $T$-particles and their triplino partners, the same gaugino exchange can result into a decay of the heavier constituent of 
the $T\tilde{T}$-meson into its lighter partner and a fermion-sfermion pair.  Depending on the kinematics, this can either destroy the bound state or produce a lighter one with its subsequent decay into the pair of gauge bosons.

On the other hand, if LSP is gravitino, the 
$T\tilde{T}$ meson can undergo a gravitational decay into it.  The most straightforward is the 
decay of $T\tilde{T}$ into a graviton and a gravitino. The corresponding lifetime, similarly 
to (\ref{TT2gg}) is given by, 
 \begin{equation} \label{TT2ggravitino}
    \tau \sim  \frac{M_{P}^4}{\alpha_s^3 M_T^5} \,.
   \end{equation}
For $M_T \sim$ TeV this gives $\tau \sim 10^{40}$ sec, which largely exceeds the age of the Universe. 

\begin{figure}[htbp]
  \centering
  \includegraphics[width=0.8\linewidth]{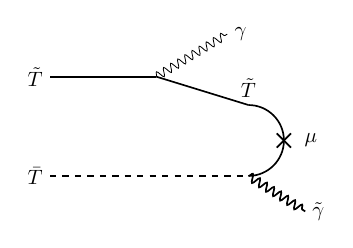}
  \caption{Superpartner diagram of the decay process shown in Fig.~\ref{fig:TTdecayComp}.}
  \label{fig:superpartner_decay}
\end{figure}

\begin{figure}[htbp]
  \centering
  \includegraphics[width=0.4\textwidth]{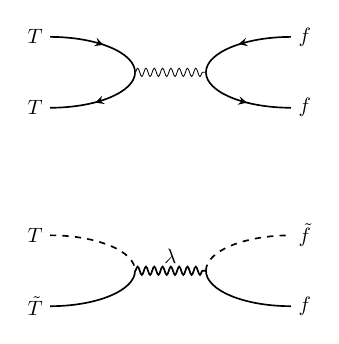}
  \caption{
  On the top, a superdiagram describing the process of super-$T^*T$-meson decay into a pair of quark or lepton superfields. 
  Below, the corresponding component diagram illustrating the decay into a sfermion-fermion final state ($\tilde{f} f$) via a virtual gaugino ($\lambda$) exchange between the scalar $T^*$ and the fermion $\tilde{T}$. The analogous diagrams describing a $T\bar{T}$ supermeson decay requires the additional $\mu$ insertions which connect $T$ and $\bar{T}$ superfields. Of course, the supersymmetry-breaking insertions, not shown explicitly, 
  must be taken into account.  }
  \label{fig:Tmesondecay}
\end{figure}

In this mass regime, the $T\tilde{T}$-meson can also decay into a gravitino and a gauge field.
 For example, the corresponding diagram can be obtained by replacing in Fig.~\ref{fig:superpartner_decay} the gaugino vertex by the gravitino one.  

\begin{figure}[htbp]
  \centering
  \includegraphics[width=0.8\linewidth]{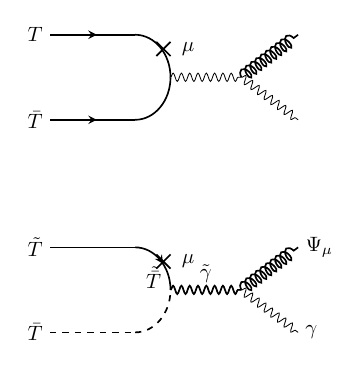}
  \caption{On the top, a superdiagram describing the processes of 
  $\tilde{T}\bar{T}$-meson decay into the gravitational (curly line) and gauge (wave line)  superfields via an intermediate gauge superfield exchange. Below, the component form of the same diagram, the intermediate line is a virtual gaugino which transits into the gravitino and the gauge boson. }
  \label{fig:T2grav}
\end{figure}

 Alternatively, the process can go via a virtual 
 gaugino exchange as described by the Fig. \ref{fig:T2grav}.
 The lifetime of this decay channel scales as, 
\begin{equation}  \label{gravilt}
\tau \sim \frac{M_P^2}{\alpha \alpha_s^3 M_T^3} \,.\end{equation}
 For $M_T \sim$ TeV, this gives  
 $\tau \sim 10^8$ sec.

 In discussion of $U(1)_T$-conserving decays we wish to comment on possible decay of 
the $TT$-meson into two $Tq$-type mesons. In this process the $TT$-meson splits into two constituent $T$-particles which then form mesons with two separate quarks.  
An example of such a process would be, 
\begin{equation}
 \bar{T}'\tilde{T} \rightarrow \bar{T}'d 
 + \tilde{T} d^c \,.
\end{equation}   
 However, for the parameter range of our interest we expect such processes to be kinematically forbidden. 
  The color-singlet  $\bar{T}'\tilde{T}$ ground state is expected to be Coulombic and compact with radius 
  \begin{equation} \label{RTradius}
   R_T^{-1} = \frac{4}{3} \alpha_s\,\bar{m} \,,  
  \end{equation}
  and binding energy $E_{bind} = \frac{8}{9} \bar{m} \alpha_s^2$, where $\bar{m} \equiv \frac{ m_{\bar{T}'} \tilde{m}_T}
 {m_{\bar{T}'} + \tilde{m}_T}$ and $\alpha_s$ is the running QCD coupling evaluated at the scale 
 $R_T^{-1}$. 
 By contrast, the masses of the heavy-light final states are $m_{\bar{T}'d} = m_{\bar{T}'} + \bar{\Lambda}_{\bar{T}'}$ and $m_{\tilde{T}d^c} = \tilde{m}_T + \bar{\Lambda}_{\tilde{T}}$,
 where  $\bar{\Lambda}_{\bar{T}'}, \bar{\Lambda}_{\tilde{T}}  \sim 
 0.5 -1$GeV. 
 Consequently, the mass difference between the initial and final states is   
$E_{bind} +  \bar{\Lambda}_{\tilde{T}} + \bar{\Lambda}_{\bar{T}'} > 0$ and decay is kinematically forbidden.

  Finally, the lightest $(T'\tilde{\bar{T}})^0$ meson can be protected against $U(1)_T$-conserving decay, 
  provided its mass is below of all the ``ordinary" supersymmetric particles including the gravitino.
In this case, the decay can take place exclusively through $U(1)_T$-violating Yukawa couplings
and is correspondingly suppressed by the powers of the small parameter 
$\sim \left (\frac{m_{3/2}}{M_{\rm GUT}} \right )^2$.  Of course, since triplino is LSP, the decay products will contain a $\tilde{T}q$-type meson.

For convenience, we summarize the above regimes of $TT$ meson decays in Table \ref{tab:TTmeson-regimes}. The above items have implications for collider phenomenology and cosmology, as shall be discussed below. 
 
\begin{table*}[htbp]
\centering
\setlength{\tabcolsep}{2.5pt}
\renewcommand{\arraystretch}{1.1}
\begin{tabular}{c|c|c|c}
\hline
Triplino number in $TT$-meson & Mass hierarchy & Decay products & Decay Times \\
\hline
\multirow{4}{*}{even} & -- & $\gamma\gamma$ & \multirow{2}{*}{$\tau<10^{-16}$ sec}\\ 
{} & $M_T>2M_{\tilde{\gamma}}$ & $\tilde{\gamma}\tilde{\gamma}$ & {} \\
{} & -- & $GG$ & \multirow{2}{*}{$\tau\sim 10^{40}$ sec} \\
{} & $M_T>2m_{3/2}$ & $\psi_\mu \psi_\mu$ & {} \\ 
\hline
\multirow{5}{*}{odd} & $M_T>M_{\tilde{\gamma}}$ & $\gamma\tilde{\gamma}$ & $\tau\sim 10^{-20}$ sec \\
{} & $M_T>m_{\tilde{f}}+m_f$ & $f\tilde{f}$ & $\tau\sim 10^{-20}$ sec \\
{} & \multirow{2}{*}{$\psi_\mu$ is LSP} & $G \psi_\mu$ & $\tau\sim 10^{40}$ sec \\ 
{} & {} & $\gamma \psi_\mu$ & $\tau\sim 10^{8}$ sec \\
{} & $M_T<M_{\rm SUSY\ particle}$ & -- & suppressed $U(1)_T$ violating decay \\
\hline
\end{tabular}
\caption{$TT$-type meson decay regimes, which are classified by the number of triplinos in the meson. The  constituency has to be understood in terms of (\ref{Ttilde2}) and (\ref{Ttilde1}), respectively. 
$G$ and $\psi_\mu$ denote the graviton and gravitino, respectively, and the other notations have been defined in the main text. The lifetimes are estimated by taking $M_T\sim M_{\tilde{\gamma}}\sim {\rm TeV}$, and the gauge coupling $\alpha \sim 10^{-2}$. 
}
\label{tab:TTmeson-regimes}
\end{table*}

\subsection{$T$-number preserving $\beta$-Decay} \label{subsec:A}

 For definiteness, let us consider a process in which the constituent quark, $q$, of the 
 $Tq$-type meson, undergoes a $\beta$-decay. 
 Here the symbol $T$ can refer to any member, scalar
($T', \bar{T}'$)  or fermion ($\psi_T$),  of 
 the $T$-sector. 
 The possibility of such a decay was already suggested in \cite{Barbieri:1993wz}. At that time, the existing precision was allowing the consideration of such a process also for the  bound states formed with the first generation quarks, 
 such as $Td$.  The idea was that in this bound state, the constituent $d$ quark 
can undergo a $\beta$-decay into the $u$-quark: 
\begin{equation}
    Td \rightarrow Tu^+ + e^- + {\nu}_e^c\;.
\end{equation}
 The crucial factor for the lifetime is the mass difference between the educt and the products, i.e., $\Delta(m_{Td}, m_{Tu})$. 
  Adopting the view of ~\cite{Barbieri:1992yy}, $\Delta$ is approximated with the analogous mass-splitting of $B$-mesons $Bd \equiv B^0$ and $Bu \equiv B^+$. However, the current precision of this mass-difference, cf \cite{ParticleDataGroup:2026mpi}, is 
\begin{equation}
\Delta(m_{Td}, m_{Tu}) \sim \Delta(m_{B^0}, m_{B^+}) \simeq (0.31 \pm 0.1) \; \text{MeV}\,.  
\end{equation}
 This abolishes the possibility of $\beta$-decay, since the mass difference is below the mass of the electron. 
 
 Thus, for the $T$-hadrons formed with the first generation of quarks, the $T$-conserving $\beta$-decays are not possible.  In contrast, the 
 $T$-hadrons formed with heavier quarks, 
 can undergo the $T$-conserving $\beta$-decays. 
 For example, for a $T$-meson formed with the participation of the $s$-quark,  we can have the following process,
 \begin{equation} \label{eq:betadecay}
  Ts \rightarrow Tu^+ + e^- + {\nu}_e^c\;.
 \end{equation}
 
 The corresponding decay rate, for $m_T \sim \;$TeV, would be
 \begin{equation}
      \Gamma^{\beta}_{Ts} = \frac{G_F^2 |V_{us}|^2}{60\pi^3} \Delta_s^5 \mathcal{F}\;, 
 \end{equation}
 where, \(\mathcal{F}\) parametrizes the hadronic form factors and spin-dependent contributions to the transition. In the absence of a dedicated calculation of the \(Tq\)-meson form factors, we assume that these contributions do not introduce any additional parametric suppression or enhancement, and take \(\mathcal{F}\sim\mathcal{O}(1)\) for the purpose of this order-of-magnitude estimate. 
 
 With $\Delta_s \approx (87.52 \pm 0.12)$ MeV, this gives a lifetime of 
 \begin{equation}
    \tau^{\beta}_{Ts} \sim  10^{-5}\;\text{sec}.
\end{equation}

 This expression suggests that if a $T$-particle is produced at relativistic velocity, the 
 decay will not be observed inside a detector. 
 However, it can be potentially observed if the $T$-particle is produced near the mass threshold and is slow.  
 In such a case, it may also be trapped inside a detector and undergo the $\beta$-decay afterward. 
 As a rough guideline, we can use the studies of LHC phenomenology of long-lived gluinos by the ATLAS collaboration in \cite{ATLAS2021}. According to it, for masses in the interval 
 $600- 2600$GeV the detector trapping probability varies $4-8$ percent.
 Of course, these are just indicative numbers since $T$-particles can be produced arbitrarily close to mass threshold, which increases the 
 trapping probability. 
 
 Notice that since the timescale of $T$-conserving  $\beta$-decay is much shorter than the minimal time of the direct $T$-decay given by
 eq. \eqref{minilt}, the $\beta$-decay signal is predicted to be followed by the one from a direct decay, which we shall discuss in the following subsection.

\subsection{$T$-number violating scalar decays} \label{subsec:B}

  Let us now discuss some experimental signatures of $T$-hadrons via a direct decay of their constituent $T$-particles.  In this subsection we focus on a scalar from $T',\bar{T}'$ sector assuming that it is a lightest member of the $T$-family.
The phenomenology of a light triplino $\psi_T$, is discussed separately in the following subsection \ref{subsec:triplino}.
 We assume that the kinematic regime is such that the dominant decay channels are through 
  $U(1)_T$-violating Yukawa couplings. 
 We therefore restrict our analysis mainly by two-body decays.
Of course, such 
a decay is only relevant if the Yukawa couplings 
of the $T$-particles are generated in the first order in $\frac{m_{3/2}}{M_{\rm GUT}}$, as this is the case in Model I. 
The corresponding lifetime is given by (\ref{minilt}).
  
This decay takes place via the Yukawa couplings (\ref{TYkawa}). Each coupling is suppressed by the factor $\frac{m_{3/2}}{M_{\rm GUT}}$. Their  precise values and correlations with the Yukawa couplings of their Higgs doublet partners depend on the structure of operators (\ref{Y6D}) which take care of the standard (wrong) fermion mass relations of the minimal $SO(10)$.  Correspondingly, at the level of the present analysis,  the Yukawa couplings (\ref{TPYukawa}) of $T',\bar{T}'$ shall be treated as the phenomenological parameters.   

These couplings can lead to various semileptonic
\begin{equation} \label{eq:TdecayQL}
    T \rightarrow l + q \;,
\end{equation}
as well as quark-quark 
\begin{equation} \label{eq:TdecayQQ}
    T \rightarrow q + q \;, 
\end{equation}
decay channels. 
 The important fact is that the produced quarks 
 have energies way above the QCD binding barrier. 
 Consequently, the process will be accompanied by the jet production.  This is the key signature of the $T$-decay.
  As it is clear from \eqref{minilt},  the timescale of the process is $\sim$\;sec. Correspondingly, whether the decay takes place inside or outside of a detector, depends on the velocity of the $T$-hadron. \\

   Thus, the following  scenarios emerge: 

 {\it 1)} If the $T$-particle decays within the detector, we observe a highly energetic jet event as a (far-)displaced vertex. In the case of the semileptonic decay channel, the jet will be accompanied by a highly energetic lepton. 

  {\it 2)}  In case of  $T$-hadrons containing the heavier quarks (such as $s$-quark), the jet event can be preceded by the $\beta$-decay process of the type \eqref{eq:betadecay}.

  {\it 3)}   Finally, we have an option in which 
  $T$-hadron simply escapes the detector. In such a case, it will share the signatures characteristic of a stable colored particle.

 A useful basis has already been prepared by various searches of signatures of long-lived colored particles. In particular, it has been
 suggested that the role of such particles can be played by colored supersymmetric particles, such as gluinos, which can be long-lived in case of $R$-parity, provided the squark masses are very high.

For a collider-stable particle, corresponding to the item {\it 3)}, there will be tracks from charged $T$-hadrons in the inner detector and possibly also in the muon chamber:
they lose energy within the electromagnetic calorimeter according to the Bethe-Bloch formula. Within the hadronic calorimeter, they will also behave like heavy versions of the known hadrons, as all components of the hadron are colored under QCD.
As described in \cite{Barnard:2015rba}, R-hadron searches at the LHC deliver further mass limits on the triplet which can be useful in the present case.

The options {\it 1)} and {\it 2)}, which predict the time-displaced vertices, produce jets from the quarks and missing energies from the neutral particles; cf. \cite{Barnard:2015rba, Bhattacherjee:2019fpt}.

   The lightest $T$-particle always undergoes a two-body decay, we can predict the energy and momentum distributions among the decay products. A quark produced from the $T$-decay will carry approximately half of its energy.  Since this energy is much larger than the confinement scale $E_q \gg \Lambda_{QCD}$, it shall escape the 
 bound state and form a highly energetic jet in the detector.

The semileptonic decay channel (\ref{eq:TdecayQL}) has two distinct possibilities, depending on whether the final-state lepton is charged or neutral. In the first case, the decay leads to a charged track in the detector, whereas in the second case the neutrino escapes undetected, giving rise to missing transverse energy.

Of course, the spectator quarks of the bound state could also absorb some of the energy of the $T$-decay, which would lead to substructures like low-energy jets or bound states that leave tracks within the detector.
 
Although a detailed analysis of the energy distributions of the
decay products and spectator quarks would be required, the total
energy carried by the resulting jet and the associated hadronic
substructures is expected to be $\simeq m_T/2$
in the
rest frame of the decaying $T$-particle.
Notice that, due to its macroscopic lifetime (\ref{minilt}),
the probability of observing a $T$-decay within the detector
depends on the boost of the $T$-particle as well as on its
lifetime. For a $T$-particle that is non-relativistic at the
time of decay, its energy is approximately given by its rest
mass, $E_T \simeq m_T$.

This gives us an additional signature of the model, which makes it distinguishable from other proposals with long-lived particles that decay within the detector. In SUSY, for example, long-lived R-hadrons which have a gluino in their bound state could decay via $\tilde{g} \rightarrow q\bar{q}\tilde{\chi}_0$, where $\chi_0$ represents a neutralino. Even though such a scenario shares many similarities with our model,
there are important differences. 
In particular, unlike a $T$-particle,  the gluino decay is a three-body decay, as it goes through a virtual heavy squark. Due to this, the energy distribution of the final states is not well-determined.

Also, the event topologies in the two scenarios are different. Since the $T$-particle undergoes a two-body decay, the two primary decay products have to be in the opposite direction from each other. This feature is absent in the three-body gluino decay, where the momentum is shared among three final-state particles. \\
In the semileptonic channel (\ref{eq:TdecayQL}), with a charged decay product, this characteristic topology can be used as an additional observable, because the emitted charged lepton as well as the accompanying jet are expected to be emitted in opposite directions, both leaving visible tracks in the detector.
Deeper insights into the experimental implementation of such observables can be found in \cite{Banerjee:2017hmw, Bhattacherjee:2019fpt}.

\begin{figure}
    \centering
    \includegraphics[width=1\linewidth]{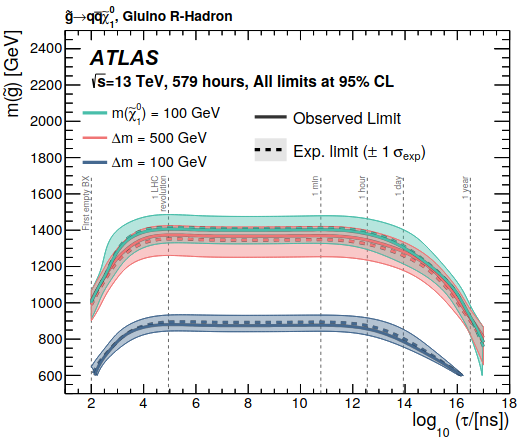}
    \caption{Expected (dashed lines) and observed (solid lines) exclusion limits at $95\%$ CL using the 2017 and 2018
datasets in the central signal regions. The limits are shown as a function of gluino mass and $\tau (\Tilde{g})$. The shaded
coloured bands represent the $\pm \sigma_{exp}$ variations from systematic and statistical uncertainties in the expected yields.
The different sets of colours represent the limits for different assumptions about the gluino–neutralino mass splitting. Reprinted from \cite{ATLAS2021}.}
    \label{fig:lifetimebounds}
\end{figure}

In fact, the ATLAS collaboration already took the gluino decay as a benchmark scenario for long-lived particles (LLP) \cite{ATLAS2021}, and we can reinterpret their results in the context of our theory.\\
The authors of this analysis observed the detector in between two bunch crossings. During this time, no collisions happen within the detector, and one can search for the signatures described in items {\it 1)} and {\it 2)}
above within a very low background environment. \\
As the production cross section scales with $m_T$ (see Fig. \ref{fig:Val}) and the probability of detecting the decay depends on the corresponding timescale, the resulting bounds will be functions of both parameters. \\
The number of signal events $N_{signal}$ in such a situation is calculated via 
\begin{equation}
    N_{signal} = L^{int} \times \sigma_{prod}\times 2\times \epsilon \times f_{\textrm{stopping}} \times f(\tau) \;,
\end{equation}
where $L^{int}$ is the integrated luminosity, $ \sigma_{prod}$ is the production cross section, $\epsilon$ is the detection efficiency, $f_{\textrm{stopping}}$ is the fraction of particles that get stopped within the detector, and $f(\tau)$ is an analysis-dependent function of the lifetime of the particle. \\
\\
In Fig. \ref{fig:lifetimebounds}
the results for $\Tilde{g} \rightarrow q \Bar{q} \Tilde{\chi}_0$ are presented: \; a mass limit of 1.4 TeV, which refers to a production cross section of $10^{-1}$ pb. \\
\\
Applying these results to our scenario, from Fig. \ref{fig:Val}  we find a minimal mass of $m_T = 630$ GeV which comes from the fact that in the triplet case $\sigma_{prod}$ is roughly $10^{-2}$ times smaller than for an 
equal-mass gluino because the triplet is in the fundamental representation unlike the adjoint representation of gluino. \\
Thus, the current situation is that the direct searches for colored scalars provide us with stronger bounds than the decay product searches; cf. section \ref{sec:theory}. \\
\\
Nevertheless, for the HL-LHC era one has to keep in mind that $\epsilon$ and $f_{\textrm{stopping}}$ will both increase with higher masses. Already in the current bound one should adjust $\epsilon$ to our scenario because the signal from two-body decays is much easier to extract than the one from a three-body decay.
Even though for the current analysis this cannot be sufficient for compensating the factor of $10^{-2}$, it can become relevant when both $\epsilon$ and $f_{\textrm{stopping}}$ increase from the $10 \%$ level up to order unity. Moreover, we expect that in the HL-LHC era the background for such searches will scale much more slowly with luminosity than for direct search analysis. Therefore, the signal-to-background ratio can be more beneficial for such decay product searches. This means that in the future a dedicated analysis of the decay product searches is rather promising for the $T$-particle searches.

\subsection{Triplino Phenomenology} \label{subsec:triplino}

 In this section we discuss some collider signatures of triplinos. If the triplino is the lightest $T$-particle state, it will have sufficient time 
to form colorless bound states with other colored particles.  Here we discuss mesons formed by triplinos and ordinary quarks, schematically denoted as $\psi_T q$. 
Similarly to analogous mesons formed by the scalar
$T$-particles, the triplino-containing bound states can undergo a $U(1)_T$-conserving 
beta-decay.  This analysis  essentially stays the same modulo the specifics of the spin.

If the triplino is not the lightest $T$-particle and is sufficiently heavy, it can  decay through a supersymmetric two-body channel.
 For example, if triplino is heavier than one of its scalar partners and a gaugino, it will decay into them, 
 \begin{equation} \label{eq:tripdecay}
     \psi_T \longrightarrow T + \lambda \;,
 \end{equation}
 where $\lambda$ denotes the relevant gaugino, for example a photino, $\lambda \equiv \tilde{\gamma}$. For
 $\tilde{m}_T \gg m_T + M_{\tilde{\gamma}}$
 The time-scale of the process is, 
 \begin{equation} \label{triplino2Tgaugino}
  \tau \sim \frac{1}{\alpha \tilde{m}_T} \,,   
 \end{equation}
and can be shorter than the hadronization time. 
It can correspondingly take place prior to 
the $\psi_{T}q$-meson formation. 

If instead the formation of $\psi_{T}q$-meson
 happens prior to the above triplino decay, after the decay, in the daughter hadron the fermionic constituent will be replaced by its bosonic partner. 
  For instance, in case of the photino-production at the hadronic level this transition looks schematically as

 \begin{equation}
     (\psi_T q)_{\rm had}
    \longrightarrow (Tq)_{\rm had}+\tilde{\gamma}\;.
 \end{equation}
 However, if the mass difference,
\begin{equation}
    \Delta m = \tilde{m}_{T}-m_T-M_{\tilde{\gamma}}\,,
 \end{equation}
 is sufficiently high, the hadron will split and the 
 decay will be accompanied by the jet production.
This is caused by the resulting highly energetic $T$-scalar that will break the originally formed bound state of the mother triplino and will start travelling through the detector once again.
  The  subsequent phenomenology of the $T$-scalar has been discussed in the previous section.

   There also exists a parameter space in which triplino can undergo a three-body decay via a virtual 
  gaugino into a $T$-scalar, a fermion (quark or a lepton) and a corresponding sfermion (squark or a slepton). In case $M_{\tilde{\gamma}} \gg \tilde{m}_T \gg m_T, m_{\tilde{f}}$, 
taking the triplet mass mixing into account and assuming that $T'$ is the lighter mass eigenstate, we have  ($g$ is the gauge coupling)
\begin{equation}
    g_{\tilde{T}T'\tilde{\gamma}}=\sqrt{2} g \cos\theta_T\;, \qquad g_{\tilde{\bar{T}}T'\tilde{\gamma}}=\sqrt{2} g \sin\theta_T\;, 
\end{equation}
such that
\begin{equation}
\begin{split}
  &  \Gamma_{\tilde{T} \rightarrow T'f\tilde{f}}
\simeq \cos^2\theta_T\frac{\alpha^2}{24\pi} \frac{\tilde{m}_{T}^3}{M_{\tilde{\gamma}}^2}\;,\\ 
&\Gamma_{\tilde{\bar{T}} \rightarrow T'f\tilde{f}}
\simeq \sin^2\theta_T\frac{\alpha^2}{24\pi} \frac{\tilde{m}_T^3}{M_{\tilde{\gamma}}^2}\;. 
\end{split}
\end{equation} 
 Here the rates refer to the decay channels through $\tilde{T}$ and $\tilde{\bar{T}}$ components, although the two are not independent mass eigenstates but form a single 
 Dirac fermion $\psi_T$ of mass $\tilde{m}_T$. In Table~\ref{tab:triplino-regimes}, the corresponding decay channel is therefore denoted schematically by $\psi_T\to T'f\tilde{f}$.

The collider phenomenology of the triplino depends on its mass, the available decay channels and the associated lifetimes. If the triplino is sufficiently long-lived to hadronize, it can form color-singlet $T$-hadrons, which may propagate through the detector and, depending on their interactions and lifetime, come to rest before decaying. If the triplino decays before hadronization, the resulting signature is instead determined by the decay products of the prompt or displaced decay. In particular, the decay of equation \eqref{eq:tripdecay} 
can produce additional hadronic activity if the mass splitting is sufficiently large. If the resulting $T$-scalar is itself long-lived, it may form a further $T$-hadron and give rise to additional displaced decays. Thus, the possible signatures range from prompt decays and jets to detector-stable $T$-hadrons and delayed decays, depending on the relevant masses, couplings and lifetimes.

An interesting regime arises when the gravitino is the LSP
and the decay of the Dirac triplino into its scalar partner
and a gravitino is kinematically allowed, whereas the
corresponding two-body decay into a $T$-scalar and a gaugino,
as well as the three-body decay into a $T$-scalar, a fermion
and a sfermion, is kinematically forbidden. 
The decay time  for this process goes as 
\begin{equation} 
\tau \sim \frac{M_P^2}{\tilde{m}_T^3}, 
\end{equation}
which for $\tilde{m}_T \sim$TeV is around 
$\sim 10^5$ sec. 

Notice that in the Model~I, if kinematically allowed, the
Dirac triplino can decay on a shorter timescale 
($\tau \sim$ sec) via the Yukawa couplings
(\ref{TYkawa}) into a fermion 
($f$) and a sfermion ($\tilde{f}'$),
\begin{equation} \label{T2fftilde}
    \psi_{T} \rightarrow f + \tilde{f}'\,,
\end{equation}
where we put prime on the sfermion in order to take into account that, as dictated by (\ref{TYkawa}),
$f$ and $\tilde{f}'$ come from different flavors, since the triplino is a leptoquark.

 If the Dirac triplino is LSP, it is stable. A stable colored LSP would hadronize into neutral and charged $T$-hadrons. Such states would behave as heavy long-lived particles rather than as ordinary prompt jets. In particular, charged $T$-hadrons could leave highly ionizing tracks and, because of their large mass, would typically have a velocity significantly below that of relativistic Standard Model particles. \\
 \\
 Overall, the triplino phenomenology can thus be organized according to the available decay channels and their associated timescales. Prompt or relatively short-lived decays arise when they are kinematically allowed via the ordinary supersymmetric Yukawa interactions of the type, $T-\tilde{T}-$gaugino. Gravitational decays can lead to much longer-lived triplino-containing hadrons. Finally, in the absence of any allowed decay channel, the triplino behaves as a stable colored particle.
 
 We summarize the described triplino phenomenology in table  \ref{tab:triplino-regimes}.

\begin{table*}[htbp]
\centering

\setlength{\tabcolsep}{2.5pt}
\renewcommand{\arraystretch}{1.1}
\begin{tabular}{c|c|c}
\hline
Mass hierarchy &Decay channel & Phenomenology \\
\hline
$\psi_T$ is LSP & no decay & stable $T$-hadrons \\
$\tilde{m}_T> m_{f} + m_{\tilde{f}'}$ (Model I) & $\psi_T\to f\tilde{f}' $ & $\tau \sim \mathcal{O}(\textrm{s})$/displaced decays\\
$\tilde{m}_T> m_{T}+m_{3/2}$ and $\psi_T$ is NLSP, $\psi_{\mu}$ is LSP & $\psi_T\to T \psi_{\mu}$ & long-lived T-hadrons \\
$\tilde{m}_T>m_T+m_{\lambda}$ & $\psi_T\to T\lambda$ & jets possible \\
$\tilde{m}_T>m_T+m_{\tilde{f}} +m_f$ & $\psi_T\to T'f\tilde{f}$ & displaced vertices/multijets\\
\hline
\end{tabular}
\caption{Triplino decay regimes. The listed signatures are indicative and depend on the
relevant couplings, precise branching 
ratios and lifetimes.
The lines $2-5$ have the supersymmetric counterpart for the decay channels of the scalar $T$-particles. However, the analog of the first line
for the scalar-$T$ must be taken with care. Although, in Model II, the lightest scalar-$T$
can be stable on a present cosmological timescale, the stability cannot be achieved within any $TT$-meson bound state, unless 
the triplino is the LSP. 
}
\label{tab:triplino-regimes}
\end{table*}

\subsection{Pure $T$-mesons at colliders} \label{subsec:C}

Finally, let us estimate the relevance of pure $T$-meson production at hadron colliders such as LHC. 
  We first discuss the mesons composed of scalar $T$-particles. 
Since, $T', \bar{T}'$ defined in 
(\ref{TTprime}) are both mass as well as color-interaction eigenstates, they can be pair-produced in gluon interactions, 
  \begin{equation}
      g  + g  \rightarrow T' + T'^*\, ~ {\rm and } ~\,  g + g  \rightarrow \bar{T}' + \bar{T}'^* 
  \end{equation}

To this end, we want to compare the total cross section of $\sigma_{gg }\rightarrow T'T'^*$ with the cross section of bound state production $\sigma_{gg \rightarrow B(T'T'^*)}$ by calculating the ratio
\begin{equation}
    R = \frac{\sigma_{gg \rightarrow B(T' T'^*)}}{\sigma_{gg \rightarrow T' T'^*}} \;.
\end{equation}
and analogously for $\sigma_{gg \rightarrow \bar{T}'\bar{T}^{'^*}}$.
For the estimate, we use the analogous cross-sections for the $s$-squarks given in \cite{Herrero:1987df}
\begin{equation}
    \sigma_{gg \rightarrow B(T' T'^*)} = \frac{ \pi^2 \alpha_s^5}{162 m_T'^2} \tau \int_{\tau}^{1} \frac{\mathrm{d}x}{x}\,
  f_{g}(x)\, f_{g}(\tau/x) \;,
\end{equation}
where $\tau = \frac{4m_T^2}{s}$ and $f_g(x)$ is the gluon partition density function as extracted from \texttt{LHAPDF} \cite{Butterworth:2015oua},
\begin{equation}
f_g(x) \approx x^{-1.3}(1 - x)^5 \;.
\end{equation}
The pair-production cross section of colored scalars which is directly applicable to the present case is given by \cite{Kramer:1997hh},
\begin{equation}
    \sigma_{gg \rightarrow TT^*} = \frac{5\pi \alpha_s^2}{48 s} \int_\tau^1 \frac{d\tau^\prime}{\tau^\prime} \int_{\tau^\prime}^1 \frac{dx}{x} f_{g}(x)\, f_{g}(\tau^\prime/x) \;,
\end{equation}
where we used the approximation $m_{\bar{T}'}^2,
m_{T'}^2\ll s$.  For illustrative purposes, we 
can estimate the expected ratio between the bound states and produced $T$-particles taking $m_T = 1 \textrm{ TeV}$ and $\sqrt{s}= 14 \textrm{ TeV}$, 
which gives,  
\begin{equation}
    R \sim \mathcal{O}(10^{-2}) \;.
\end{equation} 

This estimate gives a ratio of order $1\%$ between the direct production cross section for pure $T'T'^*$-bound states and the inclusive $T'T'^*$-pair-production cross section. It therefore suggests that directly produced pure $T$-mesons are relatively rare. By contrast, the individual $T$-particles can hadronize into $Tq$-type hadrons containing Standard Model quarks. A quantitative comparison of the final abundances of these two classes of hadrons would, however, require a more detailed treatment of hadronisation and bound state formation. Pure $T$-mesons may consequently be less relevant for collider discovery searches, although they can play an important role in cosmology, as discussed below.

\section{Cosmological bounds on $TT$-Mesons} \label{sec:cosmobounds}

The  $T$-particles as well as their composites could be formed in the early universe. 
  If the reheating temperature after inflation is above the mass of $T$-particles, their abundance will be the same as of ordinary quarks, since they will quickly thermalize 
 via gluon interactions. 

 The subsequent cosmological history is as follows.   After the temperature drops below their masses, the $T$-particles become non-relativistic and their  energy density
 red-shifts as matter.  At this point, the 
 $T$-particles can start forming various bound states. We shall focus on the meson type bound states of the form $TT$, where constituents can 
 have different spins.

  After the temperature drops below the QCD scale, the color gauge forces enter the confining phase and  the $T$-particles form colorless bound states
 with ordinary quarks as well as with other $T$-particles. 
 
 In subsequent evolution the number of $T$-hadrons can be reduced by 
 the following mechanisms:

{\it 1)} The annihilations with hadrons containing $T$-antiparticles;
 
{\it 2)}  Decay of  $T$-hadrons via some  previously discussed $T$-number conserving transitions. 
  
 
 {\it 3)} The direct $T$-number violating decays  of the constituent $T$-particles.

In this respect one must take into account that 
the interaction cross sections of $TT$ and $Tq$
hadrons are very different. 
The interaction cross sections of $Tq$ hadrons are given by the 
 QCD cross-section $\sigma_{\textrm{QCD}}$.  

The situation is different for a $TT$-meson  which is a bound state of two particles with the masses much above the QCD scale. In such a case the interaction cross section of the bound state  is not determined by $\sigma_{\textrm{QCD}}$ but rather by its Bohr radius, which is much smaller \cite{DeLuca:2018mzn}. In the present case the Bohr radius $R_T$ is given by 
d(\ref{RTradius}). 
Correspondingly, $TT$-mesons do not annihilate with each other very easily. In this situation a significant number of $TT$-mesons can survive and decay much later or never. 
    
  On the other hand,  the $T$-particles with intermediate lifetimes are constrained by cosmology.

\subsection{Bound from BBN} \label{subsec:A}

The decay products of $T T$-meson might affect the ratio between protons and neutrons in the early universe. 
In order not to disturb the standard dynamics of 
the BBN, the decays of $T$-particles must not inject too much energy in the thermal bath during the time of BBN.

 As we already discussed, the lifetimes of $TT$-mesons 
of scalar-scalar and fermion-fermion compositions are very short, $\sim 10^{-16}$ sec or so (see eqs.\eqref{TT2GaugeBosons} and \eqref{TT2Gauginos}). Such bound states are therefore irrelevant for BBN. 
The effect on BBN can potentially come from  $T\tilde{T}$-type bounds states provided they decay after the start of BBN, $t_{BBN} \sim 10$ sec.  This can happen in one of the following situations:

   {\it 1)}
 The LSP is the gravitino and the triplino is 
 NLSP. In this case the $T\tilde{T}$ meson 
can decay gravitationally with the  
lifetime given by (\ref{gravilt}).
  
  {\it 2)}  The mass of the $T\tilde{T}$-meson is below (or is almost degenerate with) the LSP and the Yukawa couplings of $T$-particles are generated in the first order in  $\frac{m_{3/2}}{M_{\rm GUT}}$.

\begin{figure}
    \centering
    \includegraphics[width=1\linewidth]{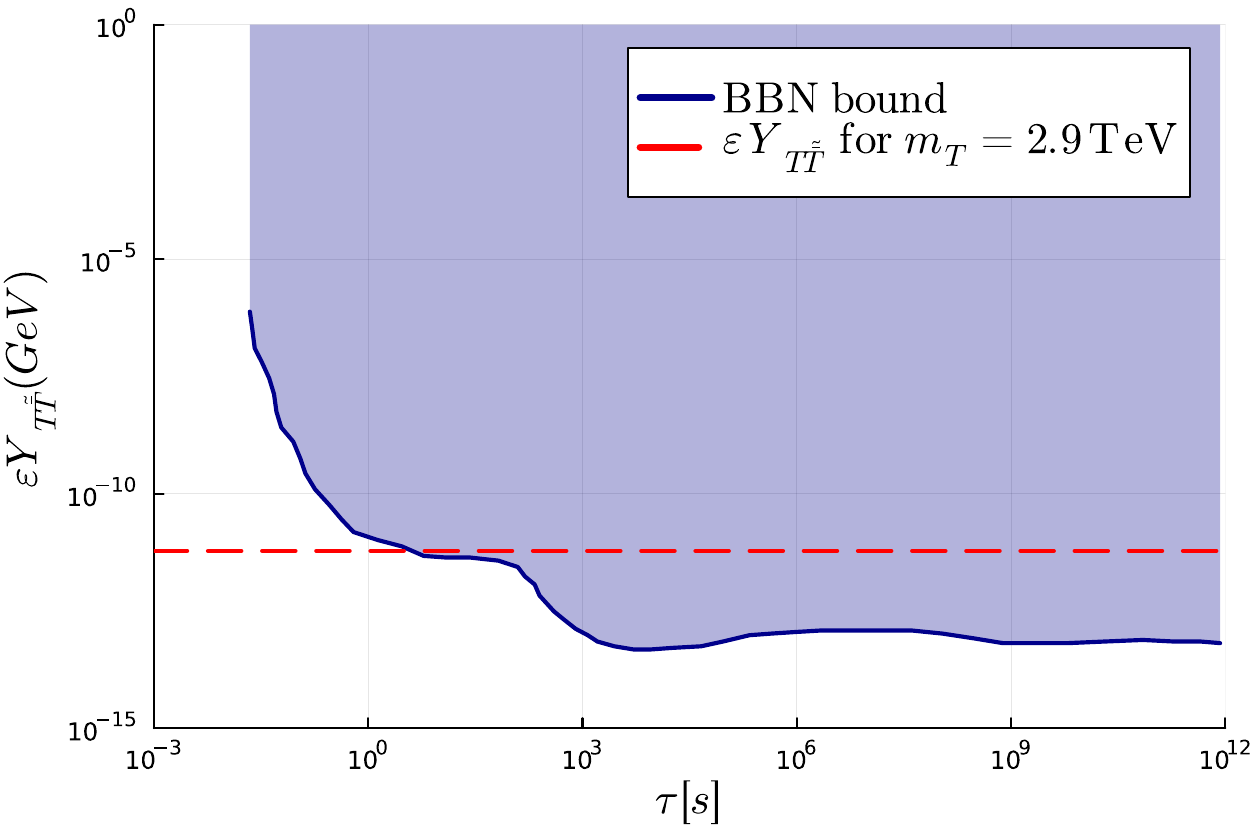}
    \caption{Upper limits (adapted from Figure 9 of Ref. \cite{Steffen:2006hw}) on hadronic energy $\epsilon$ release from hadronic decays of the $T\tilde{\bar{T}}$ as a
function of its lifetime $\tau$. The horizontal line shows the expected energy release of a thermally produced $T\tilde{\bar{T}}$ like in \cite{DeLuca:2018mzn} with exemplary  mass $m_T \approx 2.9$ TeV.}
    \label{fig:BBN}
\end{figure}

 In case of the first item, as a guideline we can use the analysis of \cite{Steffen:2006hw} which constrains the masses of the next to LSP particles. In the present case, this can be applied  to $T$-particles. The constraints are shown in 
 Fig. \ref{fig:BBN}.
Here, a relevant parameter is the ``yield" of a particle $X$, $Y_X=\frac{n_X}{s}$, with entropy density $s$ and $n_X$ is the number density of $X$, evaluated after freeze-out.

Of course, the yield for currently discussed $T$-particles, $Y_T$, depends on their exact production mechanism. For the thermal mechanism presented in \cite{DeLuca:2018mzn} $Y_T$ can be predicted;
For example, 
$Y_{T\tilde{T}} \sim [10^{-17}, \;10^{-14}]$ for a mass range of $m_T = [1.1,\; 7]$ TeV and $\epsilon_{had} \sim m_T$~\cite{DEA}. Then, for the corresponding $T$-particle with the mass $m_T \geq 1.5$ TeV the lifetime has to obey
\begin{equation}
    \tau_{T\tilde{\bar{T}}} \lesssim 1000 \; \text{s}.
\end{equation}

Now, it is clear from  the estimated structure of the lifetime (\ref{gravilt}) that BBN constraint is accommodated for the triplet masses of few tenths of TeV.  In such a case,  regardless of their abundance, the lifetime can be sufficiently short for not interfering with BBN.   

\begin{figure}
    \centering
    \includegraphics[width=1\linewidth]{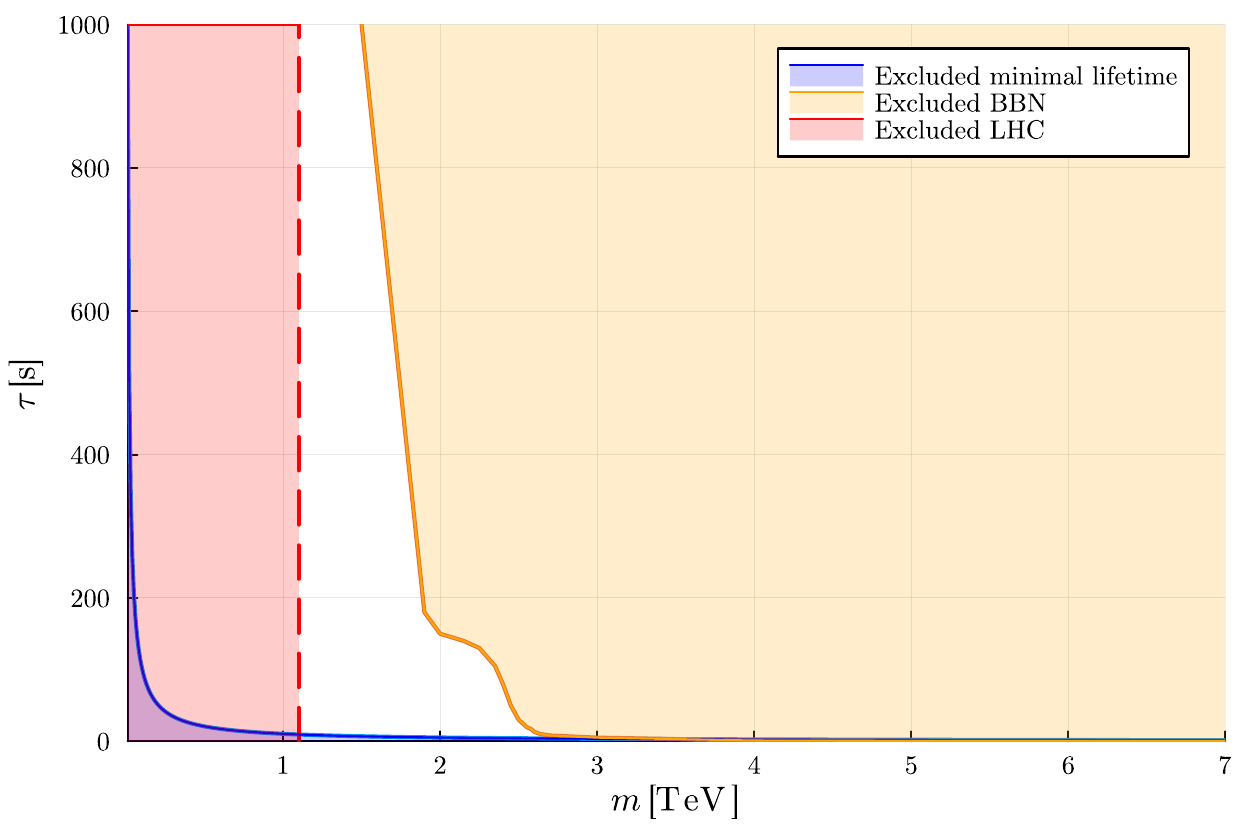}
    \caption{ An example, of available parameter space for the lightest $T$-scalar in case of item {\it 2)} in which we assume that triplino $\tilde{T}$ is LSP. Correspondingly, the only decay channel of the scalar-fermion bound state is through the decay of the scalar $T$-particle via the Yukawa interactions. The red line marks the LHC bound on production of colored scalars reported in \cite{ATLAS:2019gqq}. 
     The blue line corresponds to minimal possible lifetime in this model calculated via \eqref{lifetimeeq} by setting the Yukawa coupling to one and choosing $m_{3/2}=1\;\mathrm{TeV}$ as a benchmark value. Of course, this benchmark choice is subject to the assumed LSP mass hierarchy and should not be interpreted as a model-independent upper bound on $m_{3/2}$.
      The orange bound arises from BBN as described in the text. Notice that the plot covers only a limited parameter space. In particular, for a very long lived case (\ref{maxilt}), there is no upper bound from BBN.      
      For a detailed view of the minimal BBN bound and the minimal lifetime, see FIG. \ref{fig:zoom}. 
   }
    \label{MoneyPlot}
\end{figure}

 Let us now move to the item {\it 2)}. 
 We have seen that, if the Yukawa coupling is generated in leading order in $\frac{m_{3/2}}{M_{\rm GUT}}$, for $T$-particle masses of order TeV, the lifetime (\ref{minilt}) is around a second. This is within the validity of the BBN constraint. In Fig. \ref{MoneyPlot} we show the collection of bounds for the lightest $T$-scalar for masses of order TeV and for leading order generated Yukawa couplings of order $\sim \frac{m_{3/2}}{M_{\textrm{GUT}}}$. We thus see that the scenarios of the type {\it 2)} that are within a potential reach of future collider searches, are also consistent with the BBN bounds.  

\begin{figure}
    \centering
    \includegraphics[width=1.0\linewidth]{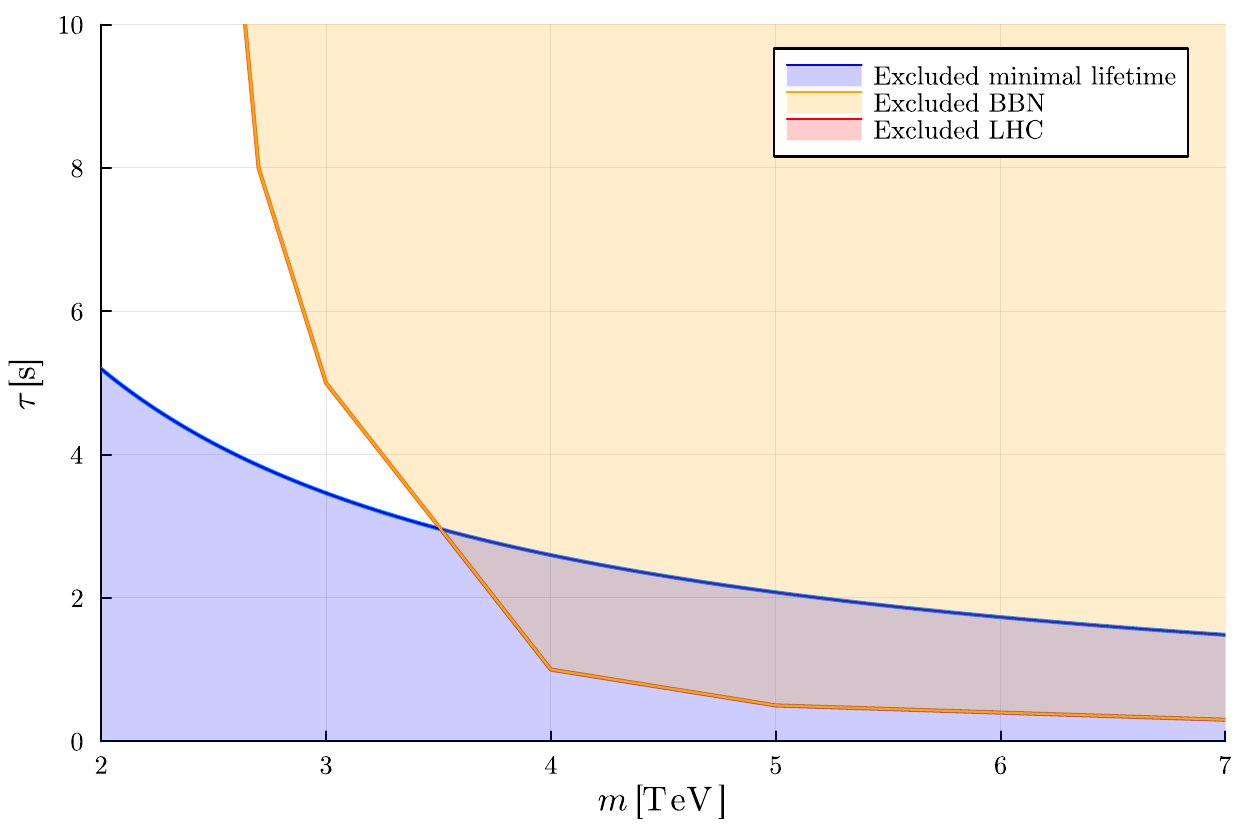}
    \caption{Available parameter space for $T$ in case of item {\it 2)}. This is a zoomed-in version of the plot \ref{MoneyPlot}.
    }
    \label{fig:zoom}
\end{figure}

On the other hand, if the Yukawa coupling is generated in higher order in $\frac{m_{3/2}}{M_{\rm GUT}}$, the lifetime increases dramatically (\ref{maxilt}) and exceeds the current age of the Universe. Obviously, in such a case, BBN imposes no constraint on $T$-particles.

\subsection{Bound from Diffuse Gamma Ray Background}
\label{subsec:A}

Cosmology provides us with another upper bound on $T$-lifetime. This bound comes from the CMB. Even if the $T\tilde{T}$ bound state evades the BBN-bound, as discussed above, the $Tq$-type bound states could still survive until later times beyond recombination. If then the $T$-particle decays, it may lead to the creation of the gamma rays that could exceed the detected diffuse gamma-ray background. 

Such analysis has already been performed for a WIMP-scenario \cite{Kamionkowski:1994je} and can be easily translated into our model. In 
particular,  in the regimes  discussed above, in which both items {\it 1)} and {\it 2)} accommodate the BBN constraints, they also satisfy the CMB bounds.

\subsection{Can $TT$-meson be a Dark Matter?}

  A potentially interesting region of parameter space is when triplino is the LSP and at the same time 
  the Yukawa coupling of the $T$-scalars
  is induced at order 
  $\frac{m_{3/2}^2}{M_{\rm GUT}^2}$ or higher. 
  In such a case, the $T\tilde{T}$-type mesons are stable on the scale of the current age of the Universe.  The question then emerges 
  whether such bound states can potentially serve as dark matter.  They do carry the two 
  immediate features desired for a ``canonical" cold dark matter: stability as well as a strongly suppressed interaction cross section. 
  
Of course, the above features are by no means sufficient and, in particular, one needs to guarantee the right 
abundance.  However, in the present case the story is more subtle due to the following issue. 

 The stable $T\tilde{T}$-type mesons are expected to be accompanied by the stable $\tilde{T}q$-type states, which can be electrically neutral or charged. Such states however are subjected to extremely severe constraints from searches for anomalously heavy isotopes in terrestrial and extraterrestrial matter, as well as from cosmological considerations, including Big Bang nucleosynthesis. These constraints can significantly restrict their allowed abundance, although converting isotope-search limits into bounds on their primordial cosmological abundance requires assumptions about the capture, transport, and accumulation of the relics. 
 
 For example, very strong constraints on this scenario are coming from searches for anomalously heavy hydrogen, 
 in which a hypothetical heavy, positively charged state would act as the nucleus of a hydrogen-like atom. In the present case, the natural candidate for such a particle is a $Tq$-meson~\cite{Yamagata:1993jq}. 
 
 These experiments have searched for such isotopes with masses up to TeV within deep sea water that has been enriched by a centrifugal treatment. Their results impose the following upper bound on the ratio between ordinary nucleons $N_n$ and $Tq$-mesons, $N_{Tq}$,
 \begin{equation}
     \frac{N_{Tq}}{N_n} \lesssim 10^{-16} \;.
 \end{equation}

Nevertheless, major uncertainties remain, whether for such experiments all the systematics have been taken into account correctly
 \cite{Ebadi:2026umu}.
For instance, quite important is the geological history of the earth. During the time of a fluid proto-earth, such heavy isotopes would sink to the core of the earth and today would not show up in appreciable abundance within the sea water. Also the enrichment process with industrial centrifuges that have been utilized for these studies is under debate \cite{Ebadi:2026umu}. 

Studies with better understood systematics as well as the charge independent ones are presented by the searches for such particles within meteoroids that possibly collect the $Tq$-mesons that are present within the dark matter  wind. The resulting bound on the ratio is \cite{Polikanov:1990sf}
\begin{equation}
    \frac{N_{Tq}}{N_n} \lesssim 10^{-14} \frac{10\textrm{ TeV}}{M_{Tq}}\;.
    \label{meteoridbound}
\end{equation}
The predicted ratio present within such meteoroid probes has been calculated in \cite{DeLuca:2018mzn} as 
\begin{multline}
\frac{N_{Tq}}{N_n}
  = n_{Tq}\,\sigma_{\mathrm{capture}}\,v_{\mathrm{rel}}\,\Delta t \approx
 \\\\ 10^{-12}
  \left(\frac{\sigma_{\mathrm{capture}}}{1/\Lambda_{\mathrm{QCD}}^{2}}\right)
  \left(\frac{10\,\mathrm{TeV}}{M_{Tq}}\right)
  \left(\frac{f_{Tq}}{10^{-5}}\right)
  \left(\frac{\Delta t}{5\,\mathrm{Gyr}}\right)
  \left(\frac{v_{\mathrm{rel}}}{10^{-3}}\right) \;,
\end{multline}
where $\sigma_{\mathrm{capture}}$ is the capture cross section of the target particle (in the present case the $Tq$-meson), $\Delta t$ is the lifetime of the meteoroid, $v_{rel}$ is its relative velocity to the DM wind and $f_{Tq} = \rho_{Tq} / \rho_{\textrm{DM}}$ is the fraction of the searched particles ($Tq$-mesons) in dark matter.  All quantities are normalized to their typical values. Comparing the above with the experimental constraint \eqref{meteoridbound} one gets the following bound, 
\begin{equation}
f_{Tq}
  \lesssim 10^{-5}
\left(\frac{0.01/\Lambda_{\mathrm{QCD}}^2}
{\sigma_{\mathrm{capture}}}\right)
\;.
\end{equation}

 Thus, in any viable scenario, one must ensure 
 that the abundance of $Tq$-mesons relative to $T\tilde{T}$-types is extremely suppressed. Now, as already said, the $TT$-mesons start forming after the temperature drops below the $T$-particle masses.   
 As the Universe cools below the $T$-particle masses, the surviving $T$-sector particles can form Coulombic $TT$ bound states once thermal dissociation becomes sufficiently suppressed. These bound states are expected to be  electrically neutral, since the constituents carry opposite electric charges.
 Also, although this requires a careful investigation, we can assume that the baryonic types, involving three 
 $T$-particles, are very rare.

 However, alongside of $TT$-mesons, the formation of some $Tq$-type states
 is inevitable. In particular, not all $T$-particles will pair up and some fraction shall remain free all the way till lower temperatures. Such free ones shall produce stable bound states with ordinary light quarks below the QCD temperatures.  Although the 
 number of $\tilde{T}q$-mesons relative to 
 $\tilde{T}T$ is expected to be reduced by 
 more efficient annihilation, very likely, such reduction is not nearly sufficient for the phenomenological viability.

 Thus, satisfying the existing bounds, requires either a sufficiently small triplino abundance before the QCD transition or efficient depletion of the stable $Tq$-type hadrons formed during hadronization.
Of course, one could also look for a mechanism 
that selectively suppresses the abundance of $Tq$-mesons while leaving the neutral $TT$ bound states sufficiently abundant to account for the dark matter.   We leave this as an open question.

 If the required relative abundance cannot be achieved, the experimental bounds must be satisfied by the uniform suppression. This can be easily achieved by constraining the reheating temperature by the triplino mass. In such a regime, both their abundance as well as their 
 contribution to dark matter will be exponentially suppressed.

The good news is that with extraordinary experimental sensitivity towards heavy stable nuclei, even with exponentially suppressed abundance, the $Tq$-type mesons can be searched for. A promising experimental proposal for such searches is given in~\cite{Ebadi:2026umu} which aims to reach a sensitivity for $Tq$-type mesons around masses of $\mathcal{O}(10) \textrm{ TeV}$ of 
\begin{equation}
    f_{Tq} \lesssim 10^{-18} \;.
\end{equation}

\section{Example of Long-Lived $T$-meson Parameter Space} \label{sec:parameterexample}

 In this section we provide an example of the parameter space  in which, after accounting the RG running,  ${T}'\tilde{\bar{T}}$-meson is stable up to decays induced by  $U(1)_T$-violating Yukawa couplings. 

Let us first focus on the channel in which $T'\tilde{\bar{T}}$ decays into a gauge boson and a gaugino. Specifically, at the low-energy scale $Q_0$ of several $\rm TeV$, the condition 
\begin{equation}\label{decayKinematicCondition}
    m_{T'\tilde{\bar{T}}}\simeq m_{T'}(Q_0)+\tilde{m}_{T}(Q_0)<m_{\rm LG}(Q_0)
\end{equation}
must be satisfied to kinematically forbid the corresponding decay. Here, $m_{\rm LG}$ denotes the lightest gaugino mass. In the following, we consider a simple case where only the $\mu$-parameter acquires a shift induced by the non-minimal K\"ahler function. Given the lower bound on the $T$-particle mass of $1.1\,\rm TeV$, we will also provide some benchmark parameter choices. 

In the presence of $T$-particles, which form a vector-like pair under the SM gauge group: 
\begin{equation}
    T=(\mathbf{3},\mathbf{1},-\frac{1}{3})\;, \qquad \bar{T}=(\bar{\mathbf{3}},\mathbf{1},\frac{1}{3})\;, 
\end{equation}
at one loop, the gauge couplings of the supersymmetric theory obey
\begin{equation}
    \frac{d g_i}{d \ln Q}=\frac{b_i}{16\pi^2}g_i^3\;, \qquad b_i=(7,1,-2)\;, 
\end{equation}
where $i=1,2,3$ correspond to $U(1)_Y, SU(2)_L,SU(3)_c$, respectively, and the GUT normalization $g_1=\sqrt{5/3}g_Y$ is used. From the one-loop RGEs of the gaugino mass $M_i$ 
\begin{equation}
    \frac{d M_i}{d \ln Q}=\frac{2b_i}{16\pi^2}g_i^2 M_i\;, 
\end{equation}
we obtain the standard relation 
\begin{equation}\label{oneLoopRelation}
    \frac{M_i(Q_0)}{M_i(M_{\rm GUT})}=\frac{g_i^2(Q_0)}{g_i^2(M_{\rm GUT})}\;. 
\end{equation}
Taking $M_{\rm GUT}\sim10^{16}{\rm GeV}$ and $g_{\rm GUT}^2/4\pi\simeq1/24$, we can calculate the numerical factors approximately: 
\begin{equation}
\begin{split}
    M_1(Q_0)\simeq0.4 M_1(M_{\rm GUT})\;, \\
    M_2(Q_0)\simeq 0.8 M_2(M_{\rm GUT})\;,\\
    M_3(Q_0)\simeq 1.6 M_3(M_{\rm GUT})\;. 
    \end{split}
\end{equation}
In our model, the $T$-particle masses are given by
\begin{equation}
    \hat{\mathcal{M}}_T=\begin{pmatrix}
        m_{3/2}^2+\mu^2_T & B_\mu \\
        B_\mu & m_{3/2}^2+\mu_T^2
    \end{pmatrix}\;, 
    \qquad 
    \tilde{m}_{T}=\mu_T\;, 
\end{equation}
where $m_{3/2}\sim \mathcal{O}({\rm few\  TeV})$ and $B_\mu \sim m_{3/2}^2$. Neglecting the small Yukawa couplings, the one-loop RGEs are 
\begin{equation}
    \begin{aligned}
        \frac{d \mu_T}{d \ln Q}&=-\frac{1}{16\pi^2}\mu_T\left(\frac{16}{3}g_3^2+\frac{4}{15}g_1^2\right)\;, \\
        \frac{d m_{T,{\rm soft}}^2}{d \ln Q}&=-\frac{1}{16\pi^2}\left(\frac{32}{3}g_3^2 M_3^2+\frac{8}{15}g_1^2M_1^2\right)\;. 
    \end{aligned}
\end{equation}
Notice that $g_2$ and $M_2$ do not enter these RGEs because $T$ is an $SU(2)_L$-singlet. Using the one-loop relation $M_i/g_i^2={\rm const.}$ shown in \eqref{oneLoopRelation}, these equations can be integrated analytically to give

\begin{flalign}
&\mu_T(Q_0)\simeq \mu_T(M_{\rm GUT})
\left(\frac{M_3(Q_0)}{M_3(M_{\rm GUT})}\right)^{\frac{4}{3}}
\left(\frac{M_1(Q_0)}{M_1(M_{\rm GUT})}\right)^{-\frac{2}{105}}
\notag\\
&\phantom{\mu_T(Q_0)}\simeq 1.9\,\mu_T(M_{\rm GUT})\;,
&&\notag\\[4pt]
&m_{T,{\rm soft}}^2(Q_0)\simeq m_{3/2}^2
-\frac{4}{3}\left(M_3^2(M_{\rm GUT})-M_3^2(Q_0)\right)
\notag\\
&\phantom{m_{T,{\rm soft}}^2(Q_0)\simeq m_{3/2}^2}
+\frac{2}{105}\left(M_1^2(M_{\rm GUT})-M_1^2(Q_0)\right)\;.
&&
\end{flalign}

  Diagonalizing $\hat{\mathcal{M}}_{T}$, we obtain the mass of the lighter triplet $T'$: 
\begin{equation}\label{T'massConstrain}
    m_{T'}(Q_0)\simeq \sqrt{m_{T, {\rm soft}}^2(Q_0)+\mu_T^2(Q_0)-B_\mu(Q_0)}\;. 
\end{equation} 
The experimental bound of the $T$-particle mass and the kinematic condition \eqref{decayKinematicCondition} now give the following constraints
\begin{equation}\label{decayConstraints}
    \begin{aligned}
        \mu_T(Q_0)&\geq 1.1{\rm TeV}\;,\\ 
        m_{T'}(Q_0)&\geq 1.1{\rm TeV}\;,\\
        m_{T'}(Q_0)+\mu_T(Q_0)&<\min_i{M_i(Q_0)}\;. 
    \end{aligned} 
\end{equation}
For a universal-gaugino-mass benchmark, we assume
\begin{equation}
    M_i(M_{\rm GUT})\equiv m_{1/2}\sim m_{3/2}\;, \quad i=1,2,3\;,
\end{equation}
and obtain
\begin{equation}
    \begin{aligned}
        m_{T, {\rm soft}}^2(Q_0)&\simeq m_{3/2}^2+2.1m_{1/2}^2\;,\\
        \min_i{M_i(Q_0)}  &=M_1(Q_0)\simeq 0.4m_{1/2}\;. 
    \end{aligned}
\end{equation}
Using the lower bound of the $T$-particle mass, the last constraint in \eqref{decayConstraints} then requires $m_{1/2}>5.5\,{\rm TeV}$. As a benchmark choice, let us take $\mu_T(Q_0)=1.1\,{\rm TeV}$. The last two inequalities in \eqref{decayConstraints} require a consistency check, \ie, 
\begin{equation}
\begin{split}
    m_{3/2}^2+1.94m_{1/2}^2+0.8\mu_T(Q_0) m_{1/2}\\<B_\mu(Q_0)\leq m_{3/2}^2+2.1m_{1/2}^2\;, 
    \end{split}
\end{equation}
which admits a solution provided that $m_{1/2}>5.0\mu_T(Q_0)\simeq 5.5\,{\rm TeV}$. Therefore, we can take $m_{1/2}=m_{3/2}=6\,{\rm TeV}$, $B_\mu(Q_0)=(10.56\,{\rm TeV})^2$, which results in $m_{T'}(Q_0)=1.14\,{\rm TeV}$. As can be checked, $m_{T'\tilde{\bar{T}}}(Q_0)\simeq 2.24\,\rm{TeV}<2.4\,{\rm TeV}=M_1(Q_0)$, which means the $T'\tilde{\bar{T}}$-meson is kinematically stable against $U(1)_T$-conserving decays. 

For completeness, other $U(1)_T$-conserving decay channels should also be considered, in particular $T'\tilde{\bar{T}}\to f\tilde{f}$ with a relatively short lifetime, as shown in Table \ref{tab:TTmeson-regimes}. This decay channel is kinematically forbidden as long as $m_{T'\tilde{\bar{T}}}(Q_0)<m_{\tilde{f}}(Q_0)+m_{f}(Q_0)$. On the other hand, the decays involving a gravitino have much longer lifetimes. Moreover, these decays can be kinematically forbidden if the gravitino is not the LSP, as in the benchmark scenario discussed above.

Since $U(1)_T$-violating decays are highly suppressed by the small Yukawa couplings, the $T'\tilde{\bar{T}}$-meson can consequently be long-lived.  In Model \textrm{II}, presented in 
Appendix \ref{appD}, in which 
the Yukawa coupling of the triplet is not generated in the leading order in $\frac{m_{3/2}}{M_{\rm GUT}}$, the lifetime of the $T$-meson 
(\ref{maxilt}) exceeds the current age of the Universe.

  \section{Status of other vacua} \label{sec:vacua} 
  
 In the unbroken global SUSY limit, the presented models (the Models I and II) 
 possess many additional vacuum states described by
 other solutions of the supersymmetric vacuum equations. 
 For example, in both models 
 there exists a trivial solution with unbroken 
 GUT symmetry in which the VEVs of all the GUT superfields are zero. 
  In the globally supersymmetric limit,  all these vacua are degenerate in energy.  
  
  The degeneracy is lifted once the supergravity as well as the supersymmetry-breaking effects are taken into account.  
   The question then emerges about the status of the alternative vacua, and whether one needs to worry about their existence. For example,  can our vacuum be metastable with an unacceptably short lifetime?  
  The answer to this question is negative, as we shall explain.
  
   Before proceeding, we wish to stress that
   the existence of multiple vacua that are  split in energy by supergravity, is a common feature of GUTs.  Due to this, the conclusions reached in our analysis are not limited to the presented models and are applicable to much wider class of supergravity GUTs.  In particular, our findings are going to validate some previously proposed consistency constraints. These are: {\it 1)} the \textit{no de Sitter} proposal~\cite{Dvali:2013eja, Dvali:2014gua, Dvali:2017eba}, which forbids vacua with positive energy densities; and {\it 2)} the criterion of \textit{Minkowski safety}  ~\cite{Dvali:2011wk}, which excludes tunneling from any Minkowski vacuum.   We shall come back to establishing the above connections in due time.

   In the globally supersymmetric limit, 
  our vacuum is separated from the rest of the degenerate vacua by the energy barriers set by the GUT scale. 
   Since the vacua are degenerate in energy, there inevitably exist domain walls that interpolate between the neighbouring
   vacua.  In the global SUSY limit, the straight infinite wall separating such vacua is topologically stable. 
The detailed analysis and explicit constructions of domain walls and vacuum 
bubbles of global supersymmetry can be found 
in \cite{Dvali:1996bg, Dvali:1996xe, Dvali:2002fi, Dvali:2020wqi}. Here we shall only use their essential features.  

In particular, since the masses of the fields involved are $\sim M_{\textrm{GUT}}$,  
 the wall tension (i.e., the energy per unit surface area) is  
    \begin{equation}  \label{Tension} 
      T_{\rm w} \geqslant  \frac{M_{\textrm{GUT}}^3}{({\rm coupling})^2} \,,
   \end{equation}  
  where the  ``coupling" in the denominator is the typical 
  non-linear coupling of the fields composing the wall. 
   The thickness of the wall is given by the inverse mass of the interpolating field, $\delta \sim M_{\textrm{GUT}}^{-1}$.  
    
    Let us now assume that the energies of the vacua are split. 
    In the standard treatment  ~\cite{Kobzarev:1974cp, Coleman:1977py}, the would-be tunneling from one vacuum into another  requires a formation of the 
   bubble of the neighboring vacuum, which then could expand classically.      
     Let us consider a spherical vacuum bubble of radius $R$
     that separates the two SUSY vacua. We shall assume that the bubble 
     radius is much larger than its thickness 
     $R \gg \delta$.  In this regime the bubble wall 
    profile is well approximated by the planar one.    
      It shall become obvious that this is an excellent approximation in all situations relevant for our studies.   
     
   In this regime we can split the contribution to the bubble energy in two parts,    
\begin{align}  \label{energy(R)} 
    E(R) \, = \, 4\pi \left ( - \frac{1}{3} R^3 \Delta V  +  T_{\rm w} R^2
    \right ).  
\end{align}       
        The first term is the volume energy of the interior and is given by the potential energy density difference between the vacua,  $\Delta V$. The second term  is the surface energy
   given by the  bubble wall tension, $T_{\rm w}$ (\ref{Tension}).
   
   In order for the tunneling to be possible,  a critical bubble 
   of zero energy must exist.   If such bubble exists, its radius, $R_c$, can be found by solving the equation $E(R_c)=0$, which 
   for  $\Delta V \, > \, 0$ gives, 
  \begin{align}  \label{Rcrit} 
    R_c \, = \, 3\, \frac{T_{\rm w} }{\Delta V}  \,. 
\end{align}  
  As usual ~\cite{Coleman:1977py},   the rate  of the bubble-nucleation is exponentially suppressed by the 
   factor ${\rm exp}(-S_E)$, where 
   $S_E$ is the action of the corresponding $O(4)$-invariant Euclidean bounce.

    Since the globally-supersymmetric vacua are exactly degenerate,  i.e., $\Delta V= 0$,  any bubble costs non-zero energy. This forbids tunneling.  Equivalently, 
    for $\Delta V= 0$ there exist no bounce with a finite $S_E$.   
  
  Let us now take into account the effects of the gravity-mediated supersymmetry breaking.
  The schematic form of the supergravity scalar potential is, 
 \begin{equation} \label{SugraPot} 
 V = {\rm e}^{\sum_Z |Z|^2/M_P^2}\left (\sum_Z |F_{Z}|^2  - 3 |W|^2/M_P^2 \right ) \,, 
  \end{equation} 
 where the sum runs over the scalar components 
 of all chiral superfields, 
 $Z$, and for definiteness we assumed a minimal 
 K\"ahler metric and ignored the $D$-terms. 
  
    As it is clear from (\ref{SugraPot}), the vacuum energy has positive contribution 
   given by the $F$-terms and the negative contribution 
   given by the VEV of the superpotential. 
 We are discussing the gravity-mediated scenario, in which supersymmetry is broken by the $F_X$-term of a hidden sector superfield $X \in Z$, which is decoupled from the GUT sector in the superpotential.  
 
  From the Goldstino Argument of~\cite{Dvali:2020uvd} 
  we know that the $F_S$-terms of the heavy GUT superfields, $S \in \,$GUT, satisfy the bound (see, the detailed discussion in the appendix \ref{appB}),
\begin{equation}  \label{GrShiftF}
   |F_{S}|  \lesssim m_{3/2}^2\;,
  \end{equation} 
 and are negligible as compared to $F_X$ which is of order, 
 \begin{equation}  \label{FXX}
   |F_{X}|  \sim M_Pm_{3/2} \;.
  \end{equation} 
 This is true in all the vacua 
    of the GUT sector, since in $M_P=\infty$ limit  
    they all preserve supersymmetry. 
   
 Now comes an important point.  In ``our"  vacuum, we must fine-tune   
  the vacuum energy to zero  via the condition 
\begin{equation} \label{VzeroC} 
 \sum_Z |F_{Z}|^2  = 3 |W_{\rm our}|^2/M_P^2 \,, 
  \end{equation} 
 where $W_{\rm our}$ implies the VEV of the superpotential in the phenomenologically desired  GUT vacuum. For example, in Model 
 I and Model II these are given by the VEVs 
 (\ref{SUSYVEVs}) and (\ref{VEVsM2}) respectively. 
  
This is usually achieved by tuning 
 the overall constant in the superpotential.  
    In each model, the fine-tuning can be done only once with the choice of a particular vacuum. 
    
Since the dominant positive contribution is coming from the $F_X$-term (\ref{FXX}), whereas all other $F_S$-terms according to the Goldstino argument~\cite{Dvali:2020uvd} 
are negligible (\ref{GrShiftF}), the tuning in  
(\ref{VzeroC}) is mainly between $W_{\rm our}$ and $F_X$, which gives, 
\begin{equation} \label{Wour} 
 W_{\rm our} \, \simeq \, |F_X|M_P/\sqrt{3} \, \sim \, M_P^2m_{3/2} \,.
  \end{equation}     
    However, other GUT vacua, while having 
  negligible $F$-terms, differ by the VEV of the superpotential.  
  The difference is, 
\begin{equation} \label{DeltaWG}
  \Delta W = |W_{\rm other} - W_{\rm our}|   
  \sim \frac{M_{\textrm{GUT}}^3}{({\rm coupling})^2}  \,.
  \end{equation} 
   For example, in Model I,  the superpotential  difference between ``our" vacuum (\ref{PATTERN}) with the VEVs (\ref{SUSYVEVs})
   and 
   the one with the unbroken GUT symmetry (all GUT VEVs zero) is: 
   \begin{equation} \label{DeltaWour}
   \Delta W = \frac{M\tilde{M}^2}{\alpha^2} 
   \frac{10}{3} \left (1 - \frac{2\lambda \tilde{M}}{9 \alpha M} \right ) 
   - \frac{1}{8} \frac{M_A M_{\chi}^2}{\beta^2} \,.  
   \end{equation} 

   Given that the mass parameters satisfy 
   $M,\tilde{M},M_A, M_{\chi} \sim M_{\textrm{GUT}}$, (\ref{DeltaWour})
   obviously confirms (\ref{DeltaWG}).
   For the sake of clarity, we shall take the dimensionless coupling constants in the superpotential to be order one (of course, without violating perturbativity).  

   The above implies that, once in our vacuum the energy is tuned to zero via (\ref{Wour}), the other vacua 
   have energy densities, 
   \begin{equation}  \label{VacSplit11}
     \Delta V \simeq - 3 \frac{|\Delta W|^2 + 2{\rm Re} [W_{\rm our} \Delta W^*]}{M_P^2}\,  
     \simeq \, - 3 \frac{|\Delta W|^2}{M_P^2} \,, 
  \end{equation} 
  where we have used (\ref{Wour}), (\ref{DeltaWG}) and the fact that, with $M_{\rm GUT} \sim 10^{16}$GeV and $m_{3/2} \sim $TeV, one has  $M_{\rm GUT}^3 \gg M_P^2m_{3/2}$.   This implies that the energy splitting in 
(\ref{VacSplit11}) is dominated by the first term.

Thus, any other GUT vacuum has the energy density
  \begin{equation}  \label{VacSplit}
     \Delta V \simeq - 3 \frac{|\Delta W|^2}{M_P^2} \sim 
     - \frac{M_{\textrm{GUT}}^6}{M_P^2} \,.  
  \end{equation} 
  
  We thus arrive at the following conclusion.  
  After taking into account the effects of supergravity and of supersymmetry breaking, 
  the GUT vacua become split in energy.
 More importantly, once we choose any desired vacuum to be a Minkowski vacuum,  
 none of the other GUT vacua can have positive energy densities. 
  This directly follows from the 
  Goldstino Argument~\cite{Dvali:2020uvd}, which 
  constrains the $F$-terms of heavy GUT superfields by (\ref{GrShiftF}). 
   This conclusion holds true for arbitrary  K\"ahler metric within the validity domain of EFT.

  This is striking, as it is an explicit manifestation of the no-de Sitter 
 proposal, put forward in~\cite{Dvali:2013eja, Dvali:2014gua, Dvali:2017eba}, which states  that a consistent theory of gravity is incompatible with any sort of stable or meta-stable de Sitter vacua.  
  As discussed in~\cite{Dvali:2020etd}, this incompatibility has a deep foundation  in the $S$-matrix consistency of  formulation of quantum gravity, the only known consistent formulation both in field theory as well as in string theory.

 Now, the GUT vacua that are not degenerate  with our Minkowski vacuum by symmetries, have negative energy densities given by (\ref{VacSplit}).  
 One may wonder whether this poses any danged 
 to our vacuum via a quantum decay. In the case of supersymmetric $SU(5)$ this question was answered negatively in~\cite{Weinberg:1982id}. 
  
 Let us analyze the stability and link it to the criterion of  \textit{Minkowski safety}. 
   First, in order to argue for stability, let us estimate the Euclidean 
  action  $S_E$ for the formation of the critical bubble.   
  Since the scale of the vacuum energy splitting is small 
  as compared to the energy barrier between the vacua, the radius of the bubble is much larger than the thickness. That is, the bubble 
  is in the thin-wall regime, with the radius, $R_c$ (\ref{Rcrit}). With our estimate of the vacuum energy splitting, this gives,   
\begin{equation}  
  R_c  \sim \frac{M_P^2}{M_{\textrm{GUT}}^3} \,. 
 \end{equation}   
   If we would ignore the gravitational effects of the  space-time curvature in the bubble interior,
   the Euclidean action would be, 
    \begin{equation} \label{SBounce}
      S_E \sim \frac{M_P^6}{M_{\textrm{GUT}}^6} \sim 10^{12}\,. 
      \end{equation}       
   This bounce already gives the enormous suppression, with the would-be lifetime exceeding the current age of the Universe  
 by many orders of magnitude.  
   
    However, in addition one must take into account the suppression effect discovered by 
  Coleman and De Luccia~\cite{Coleman:1980aw}. The point is that  
  the  gravitational curvature effects of the negative vacuum energy in the bubble interior  
  exert additional suppression 
  on the tunneling process, and can even eliminate it 
  altogether. In the thin-wall regime, this  happens when the energy splitting satisfies the following bound, 
  \begin{equation} \label{CDbound}
      |\Delta V| \leqslant \frac{3}{4} \frac{T_{\rm w}^2}{M_P^2} \,.
  \end{equation}
 In fact, the suppression of tunneling among 
$SU(5)$ vacua due to proximity to this bound was noticed in~\cite{Weinberg:1982id}.

 In the present case, we proceed in the following way. First, we notice that in case of unbroken supersymmetry, i.e., all $F$-terms zero, the tunneling from Minkowski vacuum must be strictly absent by supersymmetry.  This means that 
 in this case, the Coleman and De Luccia bound (\ref{CDbound}) must be satisfied.

 Next, let us break supersymmetry by the $F_X$-term.
 From the  Goldstino Argument~\cite{Dvali:2020uvd} the shifts of  the $F$-terms of GUT superfields and of their VEVs are bounded by (\ref{GrShiftF}) and (\ref{shiftVEV}) respectively, and the  splitting between the vacuum energies is given by (\ref{VacSplit}). Thus, both sides of the equation  (\ref{CDbound}) 
at most experience the relative corrections 
$\frac{m_{3/2}}{M_{\rm GUT}}$. 
 This means that the 
 Coleman-De Luccia bound (\ref{CDbound}) must continue to be satisfied in the leading order  even after breaking of  supersymmetry. 
 Of course, in this order, the tunneling is absent.  We thus see that, regardless 
 of model-building details, the Goldstino argument implies the absence of tunneling from the Minkowski GUT vacuum within 
 the accuracy $\frac{m_{3/2}}{M_\textrm{GUT}}$.

  The above situation strongly resonates with the  criterion  of \textit{Minkowski safety}~\cite{Dvali:2011wk}, which forbids the instability of  Minkowski vacua in a consistent theory. In particular, this criterion demands that in a supersymmetric theory, a vacuum,  even with broken SUSY, cannot be metastable \cite{Bajc:2011iu}\footnote{We must note that the Minkoswki safety ~\cite{Dvali:2011wk} fully supports the idea of so-called \textit{multiple 
 point criticality}~\cite{Bennett:1993pj, Bennett:1996vy}.}. 
  Again, similarly to \textit{no de Sitter} constraint 
 ~\cite{Dvali:2013eja, Dvali:2014gua, Dvali:2017eba}, the  
Minkoswki safety is directly rooted in the 
consistency of $S$-matrix formulation of gravity.

  In conclusion, already in the flat space analysis 
 the bounce estimate is extremely large, rendering  
 the vacuum of our interest essentially stable. 
 However, the GUT framework exhibits a much stronger tendency towards exact stability of the vacuum.  Indeed, we have seen that the \textit{Minkowski safety} requirement, which forbids tunneling, 
 is satisfied by GUTs to the leading order in $\frac{m_{3/2}}{M_\textrm{GUT}}$-expansion, without extra effort of a parameter choice.

\section{Conclusion}
\label{sec:conclusion}

Identification of the potential low energy signatures of grand unification is an outstanding task in fundamental particle physics and phenomenology. The model-independent degrees of freedom of any GUT are: {\it 1)} The gauge bosons that connect color $SU(3)_c$ and weak isospin $SU(2)_L$; and {\it 2)} Color triplet partner 
$T$ of the Higgs doublet. The particles of both categories are leptoquarks, as they can induce quark-quark as well as quark-lepton transitions. Therefore, these particles can mediate proton decay.

Due to this feature, a common lore is that the direct collider searches for the above GUT 
particles are hopeless for the current technology. 
The reason is that such particles are assumed to have masses around $M_\textrm{GUT}$, in order not to mediate fast proton decay. 

  However, it has been shown \cite{dvali1992, Dvali:1995hp}
  that the color-triplet partner $T$ of the 
 Higgs doublet, defies the above intuition. 
 The fundamental reason lies in the issue of the doublet-triplet splitting. The standard assumption is that the proton decay constraints force an extraordinary splitting  
 among the masses of the Higgs doublet and 
 its color-triplet partner. Namely, in
 order to suppress proton decay, the 
 $T$-component must gain a mass $\sim M_\textrm{GUT}$.  Of course, this would render 
 the $T$-particle irrelevant both for direct searches as well as for cosmology below 
 the GUT-scale temperatures. 

 However, as shown in  \cite{dvali1992, Dvali:1995hp},
 there exists an alternative scenario with fundamentally different phenomenological consequences in which the splitting occurs instead among the Yukawa couplings of 
 $T$ and $H$. In this scenario, the color-triplet emerges as a 
 long-lived particle, with the mass in the TeV range. Despite being light, the $T$-particle 
 cannot mediate proton decay in experimentally interesting range, since  $1/M_{\rm GUT}^4$ suppression in the proton decay rate occurs due to the suppressed Yukawa coupling rather than a high mass \footnote{ Notice that, as one of the original motivations in \cite{dvali1992}, the suppression of proton decay via the Yukawa couplings of $T,\bar{T}$ instead of their propagators, as an extra bonus, also eliminates the standard problem with  dimension-$5$ proton decay mediated by triplinos $\tilde{T}, \tilde{\bar{T}}$ \cite{Weinberg:1981wj, Sakai:1981pk}. }.

   As discussed in \cite{Dvali:2024jhs}, 
 this opens up a qualitatively new search strategy for the GUT physics, since, due to the low mass of the $T$-particle, its proton decay signatures can be correlated with the direct collider detections as well as with transition effects in cold neutron experiments, mediated by 
 $T$.   

  For phenomenological and cosmological implications of the decoupled triplet scenario
  the crucial question is at which level the Yukawa couplings of the $T$-particles are generated.  In the models proposed in the  original papers  \cite{dvali1992, Dvali:1995hp},  in the supersymmetric vacuum, the $T$-particles have strictly zero Yukawa couplings with the Standard Model superfields.
  However, it was assumed
  that non-zero couplings of order $\frac{m_{3/2}}{M_{\rm GUT}}$ will be generated due to supersymmetry breaking. 

The decoupling of the $T$-particle from the Standard Model in supersymmetric limit 
was demonstrated by the explicit derivation of the SUSY vacua in the  $SO(10)$-models constructed in \cite{Dvali:1995hp}. 
However, the generation of the coupling via supersymmetry breaking was not studied explicitly. 
In the present paper we filled this gap. 

We have constructed two versions of realistic   
supersymmetric $SO(10)$ GUTs,
which provide the decoupled color-triplet. 
 In both models, we have investigated the effects of gravity-mediated supersymmetry breaking. 
 In particular, we analysed the generation of the 
 couplings of $T$-particles due to the shift of 
 the GUT VEVs.  In this analysis we relied on the 
 general Goldstino Argument presented in \cite{Dvali:2020uvd}. This argument tells us that 
  the shifts of heavy GUT-scale VEVs, induced by 
  gravity-mediated supersymmetry breaking, are bounded by (\ref{shiftVEV}). 
We showed that this shift can induce the shift of the $T$-triplet's Yukawa coupling
endowing it with a finite macroscopic lifetime.

 This renders the $T$-particles highly relevant both for collider phenomenology as well as for the late-time cosmology. 
We have then studied various phenomenological implications and cosmological bounds. 
 We have discussed that such a long-lived, colored, and electrically charged particle would show very special signatures, which would call for alternative search strategies in experiments. 

 As proof of concept examples,  we have presented two realizations of the decoupled triplet scenario, called Model I and Model II respectively. 
 
In Model I,  the shift of the $\tilde{A}$ VEV 
that induces the Yukawa couplings of the 
$T$-particles, saturates the bound (\ref{shiftVEV}). The resulting lifetime 
of the $T$-particle (\ref{minilt}) around a second. 

In Model II, at least for the minimal K\"ahler metric, no shift of the $\tilde{A}$ VEV in order $m_{3/2}$ takes place. 
  Correspondingly, the  Yukawa coupling of the $T$-particle occurs in higher order in $\frac{m_{3/2}}{M_{\textrm{GUT}}}$. The resulting lifetime of the $T$-particle (\ref{maxilt}) exceeds the age of the Universe by many orders of magnitude. 
  This raises the question about 
 a possible dark matter candidacy of 
 $T\tilde{T}$-type mesons, which however requires a separate investigation. This is due to potentially accompanying stable $\tilde{T}q$-type mesons, abundance of which is subject to 
 very severe constraints. 
  The flip side of these  
 constraints is that they at the same time open up the experimental prospects for direct searches of stable $Tq$-type mesons. 
 The strong constraints on the abundance of stable, electrically charged $Tq$-type mesons imply that even a small relic population could have observable consequences. This motivates dedicated searches for anomalously heavy isotopes and other signatures of stable, heavy hadrons, which could provide complementary tests of the scenario considered here.

  Some conclusions reached in the current article have implications for much wider class of theories. 
  Using the explicit models constructed in the present paper as illustrative examples, we have highlighted and confirmed,
  within  supergravity GUT framework, some previously suggested phenomena that are  of general importance for the consistency of (super)gravity theories. 

  In particular, we studied the vacuum structure of the theory both before and after supersymmetry breaking.  First, using the Goldstino Argument~\cite{Dvali:2020uvd}, 
 we showed that,  
 once the energy of our vacuum is tuned to zero, 
  none of the other vacua can have positive energy densities. This finding fully supports the proposal  \cite{Dvali:2013eja, Dvali:2014gua, Dvali:2017eba} stating that 
 a consistent theory of gravity is incompatible with any sort of  stable or meta-stable de Sitter vacua. This criterion is intrinsically  linked by the $S$-matrix consistency of  formulation of quantum gravity \cite{Dvali:2020etd}. Our analysis illustrates how supergravity reinforces it in the GUT framework. 

 Next we showed that the phenomenologically interesting Minkowski vacuum is stable against tunneling for all the practical purposes. 
 However, there is more to it. Performing the  analysis based on Goldstino Argument~\cite{Dvali:2020uvd}, we showed that  within $\sim \frac{m_{3/2}}{M_{\rm GUT}}$ accuracy, the theory is in no-tunneling regime, in agreement with the \textit{Minkowski safety} criterion ~\cite{Dvali:2011wk}.  Again, this is fully aligned with the $S$-matrix consistency of (super)gravity.
  
   Another general point made in the present paper, concerns the 
 \textit{species scale} (\ref{eq:speciesscale}) which represents 
 an absolute non-perturbative gravitational cutoff of the 
 effective field theory 
~\cite{Dvali:2007hz, Dvali:2007wp, Dvali:2008fd, Dvali:2008ec, Dvali:2010vm, Dvali:2009ne, Dvali:2008jb, Dvali:2009ks, Dvali:2012uq, Dvali:2021bsy}.
Since this scale follows from certain exact properties of gravitational systems, such as black holes, it is physical and cannot be removed by any re-summation of the series. 
 We  argued that the operators 
 that transmit the effects of the GUT physics to low energies are expected to be sensitive to the species scale. 
This is due to a significant number of the particle species 
housed in GUTs and, in particular,  in supersymmetric $SO(10)$ theory. 
The sensitivity towards the species scale 
provides an indirect support for the light color-triplet scenario~\cite{dvali1992, Dvali:1995hp}, by delivering a new physical cutoff scale for high-dimensional operators of the type (\ref{Y5D}).

On phenomenological side, we have argued that the allowed parameter space
for the light color-triplet scenario 
should be within reach for future collider searches at the LHC, which makes a detection of a $T$-particle feasible. Needless to say, this is exciting news, as the $T$-particle represents a rare direct low-energy probe of GUT physics.

\begin{acknowledgments}

This work was supported in part by the Humboldt Foundation under Humboldt Professorship Award, by the European Research Council Gravities Horizon Grant AO number: 850 173-6,
by the Deutsche Forschungsgemeinschaft (DFG, German Research Foundation) under Germany's Excellence Strategy - EXC-2111 - 390814868, and Germany's Excellence Strategy under Excellence Cluster Origins and by ANR grant ANR-23-CE31-0024 EUHiggs. Furthermore, M.E. is thankful for the hospitality of the University of Chicago during the finalization of this project.

\paragraph*{Disclaimer} Funded by the European Union ERC, GRAVITES, 101071779. Views and opinions expressed are however those of the author(s) only and do not necessarily reflect those of the European Union or the European Research Council. Neither the European Union nor the granting authority can be held responsible for them.
\end{acknowledgments}

\section{Appendix A: SUSY Vacuum} \label{appA}
Let us consider the global SUSY vacuum equations
corresponding to the superpotential (\ref{WGUT}).  We immediately note that since on the desired ansatz (\ref{PATTERN}) 
$A_{ij}$ and $B_{ij}$ commute, the $F$-flatness condition for $C_{ij}$ self-consistently gives $C_{ij} =0$. consequently, all   
$C$-dependent terms drop out of the remaining equations, which take the following form:
\begin{itemize}
        \item $\frac{\partial W}{\partial \Sigma_{ij}}\overset{i\neq j}{=}2M\Sigma_{ij}+2\lambda (\Sigma^2)_{ij}+\alpha\left[\left(AB\right)_{ij}+(AB)_{ji}\right]\\\hspace*{0.75cm}=0$
        \item $\frac{\partial W}{\partial \Sigma_{jj}}=M\Sigma_{jj}+\lambda \left[\left(\Sigma^2\right)_{jj}-\frac{1}{10}\Sigma^2\right]+\\
        \hspace*{2.05cm} +\alpha\left[\left(AB\right)_{jj}-\frac{1}{10}(AB)\right]=0$
        \item $\frac{\partial W}{\partial A_{ij}}=\alpha\left[\left(B\Sigma\right)_{ji}+(\Sigma B)_{ji}\right]+2 M_A A_{ji}+\\ \hspace*{2.65cm}+2\tilde{M}B_{ji}+2\beta \bar{\chi}\sigma_{ji}\chi =0$
        \item $\frac{\partial W}{\partial B_{ij}}=\alpha(\Sigma A + A \Sigma)_{ji} + 2\tilde{M}A_{ji}=0$
        \item $\frac{\partial W}{\partial \bar{\chi}_\alpha}=(\beta A_{ij} \sigma_{ji} \chi+M_{\chi}\chi)_\alpha=0 $ 
        \item $\frac{\partial W}{\partial \chi_\alpha}=(\beta A_{ij}\bar{\chi} \sigma_{ji} +M_{\chi}\bar{\chi})_\alpha=0 $ 
          \end{itemize}
          We solve this with the following Ansatz (delivering the breaking \eqref{eq:SO10breaking}): 
          \begin{itemize}
             \item[--] $\Sigma_{ij}=\text{diag}(\stackrel{6 \times}{2,...};\stackrel{4\times}{-3,...})\Sigma$ 
             \end{itemize}
             hence $Tr\Sigma^2=\Sigma_{ij}\Sigma_{ji}=60\Sigma^2$
         \begin{itemize}
          \item[--] $A_{ij}=\text{diag}(\tilde{A}\epsilon, \tilde{A}\epsilon, \tilde{A}\epsilon, A\epsilon,A\epsilon)$ 
          \item[--] $B_{ij}=\text{diag}(\tilde{B}\epsilon, \tilde{B}\epsilon, \tilde{B}\epsilon, B\epsilon,B\epsilon)$ 
          \end{itemize}
          with $\epsilon \equiv \begin{pmatrix}
              0 & 1\\
              -1 & 0
          \end{pmatrix}$
          \begin{itemize}
              \item[--] $\chi_{\alpha}=\bar{\chi}_{\alpha}=\chi \vert +++++ \rangle$ with \\SU(5)-invariant components 
              \end{itemize}
              hence
               $\sigma_{2i-1,2i}\vert +++++ \rangle=\vert +++++ \rangle$\\ where $j\in [1,\;5]$.
          
        \noindent  Notice that
        $A_{ij}B_{jm}=-\text{diag}(\stackrel{6\times}{\tilde{A}\tilde{B},...};\stackrel{4\times}{AB,...})$ and therefore $\text{Tr}(AB)\equiv A_{ij}B_{ji}=-(6\tilde{A}\tilde{B}+4AB)$. \\
        \\
        Next, we insert this into the global SUSY vacuum equations: 
        \begin{itemize}
            \item $\frac{\partial W}{\partial \Sigma_{i\neq j}}=0$
            \item $\frac{\partial W}{\partial \Sigma_{11}}=M\Sigma -\lambda \Sigma^2+\frac{\alpha}{5}(-\tilde{A}\tilde{B}+AB)=0$
            \item $\frac{\partial W}{\partial \Sigma_{77}}=M\Sigma -\lambda \Sigma^2+\frac{\alpha}{5}(-\tilde{A}\tilde{B}+AB)=0$
            \item $\frac{\partial W}{\partial A_{21}}=2 \alpha \tilde{B}\Sigma + \tilde{M}\tilde{B}+M_A\tilde{A}+\beta \chi^2=0$
            \item $\frac{\partial W}{\partial A_{87}}=-3\alpha B\Sigma + \tilde{M}B+M_AA+\beta \chi^2=0$
            \item $\frac{\partial W}{\partial B_{21}}=2 \alpha \Sigma \tilde{A} + \tilde{M}\tilde{A}=0$
            \item $\frac{\partial W}{\partial B_{78}}=-3 \alpha \Sigma A + \tilde{M}A=0$
            \item $\frac{\partial W}{\partial \tilde{\chi}}=\left(-\beta(6\tilde{A}+4A)+M_{\chi}\right)\chi=0$
            \item $\frac{\partial W}{\partial \chi}=\left(-\beta(6\tilde{A}+4A)+M_{\chi}\right)\tilde{\chi}=0$
        \end{itemize}
        We used $A_{ij}\sigma_{ij}\vert +++++\rangle =-(6\tilde{A}+4A)\vert +++++\rangle$.\\
        \\
        We now demand that 
        \begin{equation}
            \tilde{A}=0
        \end{equation}
        We set the following equations
    \begin{enumerate}
        \item $M\Sigma -\lambda \Sigma^2+\frac{\alpha}{5}AB=0$ $\Rightarrow$ $B=\frac{5}{\alpha}\frac{4\beta}{M_{\chi}}\left(\lambda\frac{\tilde{M}^2}{9\alpha^2}-\frac{M\tilde{M}}{3 \alpha}\right)$ 
        \item $(2 \alpha \Sigma + \tilde{M})\tilde{B}+\beta \chi^2=0$
        \item $(-3 \alpha \Sigma + \tilde{M})\tilde{B}+M_AA+\beta \chi^2=0$
        \item $(-3 \alpha \Sigma + \tilde{M})A=0$ $\Rightarrow$ $\Sigma=\frac{\tilde{M}}{3 \alpha}$ 
        \item $(-4 \beta A+ M_{\chi})\chi=0$ $\Rightarrow$ $A=\frac{M_{\chi}}{4 \beta}$ 
    \end{enumerate}

    \noindent $\chi^2$ and $\tilde{B}$ are fixed from 3. and 2. \\
    \\
    Since $-3 \alpha \Sigma + \tilde{M}=0$ they become: \\
    \\
    3. $\longrightarrow$ $M_A A+\beta \chi^2 =0$ $\Rightarrow$ $\chi^2=-\frac{M_A M_{\chi}}{4\beta^2}$  \\
    2. $\longrightarrow$ $\frac{5}{3}\tilde{M}\tilde{B}+\beta \chi^2 =0$ $\Rightarrow$ $\tilde{B}=\frac{3}{20 \beta}\frac{M_A M_{\chi}}{\tilde{M}}$ \\
    \\
    Thus, all VEVs are fixed and are of order $M_{\textrm{GUT}}$. Notice that 
   since $C_{ij} = [A,B]_{ij} = 0$, 
    the term $\gamma C[A,B]$ plays no role in 
    determining the values of the above VEVs. Nevertheless, this term is important for eliminating a flat direction of relative rotations of $A_{ij}$ and $B_{ij}$. 
    \\

\section{Appendix B: Gravity Mediated Shift of GUT fields from Goldstino Argument} \label{appB}

 In this appendix, we derive the gravity-mediated shift 
 of the GUT field. We first present a general argument 
 of \cite{Dvali:2020uvd} and then apply it to the shift of $45_A$-plet in our model. 
 Let us consider the GUT theory embedded in supergravity.  
 We denote the components of the heavy Higgs superfields that break the GUT group down to the Standard Model,  $SU(3)\times SU(2)\times U(1)$, by $S_j$. 
 The corresponding superpotential 
 we denote by  $W(S)$.  In the present context, this superpotential, of course, contains 
 (\ref{WGUT}).

 We assume that the GUT sector communicates through supergravity with the hidden sector superfields that break 
 SUSY. For definiteness we take one such superfield
 and denote it by $X$.  The order parameter 
 for the spontaneous breaking of supersymmetry is the 
 VEV of the $F_X$-term of this superfield. This VEV  
 sets the SUSY breaking scale as 
  $M_{SUSY}^2 = \langle F_X \rangle $. 
  
  Ignoring other fields, the fermionic component $\Psi_X$ of the superfield $X$ is the Goldstino. This fermion becomes ``eaten up" by the gravitino, which gets the mass, 
\begin{equation} \label{gravmass}
m_{3/2}^2 \equiv \frac{|\langle W\rangle|^2}{M_P^4} = 
\frac{|\langle F_X \rangle|^2}{3M_P^2}\,, 
\end{equation}
where $\langle W\rangle$ is the expectation value of the total superpotential. The second equality in 
(\ref{gravmass}) must be satisfied in order for the vacuum energy to be zero.

We adopt the traditional scenario of minimal gravity-mediation
which assumes that the two sectors are decoupled in the superpotential,
\begin{equation} \label{WHX}
W=W(S) +W(X).
\end{equation} 
 
In the globally supersymmetric limit,  
$W(S)$ possesses the  supersymmetric vacuum in which the GUT superfields $S_j$ receive masses of order $M_{\textrm{GUT}}$. 
That is, in the limit of global supersymmetry 
($M_P \rightarrow \infty$, all other scales finite), we have: 
\begin{equation} \label{I1}
\begin{split}
& \langle \frac{\partial W}{\partial S_j}\rangle =0 \,, ~{\rm for}  ~\,  \langle S_j \rangle  \equiv S_j^{(0)} \sim M_{\textrm{GUT}}\,,
\end{split} 
\end{equation} 
 such that the corresponding 
eigenvalues of the supersymmetric mass matrix, 
\begin{equation}
{\mathcal M}_{ij} \equiv   \frac{\partial^2 W}{\partial S_i \partial S_j} |_{S^j = S_j^{(0)}} \,,
\end{equation} 
  are $ \sim M_{\textrm{GUT}} $\,.   
 At the same time, the hidden sector satisfies, 
\begin{equation} \label{I2}
0\neq\langle\frac{\partial W}{\partial X}\rangle= M^2_{SUSY}\, .
\end{equation} 
 Let us now consider the supergravity potential for $S_j$ and $X$ scalar fields
 (for simplicity, we assume the minimal K\"ahler and 
 summation over $j$ is assumed): 
 \begin{equation} \label{pot} 
 V = {\rm e}^{(|S_j|^2 + |X|^2)/M_P^2}\left ( |F_{S_j}|^2 + |F_X|^2 - 3m_{3/2}^2M_P^2 \right ) \,, 
  \end{equation} 
  where, 
\begin{equation} 
F_{S_j}\equiv \frac{\partial W}{\partial S_j} + m_{3/2}S_j^* \,,
\label{FSterm}
\end{equation}
\begin{equation}
F_X\equiv \frac{\partial W}{\partial X} +m_{3/2}X^*\,,
\label{FXterm}
\end{equation}
and as before  $m_{3/2} \equiv  W/M_P^2$\;.\\

 The Goldstino Argument of  \cite{Dvali:2020uvd} tells us that after taking 
 into account supergravity,  the $F$-terms of the  GUT sector superfields  are bounded from above by, 
 \begin{equation}  \label{GrShift}
   |F_{S_j}|  \lesssim m_{3/2}^2\;.
  \end{equation} 
  This can be understood by considering 
  various scaling regimes of the expansion in small 
  parameters $\frac{m_{3/2}}{M_{\textrm{GUT}}} \ll 1$  and 
  $\frac{m_{3/2}}{M_P} \ll 1$.

 Let us first take the double-scaling limit $M_P\rightarrow \infty\,,\, m_{3/2} \rightarrow  0$ while keeping the scale $F_X \sim M_Pm_{3/2}$ finite. 
  
 Next we specify the scaling of $M_{\textrm{GUT}}$ in this regime.  
 If we take $M_{\textrm{GUT}}/M_P \rightarrow 0$ (equivalently $m_{3/2}M_{\textrm{GUT}} \rightarrow 0$), 
 the vacuum of the GUT superfields $S_j$ reduces to a globally 
 supersymmetric case (\ref{I1}), with $\langle S_j \rangle = S_j^{(0)}$  which gives $F_{S_j} = 0$. \\

Let us now keep the ratio $M_{\textrm{GUT}}/M_P$ small but finite. 
This implies $m_{3/2}M_{\textrm{GUT}}={\rm finite}$.  However,   
as long as $m_{3/2}= 0$, 
 the supergravity $F_{S_j}$-terms of the GUT superfields 
 continue to vanish. This can be understood 
 from the following argument.   
 
 Since we have $m_{3/2} =0$, 
 gravitino is massless.  At the same time, because we have  $M_P=\infty$, it is also decoupled from the rest of the fields. 
   Thus, in this limit the Goldstino is not a longitudinal component of the gravitino.  
    Therefore, it is a physical massless fermion. 
 By the super-Goldstone theorem, Goldstino is given by the 
following superposition of $\Psi_{S_j}$ and $\Psi_X$ chiral fermions:
\begin{equation} \label{Goldstino}
\Psi_{Gold} = \frac{\langle F_{S_j} \rangle \Psi_{S_j}+\langle  F_X \rangle \Psi_X}{\sqrt{\langle  F_{S_j} \rangle^2+\langle F_X \rangle^2}}\,.
\end{equation}
 This fermion must form a zero eigenstate of the fermion mass matrix.   At the same time, the orthogonal eigenvalues are of order $M_{\textrm{GUT}}$, i.e. are infinite. 
 
   Obviously, the mixing angle between  $\Psi_X$ and GUT fermions must vanish for $m_{3/2} =0$. 
  From (\ref{Goldstino}) this mixing angle satisfies 
  \begin{equation}  
  \tan (\theta) = \frac{\sqrt{|F_{S_j}|^2}}{|F_X|} \,.
  \end{equation}
 Thus,   
  for $m_{3/2} = 0$, we have $F_{S_j} =0$, no matter how large is $M_{\textrm{GUT}}$.

From the above, it follows that for the finite values of 
$m_{3/2} \ll M_{\textrm{GUT}} \ll M_P$, 
 the VEVs of $F_{S_j}$ are controlled by the gravitino mass and not by the GUT scale.  
  Thus, the first non-zero order in which the GUT sector 
  $F$-terms appear is,  $F_{S_j} \sim m_{3/2}^2$. 

One can see this very clearly if we write the expansion of the $F$-term explicitly in powers of the parameter $\frac{m_{3/2}}{M_{\textrm{GUT}}}$
\begin{equation}
F_S = M_{\textrm{GUT}}^2
\left(
c_0
+
c_1 \frac{m_{3/2}}{M_{\textrm{GUT}}}
+
c_2 \frac{m_{3/2}^2}{M_{\textrm{GUT}}^2}
+ \cdots
\right) \;.
\end{equation}
The first coefficient, $c_0$, is zero by construction as $W(S)$ does not
break SUSY in the global limit. Now let us calculate $c_1$. To this end we calculate a solution to \eqref{FSterm} for a small shift
around the globally supersymmetric value $ S = S_0 + \delta S$. We get 
\begin{equation}
\left.
\frac{\partial^2 W}{\partial S^2}
\right|_{S = S_0} \, \delta S
+ m_{3/2} S_0^{*} = 0 \;,
\end{equation}
and because $\left(\frac{\partial^2 W}{\partial S^2}\right)_{S = S_0} \sim S_0 \sim M_\textrm{GUT}$ we are left with
\begin{equation}
    \delta S = m_{3/2}\;.
\end{equation}
  
   We thus conclude that in gravity-mediation the 
   VEVs of the GUT sector fields are shifted from their globally supersymmetric minima in a way that obeys the condition (\ref{GrShift}).

  This information is sufficient for our estimate of the 
  shift of $\tilde{A}$-VEV in $45_A$.
  Applying the condition (\ref{GrShift}) to the 
   $F$-term of $\tilde{B}$, (i.e., $B_{12}$, $B_{34}$, $B_{56}$), e.g.,  
    \begin{equation} \label{SugraFFF}
        F_{B_{21}}= \frac{\partial W}{\partial B_{21}} + m_{3/2}B^*_{21} \;, 
    \end{equation}
    we get 
    \begin{equation} \label{SugraFB}
       F_{\tilde{B}}  = (4\alpha \Sigma + 2\tilde{M}) \langle \tilde{A} \rangle -m_{3/2}\tilde{B}  \lesssim m_{3/2}^2 \;.
    \end{equation}
     Inserting the globally-supersymmetric VEVs 
     $\Sigma=\frac{\tilde{M}}{3 \alpha}$ and $\tilde{B}=\frac{3}{20 \beta}\frac{M_A M_{\chi}}{\tilde{M}}$, we find 
    $\frac{10}{3}\tilde{M} \langle \tilde{A} \rangle -m_{3/2}\frac{3}{20 \beta}\frac{M_A M_{\chi}}{\tilde{M}} \sim m_{3/2}$\;. Therefore, in the leading order, $\langle \tilde{A} \rangle$  satisfies  
    \begin{equation}
         \langle \tilde{A} \rangle \sim
          m_{3/2} \frac{\mathcal{O}(M_{\textrm{GUT}})}{\mathcal{O}(M_{\textrm{GUT}})} \sim m_{3/2}\;.
    \end{equation}
   Thus, we discover that  the gravity-mediated 
  supersymmetry breaking corrections induce the following 
   shift of the $\tilde{A}$-component of the  $45_A$ VEV, 
   \begin{equation}
   \delta \tilde{A} = \langle \tilde{A} \rangle \sim m_{3/2} \,.
   \end{equation} 
 As discussed in the main text, this generates the Yukawa coupling of the color-triplet 
 of order $\frac{m_{3/2}}{M_{\textrm{GUT}}}$.

  \section{Appendix C: Generation of Effective High Dimensional Operators} \label{appC}
   
   In this appendix, we discuss how the effective high-dimensional operators of the type (\ref{Y5D}) and (\ref{Y6D}) can be generated by the exchange of the heavy fields with 
   renormalizable interactions.  We shall follow 
  \cite{dvali1992} in which two possible exchanges 
  have been outlined. These have been called 
  scalar and fermion exchanges, respectively. 
  Both exchanges take place  among the superfields
  the names of which refer to the relevant virtual components.

 \subsection{Fermion exchange}  
     We start with the operator (\ref{Y5D}). As 
     shown in \cite{dvali1992}, this operator can be generated by the exchange of heavy $144, \overline{144}$-dimensional superfields  with the following couplings 
 in the superpotential:
 \begin{equation} \label{W144}
 W_{\rm F} =  16_F\gamma_i\overline{144}_j  A_{ij} + M_{144} 
 144_j\overline{144}_j + 10_{Hj} 144_j16_F \,,
 \end{equation}
 where $j = 1,2, ..., 10$ is a $SO(10)$ tensor index, 
 $\gamma_j$ are $SO(10)$ gamma matrices and the spinor indices are not shown explicitly. $M_{144} \sim M_{\textrm{GUT}}$ is a mass term, and for simplicity, we set the dimensionless coupling constants equal to one. 

  Of course, for more than one generation the 
  couplings and masses must be supplemented by family indexes and promoted into the matrices in generation space.  
  As discussed in \cite{Dvali:1995hp} in  such a case the corresponding matrices must be symmetric in generation indices, which shall be assumed.

\begin{figure}[ht]
\centering
\resizebox{\columnwidth}{!}{
\begin{tikzpicture}
  \begin{feynman}
    
    \vertex (m) [insertion=0, scale=1.8, line width=1.2pt, inner sep=0pt] at (0,0) {}; 
    \node [below=0.4cm of m] {\(\scriptstyle M_{144}\)};
    
    \vertex (vL1) at (-1.4, 0); 
    \vertex (vR1) at (1.4, 0);
    \vertex (i1) at (-2.4, 0) {\(16_F\)};
    \vertex (f1) at (2.4, 0) {\(16_F\)};
    \vertex (h1) at (-1.4, 1.4) {\(10_{Hj}\)};
    \vertex (a1) at (1.4, 1.4) {\(A_{ij}\)};

    \diagram* {
      (i1) -- [fermion] (vL1), 
      (m) -- [fermion, edge label=\(\scriptstyle \overline{144}_j\)] (vL1), 
      (f1) -- [fermion] (vR1), 
      (m) -- [fermion, edge label'=\(\scriptstyle 144_j\)] (vR1), 
      (h1) -- [charged scalar] (vL1), 
      (a1) -- [charged scalar] (vR1), 
    };

    \draw [double, distance=2.5pt, -{Implies}, line width=1.0pt] (3.2, 0) -- (4.2, 0);

    \vertex (eff) [dot] at (6.5, 0) {};
    
    \vertex (i2) at (5.3, 0) {\(16_F\)};
    \vertex (f2) at (7.7, 0) {\(16_F\)};
    \vertex (h2) at (5.8, 1.4) {\(10_{Hj}\)};
    \vertex (a2) at (7.2, 1.4) {\(A_{ij}\)};

    \diagram* {
      (i2) -- [fermion] (eff),
      (f2) -- [fermion] (eff),
      (h2) -- [charged scalar] (eff),
      (a2) -- [charged scalar] (eff),
    };
    
    \node [below=0.4cm of eff] {\(\scriptstyle \gamma_i\)};

  \end{feynman}
\end{tikzpicture}
}
\caption{A diagrammatic description of emergence of the effective operator \eqref{Y5D} after integrating out the heavy $144, \overline {144}$ superfields. Adapted from Fig.1 in \cite{dvali1992}.}
\label{fig:dvali1992}
\end{figure}
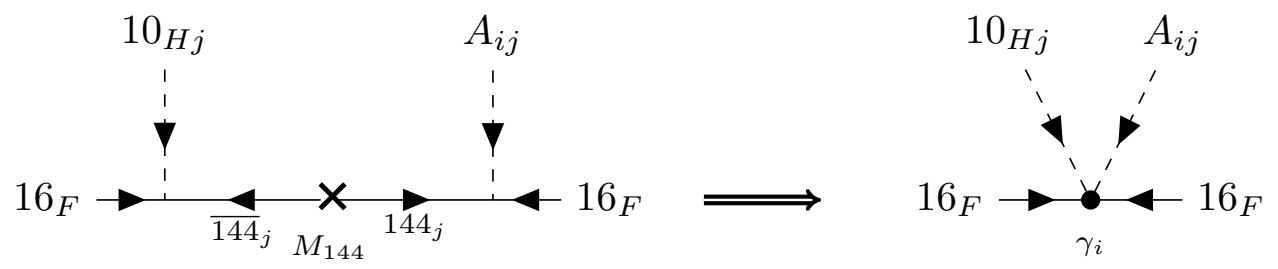

   It is easy to see that the exchange of the heavy superfields
  (see diagram in Fig. \ref{fig:dvali1992})  generates the
 effective operator  (\ref{Y5D}) with 
 $\Lambda \sim M_{144}$
 in the superpotential. 
The mechanism of generation of the split 
 couplings from the superpotential 
 (\ref{W144}) can be understood in 
 the language of mass matrices.  
 It is most convenient to use  the decomposition of the participating  $SO(10)$ representations in terms 
 of the Pati-Salam subgroup  $SU(4)\times SU(2)_L\times SU(2)_R \subset SO(10)$, which has the form, 
\begin{eqnarray}
16& = & (4.2.1) + (\bar{4}.1.2) \nonumber \\
10 & = & (6.1.1) + (1.2.2)  \nonumber \\
45 & = & (15.1.1) + (1.3.1) + (1.1.3) + (6.2.2)  \nonumber  \\
144 & = & (4.2.1) + (\bar{4}.1.2) + (20.2.1) + \nonumber  \\
 &&(\bar{20}.1.2) + (4.2.3) + (\bar{4}.3.2)  \,.
\end{eqnarray}
  
     Reducing the $A_{ij}$ and $10_{Hj}$ Higgs fields to the
     fragments with non-zero VEVs,  
     $(1.1.3)_A$ and   $(1.2.2)_{10_H}$, and inserting them in the superpotential (\ref{W144}), we get the following mass matrices    
 \begin{equation} \label{mass1}
\bordermatrix{
	& (4.2.1)_{16_F}	&  (4.1.2)_{\bar{144}} 	\cr
(\bar{4}.1.2)_{16_F}	& 0	& (1.1.3)_A	\cr
(\bar{4}.1.2)_{144}	& (1.2.2)_{10_H}	& M_{144} 	\cr
}
\end{equation}
 and 
 \begin{equation} \label{mass2}
\bordermatrix{
	& (\bar{4}.1.2)_{16_F}	&  (\bar{4}.2.3)_{\bar{144}} 	\cr
(4.2.1)_{16_F}	& 0	& (1.1.3)_A	\cr
(4.2.3)_{144}	& (1.2.2)_{10_H}	& M_{144} \cr
}\,.
\end{equation}
 Setting $(1.2.2)_{10_H} =0$, 
and taking into account that the VEV  $\langle (1.1.3)_A \rangle \sim 
M_{\textrm{GUT}}$, we see that 
 the following components, denoted by the subscript l for light, remain massless 
 after GUT symmetry breaking: 
 \begin{eqnarray}
(\bar{4}.1.2)_{l} & = &    \frac{M_{144}  (\bar{4}.1.2)_{16_F} -     \langle (1.1.3)_A \rangle (\bar{4}.1.2)_{144}}{
\sqrt{\langle (1.1.3)_A \rangle^2 + M_{144}^2 }} \nonumber \\
(4.2.1)_{l} & = & \frac{M_{144}  (4.2.1)_{16_F} -     \langle (1.1.3)_A \rangle (4.2.3)_{144}}{
\sqrt{\langle (1.1.3)_A \rangle^2 + M_{144}^2 }} \,.
\end{eqnarray}
  All other components gain masses of order 
  the GUT scale.  These light fragments form the ordinary 
  Yukawa couplings with the Higgs doublets residing 
  in $(1.2.2)_{10_H}$. 
  
   On the other hand, the color-triplet partners reside in  $(6.1.1)_{10_H}$. 
    This field acquires the following Yukawa coupling with the above light quark and lepton superfields: 
    \begin{equation} 
    (6.1.1)_{10_H}  \langle (1.1.3)_A \rangle (\bar{4}.1.2)_{l} (\bar{4}.1.2)_{l}
   \end{equation}  
    This coupling, however, vanishes since    
    it is antisymmetric 
   in $SU(4)$ indices and symmetric in $SU(2)_R$.   As already said, for more than one family, we impose the prescription of~\cite{Dvali:1995hp} demanding that the  couplings are symmetric in generation indices.
   Consequently, the triplet decouples.

  Next, the operators of the type (\ref{Y6D}) can be generated by  splitting the masses of the components 
 of the $144$-plet through its couplings to 
 $45$-dimensional Higgs representations in the superpotential. For example, 
 \begin{equation}  
 A_{kl}  144_j \gamma_k\gamma_l\overline{144}_j 
 \end{equation}
 After integrating out the heavy $144$-plet, we obtain  
 the series of operators of the type 
  \begin{equation} \label{Y10}
  \begin{split}
 &\frac{10_{Hi} A_{Hij} A_{kl}}{M_{144}^2}16_F \gamma_j \gamma_{k} \gamma_{l} 16_F \, + \\ +\,
  &\frac{10_{Hi} A_{Hij} A_{kl} A_{mn}}{M_{144}^3}
  16_F \gamma_j \gamma_{k} \gamma_{l} \gamma_{m} \gamma_{n}16_F \, + \,...   
  \end{split}
 \end{equation} 
  Of course, the analogous operators can be generated with the participation of the $B_{ij}$ field,
  \begin{equation}  
 B_{kl}  144_j \gamma_k\gamma_l\overline{144}_j \,,
 \end{equation}
  which gives, 
  \begin{equation} \label{Y10}
  \frac{10_{Hi} A_{Hij} B_{kl}}{M_{144}^2}16_F \gamma_j \gamma_{k} \gamma_{l} 16_F \, + \,...  
 \end{equation} 
  Notice that since the VEVs of the GUT fields 
  $A_{ij}, B_{ij}$ are of order $M_{\textrm{GUT}}$, their contribution to the mass 
  of the $144$-plet is not necessarily subdominant 
  as compared to the $SO(10)$-invariant mass 
  $M_{144}$.  The GUT VEVs therefore can appear in the 
  denominator of the effective operators.  For example, 
  for $\langle B \rangle \gg M_{144}$, the dominant operators 
  have the form
  \begin{equation} 
  \begin{split}
    &  \frac{10_{Hi} A_{Hij}}{\langle B_{kl} \rangle}
  16_F \gamma_j \gamma_{k} \gamma_{l} 16_F \, + \,...\\  
    ...\;+\;&\frac{10_{Hi} A_{Hij} M_{144}}{\langle B_{kl} \rangle 
    \langle B_{mn} \rangle}
  16_F \gamma_j \gamma_{k} \gamma_{l} \gamma_{m} \gamma_{n}16_F \, + \,...  
  \end{split}
 \end{equation} 

 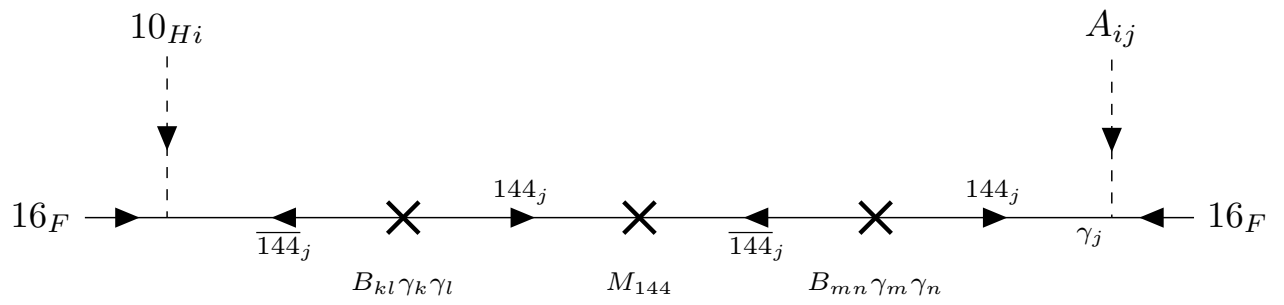
\begin{figure}[ht]
\centering
\resizebox{\columnwidth}{!}{
\begin{tikzpicture}
  \begin{feynman}
   
    \coordinate (vL) at (-4.5, 0); 
    \vertex (vR) [label={[xshift=-0.2cm, yshift=-0.4cm]\(\scriptstyle \gamma_j\)}] at (4.5, 0);
    
    \vertex (i1) at (-5.7, 0) {\(16_F\)};
    \vertex (f1) at (5.7, 0) {\(16_F\)};
    
    \vertex (h1) at (-4.5, 1.8) {\(10_{Hi}\)};
    \vertex (a1) at (4.5, 1.8) {\(A_{ij}\)};

    \coordinate (b1_pos) at (-2.25, 0);
    \coordinate (m_pos) at (0,0);
    \coordinate (b2_pos) at (2.25, 0);

    \diagram* {
      (i1) -- [fermion] (vL),
      (f1) -- [fermion] (vR),
      (h1) -- [charged scalar] (vL),
      (a1) -- [charged scalar] (vR),

      (b1_pos) -- [fermion, edge label=\(\scriptstyle \overline{144}_j\)] (vL), 
      (b1_pos) -- [fermion, edge label=\(\scriptstyle 144_j\)] (m_pos),
      (b2_pos) -- [fermion, edge label=\(\scriptstyle \overline{144}_j\)] (m_pos),
      (b2_pos) -- [fermion, edge label=\(\scriptstyle 144_j\)] (vR),
    };

    \foreach \p in {b1_pos, m_pos, b2_pos} {
        \draw [line width=1.2pt] ($(\p) + (-0.15, -0.15)$) -- ($(\p) + (0.15, 0.15)$);
        \draw [line width=1.2pt] ($(\p) + (-0.15, 0.15)$) -- ($(\p) + (0.15, -0.15)$);
    }

    \node [below=0.4cm of b1_pos] {\(\scriptstyle B_{kl} \gamma_k \gamma_l\)};
    \node [below=0.4cm of m_pos]  {\(\scriptstyle M_{144}\)};
    \node [below=0.4cm of b2_pos] {\(\scriptstyle B_{mn} \gamma_m \gamma_n\)};

  \end{feynman}
\end{tikzpicture}
}
\caption{ Generation of an effective operator via a higher-order chain of exchanges of the heavy  $144,\overline{144}$-multiplets.}
\label{fig:newdiagram}
\end{figure}

Of course, in each concrete realization of the flavor structure, one must make sure that additional mixing terms between $144, \bar{144}$ and $16_F$ superfields
do not induce a direct coupling between $T$-particles residing in 
$10_{H}$ and the light quark and lepton superfiels that shall represent the mixtures of 
components from $144, \bar{144}$ and 
$16_F$.

\subsection{Scalar exchange}

 In order to generate the effective operator 
 (\ref{Y5D}) in a scalar exchange channel, following \cite{Dvali:1995hp}, we introduce three $10$-plet superfields  $10_{H}, 10_{H}^{'}, 10_{H}^{''}
 $. 
 The relevant part of the superpotential has the following form,  
 \begin{equation} \label{W1010}
 W_{\rm F} =  16_F\gamma_j 16_F\, 10^{''}_{Hj} 
 + 10_{Hj}^{'} (M_{10} 10_{Hj}^{''}   +  A_{ji} 10_{Hi} )\,,
 \end{equation}
 where $M_{10} \sim M_{\textrm{GUT}}$ and for 
 simplicity of illustration we put the coupling constants equal to one.

     Since in the SUSY vacuum (\ref{SUSYVEVs}) the VEV (\ref{AtildeVEV}) of $45_A$ vanishes on the $\tilde{A}$ component, the color-triplet components 
     $10_{Hj},~j \in [1,2,...,6]$,  which transform as $(6.1.1)$ representation under the Pati-Salam subgroup, 
     are masless, whereas the analogous color-triplets from 
    $10_{Hj}^{'}, 10_{Hj}^{''}$,     
    gain masses 
    $=M_{10}$.  This effect is displayed by the following supersymmetric mass matric, 
       \begin{equation} \label{mass10TT}
\bordermatrix{
	& (6.1.1)_{10_H''} &  (6.1.1)_{10_H} 	\cr
(6.1.1)_{10_H^{'}}	& M_{10}	&  0	\cr
}\,.
\end{equation}  
  Correspondingly, there is no admixture from the light triplets in $10_{Hj}^{''}$, which interacts with $16_F$-s.     
 Instead, the light triplets 
    $T,\bar{T}$ entirely reside in $10_{H}$ and are  decoupled from the $16_F$-multiplets.
   
   The story for the weak doublets is very different. 
   Plugging the VEV of 
   $A_{ij}$ which is non-zero on the $(1.1.3)_A$-component in (\ref{W1010}), we get the following mass matrix for the electroweak doublet components of the three superfields: 
   \begin{equation} \label{mass1010}
\bordermatrix{
	& (1.2.2)_{10_H''} &  (1.2.2)_{10_H} 	\cr
(1.2.2)_{10_H^{'}}	& M_{10}	& \langle (1.1.3)_A \rangle	\cr
}\,.
\end{equation}  
   We see that the Higgs doublet components 
   $(1.2.2)_{10_H}$ and $(1.2.2)_{10_H^{''}}$ mix  
   and one combination   
   \begin{equation}
(1.1.2)_{l}  =  \frac{M_{10} (1.2.2)_{10_H} - 
\langle (1.1.3)_A \rangle (1.2.2)_{10_H''}}{
\sqrt{\langle (1.1.3)_A \rangle^2 + M_{10}^2 }} 
\end{equation}
remains massless, whereas the orthogonal one gains a GUT-scale mass and decouples. 
 The light doublet partially resides in 
   $(1.2.2)_{10_H''}$ and therefore has unsuppressed Yukawa coupling of order, 
   \begin{equation} \label{indH}
Y_{H}  =  \frac{A}{
\sqrt{A^2 + M_{10}^2 }} \,,
\end{equation}  
   with $16_F$ superfields. 

  If supersymmetry breaking effects generate a small VEV on the $\tilde{A}$ component which transforms as $(15.1.1)$ under the Pati-Salam subgroup and breaks the $SU(4)$-quark-lepton color down to $SU(3)_c\times U(1)$, the structure of the color-triplet mass matrix 
  changes to 
       \begin{equation} \label{mass10TT}
\bordermatrix{
	& (6.1.1)_{10_H''} &  (6.1.1)_{10_H} 	\cr
(6.1.1)_{10_H^{'}}	& M_{10}	&  \tilde{A}	\cr
}\,.
\end{equation} 
  Due to this, the $(6.1.1)_{10_H^{''}}$ acquires a small admixture, $\sim \frac{\tilde{A}}{M_{10}}$, of the light 
  color-triplets $T,\bar{T}$.   Therefore, the 
  $T$-particles acquire the Yukawa couplings with  $16_F$-superfields of order, 
     \begin{equation} \label{indT}
Y_{T}  =  \frac{\tilde{A}}{
\sqrt{\tilde{A}^2 + M_{10}^2 }} \simeq \frac{\tilde{A}}{M_{\rm GUT}}\,.
\end{equation}

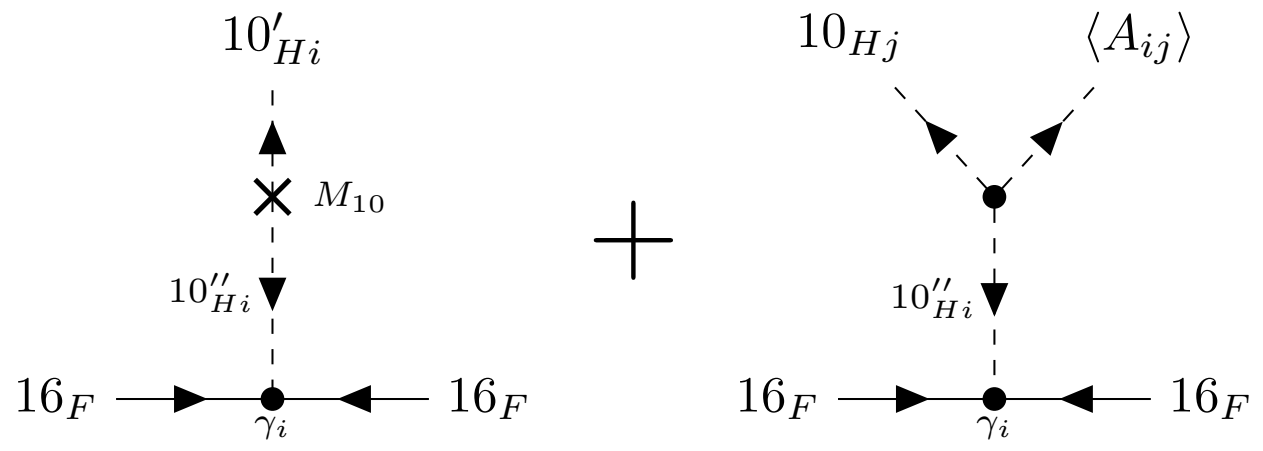
\begin{figure}[ht]
\centering
\resizebox{\columnwidth}{!}{
\begin{tikzpicture}
  \begin{feynman}

    \vertex (i1) at (-2.5, 0) {\(16_F\)};
    \vertex (f1) at (0.5, 0) {\(16_F\)};
    \vertex (v_yuk1) [dot] at (-1, 0) {};
    \node [below=0.2cm of v_yuk1] {\(\scriptstyle \gamma_i\)};
    
    \vertex (m_pos1) [coordinate] at (-1, 1.4); 
    \vertex (v_top1) at (-1, 2.5) {\(10'_{Hi}\)};

    \diagram* {
      (i1) -- [fermion] (v_yuk1),
      (f1) -- [fermion] (v_yuk1),
      
      (m_pos1) -- [charged scalar, edge label'=\(\scriptstyle 10''_{Hi}\)] (v_yuk1),
      (m_pos1) -- [charged scalar] (v_top1),
    };

    \draw [line width=1.0pt] ($(m_pos1) + (-0.12, -0.12)$) -- ($(m_pos1) + (0.12, 0.12)$);
    \draw [line width=1.0pt] ($(m_pos1) + (-0.12, 0.12)$) -- ($(m_pos1) + (0.12, -0.12)$);
    \node [right=0.15cm of m_pos1] {\(\scriptstyle M_{10}\)};

    \node at (1.5, 1.1) {\Huge \(+\)};

    \vertex (i2) at (2.5, 0) {\(16_F\)};
    \vertex (f2) at (5.5, 0) {\(16_F\)};
    \vertex (v_yuk2) [dot] at (4, 0) {};
    \node [below=0.2cm of v_yuk2] {\(\scriptstyle \gamma_i\)};
    
    \vertex (v_mix2) [dot] at (4, 1.4) {};
    \vertex (h_field2) at (3, 2.5) {\(10_{Hj}\)};
    \vertex (a_field2) at (5, 2.5) {\(\langle A_{ij} \rangle\)};

    \diagram* {
      (i2) -- [fermion] (v_yuk2),
      (f2) -- [fermion] (v_yuk2),
     
      (v_mix2) -- [charged scalar, edge label'=\(\scriptstyle 10''_{Hi}\)] (v_yuk2),
      (v_mix2) -- [charged scalar] (h_field2),
      (v_mix2) -- [charged scalar] (a_field2),
    };

  \end{feynman}
\end{tikzpicture}
}
\caption{Generation of the effective operator (\ref{Y5D}) in a scalar exchange channel.}
\label{fig:ScalarExchange}
\end{figure}

Let us also comment on the flavor structure of the theory.   First, the elimination of unwanted minimal $SO(10)$-relations among the fermion masses can be achieved via mixing $16_F$-s with heavy $16_F^{'},\bar{16}_F^{'}$ superfields with their masses split by the  couplings to Higgs $45$-plets. For example, 
 \begin{equation} 
 W_{\rm F} =  g_{ab}16_F^{'a}\gamma_j 16_F^{b}\, 10^{''}_{Hj} + \tilde{g}_{ab}\bar{16}_F^{'a}\sigma_{ij} 16_F^{'b} \,B_{ji}\, + ... \,.
 \end{equation}
  Here we have explicitly introduced the generation indices 
  $a,b,=1,2,3$ and the corresponding coupling 
  matrices $g_{ab}, \tilde{g}_{ab}$. 
  Since the VEV of $B_{ij}$, given by 
  (\ref{SUSYVEVs}) breaks both $SU(2)_R$ as well as $SU(4)_c$ symmetries of the Pati-Salam group, the unwanted relations among the 
  Yukawa couplings of the electroweak 
  Higgs doublets residing in $10_{H}^{''}$ are eliminated.  Of course, as usual in GUTs, 
  due to the freedom 
  in theory parameters such as $g_{ab}, \tilde{g}_{ab}$ or the VEVs of the GUT Higgs fields, the masses of quarks and leptons cannot be predicted and instead must be treated as the emerging effective parameters.

\section{ Appendix D: Model \textrm{II}, Variations on a Theme} \label{appD}
  We wish to show that the framework of 
  decoupled color-triplet \cite{dvali1992}
 permits variations with additional symmetries that protect the smallness of the  Yukawa coupling of the $T$-particles beyond the order
 $\frac{m_{3/2}}{M_{\textrm{GUT}}}$. A particular importance of such realizations is that the $T$-particle appears to be stable on the timescales 
 longer than the age of the Universe.

 The heavy GUT Higgs sector is represented by the same content as Model I, with the difference that we exclude $54$-plet $\Sigma$. 
 We are thus left with $A_{ij}$ and $B_{ij}$ chiral superfields in 
 $45$ representations of $SO(10)$ GUT and a pair of $\bar{\chi}, \chi$ in $\bar{16}, 16$ representations. 
 
 The superpotential
 has the following form, 
\begin{equation} \label{W2GUT}
\begin{split}
    W= & \frac{M_A}{8} {\rm tr} A^2 + \frac{a}{16} {\rm tr} A^4 + \frac{b}{8} {\rm tr} A^2B^2 \, + 
    \\ 
    & \frac{M_B}{4} {\rm tr} B^2 + \frac{\beta}{2} B_{ij} \bar{\chi}\sigma_{ij}\chi + M_{\chi} \bar{\chi}\chi \, +  \\
& c \, {\rm tr} [A,B]^2 \,
\end{split}
\end{equation}
Here the parameters $M_A,M_B,M_{\chi} \sim M_{\rm GUT}$ have mass-dimensionalities, the 
parameter $\beta$ is dimensionless and
the parameters $a,b,c \sim M_{GUT}^{-1}$ have inverse-mass dimensionalities. The term in the last line represents a square of the commutator between $A_{ij}$ and $B_{ij}$. This term ensures that in the desired vacuum, there exist no extra $F$-flat or $D$-flat  
directions beyond the $SO(10)$ gauge orbit.

   This model possess a SUSY vacuum with the following VEVs: 
         
         \begin{itemize}
          \item $A_{ij}=\text{diag}(0, 0, 0, A\epsilon,A\epsilon)$ 
          \item  $B_{ij}=\text{diag}(\tilde{B}\epsilon, \tilde{B}\epsilon, \tilde{B}\epsilon, B\epsilon, B\epsilon)$ 
          \end{itemize}
          with $\epsilon \equiv \begin{pmatrix}
              0 & 1\\
              -1 & 0
          \end{pmatrix}$   
           and 
          \begin{itemize}
         \item $\chi_{\alpha}=\bar{\chi}_{\alpha}=\chi \vert +++++ \rangle$,  \\
         the  SU(5)-invariant components 
              \end{itemize}
         where, 
         \begin{equation} \label{VEVsM2}
\begin{split}
  A^2 & = \frac{1}{b} \left({\mathcal Y} + \frac{10}{3} M_B\right)
  \\
    B  & = \frac{2M_B M_{\chi}}{3\beta} \frac{1}{{\mathcal Y}}
\,, \\
    \tilde{B}  & = -\frac{M_{\chi}}{3\beta} \left (1 + 
     \frac{4}{3}M_B \frac{1}{{\mathcal Y}} \right ) 
\,, \\
  \chi^2  & = \frac{M_B}{\beta} \tilde{B} \,,
\end{split}
\end{equation}
 where the quantity ${\mathcal Y}$ is the solution of the following cubic equation:
         \begin{equation} 
{\mathcal Y}^3 \, \frac{a}{b} \, + \, {\mathcal Y}^2 \,\left(-M_A + \frac{10a}{3b}  M_B\right )\, 
+ \, b \, \left ( \frac{ 2 M_BM_{\chi}}{3\beta} \right )^2 = 0 \,.     
\end{equation}
Notice that there is no dependence of the parameter $c$ since the last term in (\ref{W2GUT}) as well as all its first derivatives vanish for the VEVs given by 
(\ref{VEVsM2}).

 It is easy to see that the 
 soft gravity-mediated supersymmetry breaking 
 induced no shift in the components 
 $A_{ij},~~i,j \in [1,2,...,6]$. In order to see this, let us parameterize the VEV of  
 $A_{ij}$ as  
 \begin{equation} \label{AtildeA}
 A_{ij}=\text{diag}(\tilde{A}\epsilon, \tilde{A}\epsilon, \tilde{A}\epsilon, A\epsilon, A\epsilon) \,,    
 \end{equation} 
 and evaluate the validity of the VEV  
 $\tilde{A} =0$ on the vacuum (\ref{VEVsM2})
 after taking into account the effects of gravity-mediated supersymmetry breaking. Of course, it is crucial that in global SUSY limit 
 the vacuum (\ref{VEVsM2}) supports no flat directions (inclusion of the commutator $[A,B]^2$ in the superpotential lifts all).
 In particular, the mass of the canonically normalized  
 superfield $\tilde{A}_c \equiv \sqrt{3} \tilde{A}$
 in this SUSY vacuum is 
\begin{equation}
 M_{\tilde{A}_c} = \frac{1}{3} \langle \frac{\partial^2 W}{\partial\tilde{A}^2}\rangle  = \frac{1}{2}\left(b\tilde{B}^2-M_A\right ) \sim M_{\rm GUT} \,,
\end{equation}
 where the VEV of $\tilde{B}$ is given by 
 (\ref{VEVsM2}) and we exclude any accidental fine tuning of the parameters that would lower $M_{\tilde{A}}$. 
 
 Correspondingly the shift of VEVs is at most $\sim m_{3/2}$, as imposed by the Goldstino argument.  
 However, for $\tilde{A}$ no shift takes place in this order. Indeed, according to Goldstino argument ~\cite{Dvali:2020uvd} the VEV of the $F$-term of any heavy superfield induced by gravity-mediated breaking 
 is bounded from above by $\sim m_{3/2}^2$.  
 For the $\tilde{A}_c$-supefield this gives
 \begin{equation}  \label{FtildeA}
   |F_{\tilde{A}}| = |M_{\tilde{A}_c} \tilde{A}_c  + m_{3/2} \tilde{A}_c^*|
   \lesssim m_{3/2}^2\;.
  \end{equation}
   Since $M_{\tilde{A}}\sim M_{\rm GUT}$, this excludes the shift of $\tilde{A}$ in order $m_{3/2}$. Instead, the maximal shift that can satisfy (\ref{FtildeA}) is 
  \begin{equation}  \label{Shift-tildeA}
  \tilde{A} 
   \lesssim \frac{m_{3/2}^2}{M_{\rm GUT}}\;.
  \end{equation}  

  This is fully confirmed by the explicit minimization of the effective potential, which shows that (at least at the tree-level) the shift is zero.  Indeed,  
 evaluating the potential for the scalar component $\tilde{A}$, we find that it has no tadpoles.  The bilinear term has the form, 
\begin{equation}  \label{FtildeA111}
 |M_{\tilde{A}_c} \tilde{A}_c  + m_{3/2} \tilde{A}_c^*|^2 +  c_{\rm soft}\,m_{3/2}\,  M_{\tilde{A}_c} \tilde{A}_c^2/2 \, + \, {\rm h.c.}\,,
  \end{equation}
 where we assumed minimal K\"ahler with  soft 
 parameter $c_{\rm soft} \sim 1$. It is obvious that the positive-definite mass term $|M_{\tilde{A}_c}|^2 \tilde{A}_c^*\tilde{A}_c$, which is of order $M_{\rm GUT}^2$, dominates over $m_{3/2}$-corrections and the VEV of $\tilde{A}$ is zero. 

 We must note that while the Goldstino argument 
 of~\cite{Dvali:2020uvd} which enforces the upper bound (\ref{shiftVEV}) of the heavy VEVs
 holds beyond a tree-level minimal K\"ahler, 
 the non-minimal corrections to the K\"ahler metric can induce the non-zero shift of  $\tilde{A}$ up to the order $\sim m_{3/2}$.
  For example, such a shift can be triggered by 
  the operators of the type $\Delta K = \frac{1}{\Lambda^2}|{\rm tr} (AB)|^2$.

  Modulo such non-minimal corrections, in this version of the theory, the coupling of
 the color-triplets to the Standard Model particles in order $\frac{m_{3/2}}{M_{\textrm{GUT}}}$ is absent. Correspondingly, the decay is absent 
 in the order $\frac{m_{3/2}^2}{M_{\textrm{GUT}}^2}$. 
 
 If the Yukawa coupling of the $T$-particle is generated in order $\frac{m_{3/2}^2}{M_{\textrm{GUT}}^2}$,
 the lifetime is increased by a factor $10^{26}$ as compared 
 to (\ref{lifetimeeq}), making it longer than 
 \begin{equation}
     \tau \sim 10^{26} {\rm sec} \,.
 \end{equation}
 This timescale exceeds the current age of the Universe by 
 $\sim 10^{9}$.

\newpage

\setlength{\bibsep}{5pt}

\bibliographystyle{utphys}
\bibliography{refs}

\end{document}